\documentclass[aps,prb,citeautoscript,twocolumn,longbibliography,superscriptaddress]{revtex4-1}

\usepackage[T1]{fontenc}
\usepackage{latexsym}
\usepackage{graphicx} 
\usepackage{epstopdf}
\usepackage{amsmath}  
\usepackage{amssymb}
\usepackage{comment}
\usepackage{cases}
\usepackage{lipsum}
\usepackage{float}
\usepackage{tikz}
\usepackage{makecell}
\usepackage{bbold}

\usepackage{simplewick}

\usepackage[breaklinks=true]{hyperref}

\begin{document}

\title{Universal collective modes from strong electronic correlations: Modified $1/\mathcal{N}_f$ theory with application to high-$T_c$ cuprates}

\author{Maciej Fidrysiak}
\email{maciej.fidrysiak@uj.edu.pl}
\affiliation{Institute of Theoretical Physics, Jagiellonian University, ul. {\L}ojasiewicza 11, 30-348 Krak{\'o}w, Poland }
\author{J{\'o}zef Spa{\l}ek}%
\email{jozef.spalek@uj.edu.pl}
\affiliation{Institute of Theoretical Physics, Jagiellonian University, ul. {\L}ojasiewicza 11, 30-348 Krak{\'o}w, Poland }

\begin{abstract}
A nonzero-temperature technique for strongly correlated electron lattice systems, combining elements of both variational wave function (VWF) approach and expansion in the inverse number of fermionic flavors ($1/\mathcal{N}_f$), is developed. The departure point, VWF method, goes beyond the renormalized mean-field theory and provides semi-quantitative description of principal equilibrium properties of high-$T_c$ superconducting cuprates. The developed here scheme of VWF+$1/\mathcal{N}_f$, in the leading order provides dynamical spin and charge responses around the VWF solution, generalizing the weak-coupling spin-fluctuation theory to the regime of strong correlations. Thermodynamic corrections to the correlated saddle-point state arise systematically at consecutive orders. Explicitly, VWF+$1/\mathcal{N}_f$ is applied to evaluate dynamical response functions for the hole-doped Hubbard model and compared with available determinant quantum-Monte-Carlo data, yielding a good overall agreement in the regime of coherent collective-mode dynamics. The emergence of well-defined spin and charge excitations from the incoherent continua is explicitly demonstrated and a non-monotonic dependence of the charge-excitation energy on the interaction magnitude is found. The charge-mode energy saturates slowly when approaching the strong-coupling limit, which calls for a reevaluation of the $t$-$J$-model approach to the charge dynamics in favor of more general $t$-$J$-$U$ and $t$-$J$-$U$-$V$ models. The results are also related to recent inelastic resonant $X$-ray and neutron scattering experiments for the high-$T_c$ cuprates.
\end{abstract}

\maketitle

\section{Introduction}
\label{section:introduction}

Strong electronic correlations in condensed-matter systems support formation of exotic states of matter, such as high-temperature superconducting and pseudogap phases in doped layered copper oxides or heavy-fermion superconducting phase in $4f$/$5f$-electron systems. The simplest theoretical frameworks to study these phenomena are based on the Hubbard, $t$-$J$, $t$-$J$-$U$, and Anderson-lattice models, with possible multi-orbital extensions. A number specific effects observed in high-$T_c$ superconductors and heavy-fermion systems have been satisfactorily interpreted using these models, within schemes specifically designed to incorporate the effects of \emph{strong local correlations}, such as variational wave function (VWF) approach (in diagrammatic\cite{MetznerPhysRevLett1989} or Monte-Carlo\cite{CeperleyPhysRevB1977} form), or dynamical mean-field theory (DMFT).\cite{GeorgesRevModPhys1996} However, new spectroscopic evidence\cite{DeanNatMater2013,IshiiNatCommun2014,LeeNatPhys2014,GuariseNatCommun2014,WakimotoPhysRevB2015,MinolaPhysRevLett2017,IvashkoPhysRevB2017,PengNatPhys2017,MeyersPhysRevB2017,ChaixPhysRevB2018,Robarts_arXiV_2019,ZhouNatCommun2013,GretarssonPhysRevLett2016,IshiiPhysRevB2017,HeptingNature2018,IshiiJPhysSocJapan2019,FumagalliPhysRevB2019,LeTaconNatPhys2011,JiaNatCommun2014,PengPhysRevB2018,LinNPJQuantMater2020,NagArxiv2020} suggests that frameworks, based solely on local-correlation effects, are usually insufficient to provide a satisfactory account of magnetic and charge dynamics in those materials. Namely, resonant inelastic $X$-ray scattering (RIXS) and inelastic neutron scattering (INS) experiments have revealed well defined and \emph{universal} magnetic excitations (paramagnons) in metallic phase of high-temperature (high-$T_c$) copper oxides,\cite{DeanNatMater2013,IshiiNatCommun2014,LeeNatPhys2014,GuariseNatCommun2014,WakimotoPhysRevB2015,MinolaPhysRevLett2017,IvashkoPhysRevB2017,PengNatPhys2017,MeyersPhysRevB2017,ChaixPhysRevB2018,Robarts_arXiV_2019} iron pnictides\cite{ZhouNatCommun2013}, and in iridates.\cite{GretarssonPhysRevLett2016} The universality of those collective modes relies on their appearance independently of the circumstance whether the corresponding system is in the broken-symmetry state or not. In the former case, they represent usual Goldstone bosons. RIXS has also provided a detailed account of previously overlooked high-energy discrete charge modes in the cuprates  that are currently under close scrutiny.\cite{IshiiPhysRevB2017,HeptingNature2018,IshiiJPhysSocJapan2019,FumagalliPhysRevB2019,LeTaconNatPhys2011,JiaNatCommun2014,PengPhysRevB2018,LinNPJQuantMater2020,NagArxiv2020}

Regarding that weak-coupling theory predicts a rapid overdamping of magnetic excitations in the metallic state, in disagreement with experiment,\cite{FidrysiakPhysRevB2020} an interplay between the long-wavelength collective modes and local correlations becomes the factor that must be taken into account in order to successfully reproduce collective dynamics. Since none of the approximation schemes, mentioned above, captures local-correlations and long-wavelength collective excitations \emph{on the same footing}, one is urged to resort to inherently unbiased techniques, such as determinant quantum Monte-Carlo (DQMC). However, DQMC is restricted to relatively small systems and suffers from the sign problem.\cite{LohPhysRevB1990} Moreover, extraction of real-time-dependent properties from imaginary-time DQMC results involves an ill-conditioned analytic continuation of numerical data,\cite{TripoltComPhysCommun2019} limiting its accuracy in regard to the dynamical effects, particularly at higher temperature. Remarkably though, comparative studies of the Blankenbecler-Scalapino-Sugar quantum Monte-Carlo, DMFT, and cluster DMFT phase diagrams for the two-dimensional Hubbard model indicate that long-range fluctuations (such as paramagnons) are crucial for a proper description of the metal-insulator transition, even in relatively small systems.\cite{SchaferPhysRevB2015} As a step towards understanding the dynamics of strongly correlated materials, development of alternative techniques, applicable in the thermodynamic limit and capable of describing local-correlations and collective-mode effects on the same footing, is thus highly needed. 

In this paper, we present a comprehensive theoretical framework which is appropriate for strongly correlated lattice electron systems and combines variational wave function (VWF) method with field-theoretical expansion in inverse number of fermionic flavors ($1/\mathcal{N}_f$), hereafter dubbed as VWF+$1/\mathcal{N}_f$ approach. In the leading order, the VWF+$1/\mathcal{N}_f$ method allows to study collective spin- and charge excitations around the correlated ground state and systematically incorporate fluctuation-induced corrections to thermodynamics at higher orders, suggesting a possibility of generalizing the Moriya-Hertz-Millis spin-fuctuation theory\cite{HertzPhysRevB1976,MoriyaBook1985,MillisPhysRevB1993,TakahashiBook2013} to the situation with strongly-correlated fermions. At the same time, VWF+$1/\mathcal{N}_f$ is relatively lightweight computationally and applicable to large systems ($> 10^5$ orbitals), and also to broken-symmetry states. The price paid for those advantages is the accuracy loss due to inevitable truncation of the $1/\mathcal{N}_f$ series and approximate diagrammatic treatment of variational wave functions. In a separate brief contribution,\cite{FidrysiakPhysRevB2020} we have already applied the simplified version of this method to selected hole-doped cuprates and achieved a satisfactory agreement with the experimentally obtained paramagnon spectra across the phase diagram. The charge-mode description requires still a refined analysis.

Both VWF and $1/\mathcal{N}_f$ approaches, as treated separately, have received considerable attention so that their applicability range and limitations are well understood. The former provides a good description of local properties and static correlations in wide range of doping and interaction strength, by going beyond the renormalized mean-field theory.\cite{ParamekantiPhysRevB2004,SpalekPhysRevB2017,ZegrodnikPhysRevB2017_2,FidrysiakJPhysCondensMatter2018,ZegrodnikPhysRevB2018_2,ZegrodnikPhysRevB2019} The latter allows for systematic evaluation of dynamical response functions and thermodynamic fluctuation-corrections in the weak- and intermediate-coupling regimes. At the strong coupling, the plain $1/\mathcal{N}_f$ expansion is severely limited by Fierz ambiguity,\cite{CastellaniPhysLettA1979} originating from multiple equivalent ways of carrying out the Hubbard-Stratonovich transformation of fermionic interaction vertices. This results in a problematic dependence of calculated phase diagrams on unphysical parameters.\cite{BaierPhysRevB2004} For common models, several physically motivated decoupling schemes have been developed over the years (see, e.g., Refs.~\citenum{SchulzPhysRevLett1990,JohnPhysRevB1991}), yet it is not clear how to systematically generalize them to arbitrary Hamiltonians.  As one of the elements of VWF+$1/\mathcal{N}_f$ implementation, we propose a conceptually different route to mitigate the Fierz problem by means of a specialized resummation of electronic interactions in the $1/\mathcal{N}_f$ series. Finally, the construction of the combined VWF+$1/\mathcal{N}_f$ is completed by carrying out a \textit{second resummation}, ensuring that the the large-$\mathcal{N}_f$ (saddle point) free energy coincides with that obtained using VWF approach. The procedure, outlined above, provides a systematic way to improve the plain VWF solution, as well as to calculate dynamical response functions to a desired accuracy.

After setting up the formalism, we apply the VWF+$1/\mathcal{N}_f$ approach to the hole-doped Hubbard model at the strong coupling ($U/|t| = 8$) and compare both static and dynamic spin and charge susceptibilities with available DQMC data. The static spin susceptibility profiles agree semi-quantitatively for the two techniques if an additional $\mathbf{k}$-independent renormalization factor $Z$ for large-$\mathcal{N}_f$ VWF+$1/\mathcal{N}_f$ susceptibilities is introduced, in a direct analogy to that invoked in linear spin-wave theory calculations. We find that $Z$ increases towards unity in the hole-overdoped case, indicating a gradual loss of multi-paramagnon scattering processes significance, as intuitively expected. The calculated energies of charge and paramagnon excitations match quantitatively those obtained by DQMC as long as the width of those modes is not exceedingly large. In the regime, where the collective excitations become highly incoherent, the VWF+$1/\mathcal{N}_f$ and DQMC peak-intensity energies depart progressively.

Finally, we demonstrate explicitly the gradual emergence coherent quasiparticles from the particle-hole continuum as the system evolves from the weakly-interacting Fermi liquid to strongly correlated metal with the increasing on-site Coulomb repulsion magnitude. This progressive development is of basic interest in the context of the robust paramagnon and charge modes observed in a variety of correlated compounds, but it is difficult to describe solely within $1/\mathcal{N}_f$ expansions developed previously.\cite{FoussatsPhysRevB2002} This is because those techniques are constructed based on Hubbard operators and thus intended for strongly-coupled ($t$-$J$-model\cite{ChaoJPCM1977,SpalekPRB1988}) limit. In effect, the impact of the finite-$U$ effects on charge dynamics may have not received an appropriate attention so far. By employing the particle-hole non-symmetric Hubbard model to reproduce the fermiology of high-$T_c$ superconductors, we find a sharp crossover behavior, manifesting itself as a non-monotonic dependence of charge mode energy on the magnitude of electron-electron interaction.  For overdoped system ($\delta = 20\%$), charge mode hardens with the increasing interaction for $U \lesssim 0.5 W$ (where $W$ is the bare bandwidth) and exhibits a gradual softening above this threshold. Additionally, in the crossover regime, a sharp peak in charge response emerges from the continuum as a consequence of single-particle bandwidth renormalization due to strong correlations. We also demonstrate that the charge mode energy undergoes large renormalization (by a factor of $\sim 2$) as the interaction increases from $U \sim W$ to $U = \infty$. The charge-mode dynamics thus turns out to be governed, to a large extent, by the finite-$U$ effects, which calls for a reconsideration of the strong-coupling ($t$-$J$ model limit) approaches to quantitative study of charge excitations in the high-$T_c$ cuprates in favor of more general $t$-$J$-$U$ and $t$-$J$-$U$-$V$ models.\cite{SpalekPhysRevB2017,ZegrodnikPhysRevB2017_2,ZegrodnikPhysRevB2019,ZegrodnikPhysRevB2020} This is one of the principal findings of the present and former\cite{FidrysiakPhysRevB2020} VWF+$1/\mathcal{N}_f$ analyses. The spin-excitation peak-intensity energy, on the other hand, decreases systematically as $U$ is increased, with a resonance-like feature building up on top of the continuum.

The paper is organized as follows. In Sec.~\ref{section:variational_approach} we overview the \textbf{D}iagrammatic \textbf{E}xpansion of the \textbf{G}utzwiller \textbf{W}ave \textbf{F}unction (DE-GWF) formulation of the VWF technique, that combines well with the $1/\mathcal{N}_f$ expansion and is applicable also to non-zero temperature situation. In Sec.~\ref{section:large_n_expansion} we propose a resummation scheme for $1/\mathcal{N}_f$ expansion, based on two-channel Hubbard-Stratonovich decoupling of the interaction term. In Sec.~\ref{section:variational+fluctuations} we construct hybrid VWF+$1/\mathcal{N}_f$ technique, unifying the developments of Secs.~\ref{section:variational_approach} and \ref{section:large_n_expansion} and encompassing the two source methods as its constituents. Essentially, in Secs.~\ref{section:variational_approach}-\ref{section:variational+fluctuations}, three expressions for the action are constructed in the order of increasing complexity, starting from the approximate variational wave function saddle point solution in Sec.~\ref{section:variational_approach}, going trough the modified weak-coupling version of $1/\mathcal{N}_f$ expansion as an intermediate step in Sec.~\ref{section:large_n_expansion}, and ending up with rigorous $1/\mathcal{N}_f$ series around the fully correlated variational state in Sec.~\ref{section:variational+fluctuations}. In Sec.~\ref{section:benchmark_Hubabrd_model} we apply the VWF+$1/\mathcal{N}_f$ to the Hubbard model with a variable doping level and compare calculated static- and dynamic susceptibilities with available DQMC data. In Sec.~\ref{section:robust_collective_excitations} we discuss the emergence of robust spin and charge excitations from the corresponding incoherent particle-hole continua. Finally, in Sec.~\ref{section:discussion} we provide summary and discussion. Technical details and supplementary analysis are shifted to Appendices~\ref{appendix:constraint_representation}-\ref{appendix:computational_aspects}.

\section{Variationally obtained state as the saddle-point solution}
\label{section:variational_approach}

The starting point of the VWF+$1/\mathcal{N}_f$ technique is the choice of variational wave function $|\Psi_\mathrm{var}\rangle \equiv \hat{P}(\boldsymbol{\lambda}) |\Psi_0\rangle$, where $\hat{P}$ is an operator dependent on the vector of parameters, $\boldsymbol{\lambda}$, and $|\Psi_0\rangle$ is an ``uncorrelated'' wave function that is defined as the ground state of effective tight-binding Hamiltonian with physically relevant correlations included, i.e., from $\hat{\mathcal{H}}_\mathrm{eff} |\Psi_0\rangle = E |\Psi_0\rangle$, and adjusted variationally by selecting $\hat{\mathcal{H}}_\mathrm{eff}$. The precise form of $\hat{\mathcal{H}}_\mathrm{eff}$ is not specified at this point as this effective single-particle Hamiltonian becomes a variationally adjusted object itself, see, e.g., Ref.~\citenum{KaczmarczykPhilMag2014}. Numerous physically motivated choices for $\hat{P}$ are possible, including those of Gutzwiller\cite{BiborskiPhysRevB2020,GutzwillerPhysRevLett1963,SpalekPhysRevB2017,BunemannEPL2012} and Jastrow\cite{JastrowPhysRev1955} types. In this manner, the complete description of the correlated state, based on $\hat{\mathcal{H}}$ and $|\Psi_\mathrm{var}\rangle$, contains, as intrinsic ingredient, also effective single-particle (``quasiparticle'') dynamics controlled by $\hat{\mathcal{H}}_\mathrm{eff}$ and $|\Psi_0\rangle$.\cite{KaczmarczykPhilMag2014,BunemannEPL2012,KaczmarczykPhysRevB2013,GebhardPhysRevB1990}

The plain VWF method reduces to minimization of the energy functional

\begin{align}
  \label{eq:variational_energy}
  E_\mathrm{var} \equiv \langle \hat{\mathcal{H}} \rangle_\mathrm{var} \equiv \frac{\langle\Psi_\mathrm{var}| \hat{\mathcal{H}} |\Psi_\mathrm{var}\rangle}{\langle\Psi_\mathrm{var}|\Psi_\mathrm{var}\rangle} \equiv \frac{\langle\Psi_0| \hat{P}\hat{\mathcal{H}}\hat{P} |\Psi_0\rangle}{\langle\Psi_0|\hat{P}^2|\Psi_0\rangle}
\end{align}

\noindent
with respect of both $\boldsymbol{\lambda}$ and $\hat{\mathcal{H}}_\mathrm{eff}$, under the constraint $\langle \hat{N}_e \rangle_\mathrm{var} \equiv N_e$. The operator $\hat{\mathcal{H}}$ is the model Hamiltonian, whereas $\hat{N}_e$ and $N_e$ denote particle number operator their total number, respectively. We have also introduced correlated variational averages, marked by the subscript ``$\mathrm{var}$''. With the use of Wick's theorem, this time in real space, for the uncorrelated $|\Psi_0\rangle$ state, both the numerator and denominator can be factorized in terms of two-point correlation functions (``lines'') of the form $P_{\alpha\beta} \equiv \langle \hat{c}^\dagger_{\alpha} \hat{c}_{\beta} \rangle_0 \equiv \langle \Psi_0| \hat{c}^\dagger_{\alpha} \hat{c}_{\beta} | \Psi_0 \rangle$, where  $\alpha, \beta$ are indices combining lattice-site position and local degrees of freedom (e.g., spin and/or orbital). Note that, in the superconducting (paired) state, additional off-diagonal lines $S_{\alpha\beta} \equiv \langle\hat{c}_\alpha^\dagger \hat{c}_\beta^\dagger\rangle_0$ appear. A vector composed of all the emerging lines will be hereafter denoted as $\mathbf{P}$. In turn, the energy functional~\eqref{eq:variational_energy} depends on line- and correlator-parameter vectors, i.e., $E_\mathrm{var} = E_\mathrm{var}(\mathbf{P}, \boldsymbol{\lambda})$. The variational problem reduces to minimization of $E_\mathrm{var}$ over $\mathbf{P}$ and $\boldsymbol{\lambda}$ under the constraints $P_{\alpha\beta} \equiv \langle \Psi_0| \hat{c}^\dagger_\alpha \hat{c}_\beta | \Psi_0 \rangle$ and $\langle \hat{N}_e \rangle_\mathrm{var} \equiv N_e$. As we show below, for computational purposes, it is useful to restrict further the correlator variational space by means of a vector composing the constraints, $\mathbf{C}(\mathbf{P}, \boldsymbol{\lambda}) \equiv \mathbf{0}$ which will be concretized in model situation at hand.

The plain WVF method, outlined above, is a zero-temperature formalism that cannot account for the thermal effects and phase transitions at temperature $T > 0$. In the following we employ its more refined  finite-temperature extension that reduces to the plain WVF solution in the $T \rightarrow 0$ limit. The starting point of the improved approach is the generalized Landau-type functional taken in the form

\begin{widetext}
\begin{align}
  \label{eq:landau_functional_variational_method}
  \mathcal{F}(\boldsymbol{\lambda}, \mathbf{P}, \boldsymbol{\xi}, \boldsymbol{\rho}, \mu) = - \frac{1}{\beta} \cdot \ln \mathrm{Tr} \exp\left(-\beta E_\mathrm{var}(\mathbf{P}, \boldsymbol{\lambda}) + i \beta \sum_{\alpha\beta} {\xi}_{\alpha\beta}^{*} (P_{\alpha\beta} - \hat{c}^\dagger_\alpha \hat{c}_\beta) + i \beta \boldsymbol{\rho}^T \cdot \mathbf{C}(\mathbf{P}, \boldsymbol{\lambda}) + \beta \mu \left(\sum_\alpha P_{\alpha\alpha} - N_e\right)\right), 
\end{align}

\end{widetext}

\noindent
where $\boldsymbol{\xi}$ and $\boldsymbol{\rho}$ are (column) vectors composed of Lagrange multipliers to be defined subsequently, trace is taken over the electronic degrees of freedom, and $\beta \equiv (k_B T)^{-1}$ is the inverse temperature. The chemical  potential, $\mu$, ensures that the total number of electrons in the system equals to $N_e$. The first term represents the functional~\eqref{eq:variational_energy}, here expressed via $\mathbf{P}$ and $\boldsymbol{\lambda}$. The second term enforces the constraint that the variational optimization of $P_{\alpha\beta}$ and their self-consistent calculation in the $|\Psi_0\rangle$ state coincide. The third represents extra constraints to be defined, and the last is the condition for $\mu$. Note that this functional is defined for non-paired phases ($S_{\alpha\beta} \equiv 0$).

The whole methodology of the approach is as follows. We start from the model Hamiltonian $\hat{\mathcal{H}}$ (Hubbard, $t$-$J$, $t$-$J$-$U$, etc.) so that $E_\mathrm{var}$ is given by Eq.~\eqref{eq:variational_energy}.\cite{KaczmarczykPhysRevB2013,SpalekPhysRevB2017} To assure that the quantities calculated variationally coincide with those computed self-consistently, i.e., the Bogoliubov theorem is fulfilled (as discussed earlier, see Ref.~\cite{JedrakPhsRevB2011}), we supplement $E_\mathrm{var}$ determination with the Lagrange constraints (the following three terms). The Langrange multipliers are to be determined self-consistently (cf. also Appendix~\ref{appendix:constraint_representation}). Finally, the nonzero temperature single-particle fluctuations are incorporated by means of the trace on the right-hand-side of Eq.~\eqref{eq:landau_functional_variational_method}, as taking $\mathrm{Tr} \exp$ means averaging over the excited states. The physical free energy is determined as the saddle-point of the functional \eqref{eq:landau_functional_variational_method}, leading to the set of self-consistent equations

\begin{align}
  & i \xi_{\alpha\beta}^{*} =  \frac{\partial E_G(\mathbf{P}, \boldsymbol{\lambda})}{P_{\alpha\beta}} - i \boldsymbol{\rho}^T \cdot \frac{\partial \mathbf{C} (\mathbf{P}, \boldsymbol{\lambda})}{\partial P_{\alpha\beta}} - \mu \delta_{\alpha\beta}, \label{eq:xi_eq} \\
  & \mathbf{C}(\mathbf{P}, \boldsymbol{\lambda}) = \mathbf{0}, \\
  & P_{\alpha\beta} = \frac{\mathrm{Tr} \hat{c}^\dagger_\alpha \hat{c}_\beta \exp(-\beta \hat{\mathcal{H}}_\mathrm{eff})}{\mathrm{Tr} \exp(-\beta \hat{\mathcal{H}}_\mathrm{eff})},\\
  & \hat{\mathcal{H}}_\mathrm{eff} = \sum_{\alpha\beta} i \xi^{*}_{\alpha\beta} \hat{c}^\dagger_\alpha \hat{c}_\beta, \label{eq:heff}\\
  & \sum_\alpha P_{\alpha\alpha} = N_e. \label{eq:n_tot_eq}
\end{align}

\noindent
The free energy may be then written as

\begin{align}
  \label{eq:free_energy_variational}
  F = & -\frac{1}{\beta} \ln \mathcal{F}(\boldsymbol{\lambda}, \mathbf{P}, \boldsymbol{\xi}, \boldsymbol{\rho}, \mu)|_\text{at saddle point} = \nonumber \\ & E_G(\mathbf{P}, \boldsymbol{\lambda}) - \sum_{\alpha\beta} \hat{\mathcal{H}}_\mathrm{eff}^{\alpha\beta} P_{\alpha\beta} - \frac{1}{\beta} \ln\mathrm{Tr} \exp(-\hat{\mathcal{H}}_\mathrm{eff}) \equiv \nonumber \\ &
E_G(\mathbf{P}, \boldsymbol{\lambda}) - T S,
\end{align}

\noindent
where all the variables acquire the saddle-point values defined by Eqs.~\eqref{eq:xi_eq}-\eqref{eq:n_tot_eq}, and the entropy reads

\begin{align}
S = -k_B \sum_l n_l \ln n_l - k_B \sum_l (1 - n_l) \ln (1 - n_l),\label{eq:9}
\end{align}

\noindent
where  indices $l$ runs over the complete set of eigenvalues of $\hat{\mathcal{H}}_\mathrm{eff}$. At $T = 0$, the free energy~\eqref{eq:free_energy_variational} reduces thus to the minimum of the variational energy, as stated above. It is essential to note that eigenenergies are obtained by diagonalization of the effective \emph{single-particle} Hamiltonian $\hat{\mathcal{H}}_\mathrm{eff}$, defined by Eq.~\eqref{eq:heff}, which describes the interaction-renormalized Landau-type quasiparticles. For a detailed discussion of $\hat{\mathcal{H}}_\mathrm{eff}$ as the object controlling effective quasiparticle dynamics, see Refs.~\citenum{BunemannPhysRevB2003}~and \citenum{FidrysiakJPhysCondensMatter2018}. The structure of the free energy thus implies that the variational technique, formulated above, incorporates single-particle excitations around the correlated ground-state, but lacks the collective mode contribution to thermodynamics. The aim of the next two sections is to provide an extension of the variational scheme which becomes capable of describing those excitations (fluctuations).

We point out that various distinct schemes for finite-temperature extensions of the variational approach (of the Gutzwiller type) have been proposed previously,\cite{WangPhysRevB2010,LanataPhysRevB2015} providing a way to incorporate effects of correlations on single-particle entropy that are of importance at extremely high temperatures ($k_BT$ of the order of half-single-particle bandwidth, $W/2$). Those techniques should thus constitute a better approximation to the exact free energy than the finite-temperature method of Eq.~\eqref{eq:landau_functional_variational_method} at $k_BT \sim W/2$, but yield a negative value of the entropy in the $T \rightarrow 0$ limit, as well as do not incorporate collective-fluctuation effects. As we demonstrate below by comparing our results with DQMC simulations, Eq.~\eqref{eq:landau_functional_variational_method}, supplemented with the fluctuation-corrections, provides a reliable approximation in the regime of moderate temperatures, $k_BT \ll W/2$.

\subsection{Linked cluster expansion of the correlated energy functional}

The main difficulty involved in the variational procedure is evaluation of the correlated energy functional $E_\mathrm{var}(\mathbf{P}, \boldsymbol{\lambda})$. In the following we restrict ourselves to the intra-orbital correlator of the form $\hat{P} = \hat{P}_G \equiv \prod_{il} \hat{P}_{Gil} = \prod_I \hat{P}_{G, I}$, for which efficient computational schemes have been developed.\cite{BunemannEPL2012} Here ``$i$'' and ``$l$'' run over lattice-site positions and orbital indices, respectively. For brevity, we have introduced also joint position and orbital index, $I \equiv (i, l)$. The correlated expectation values, evaluated using $\hat{P}_G$, are labeled with by the subscript ``$G$''. To make sure that the VWF solution interfaces well with the fluctuation extensions introduced in Sec.~\ref{section:variational+fluctuations}, already at this stage one needs to ensure that the employed correlator is compatible with the spin rotational symmetry. In particular, spin quantization axis cannot be arbitrarily fixed as this would discard the spin precession associated with the magnetic excitations (spin-waves/paramagnons) that are operative at low-energies. Violating this symmetry requirement at the correlator level typically results in breakdown of the Goldstone's theorem in ordered states. We thus postulate $\hat{P}_{G, I}$ in the spin-rotationally-invariant form, i.e., take

\begin{align}
  \label{eq:GutzwIler_correlator}
  \hat{P}_{G, I} \equiv& \lambda_{I0} \lvert 0\rangle_{I}{}_{I}\langle 0\rvert + \lambda_{I\uparrow\uparrow} \lvert\uparrow\rangle_{I}{}_{I}\langle \uparrow\rvert + \lambda_{I\downarrow\downarrow} \lvert\downarrow\rangle_{I}{}_{I}\langle \downarrow\rvert + \nonumber\\& \lambda_{I\uparrow\downarrow} \lvert\uparrow\rangle_{I}{}_{I}\langle \downarrow\rvert + \lambda_{I\downarrow\uparrow} \lvert\downarrow\rangle_{I}{}_{I}\langle \uparrow\rvert + \lambda_{Id} \lvert\uparrow\downarrow\rangle_{I}{}_{I}\langle \uparrow\downarrow\rvert.
\end{align}

\noindent
Furthermore, we require that $\hat{P}_{G, I}$ is Hermitian, which implies that $\lambda_{I\uparrow\downarrow} = \lambda^{*}_{I\downarrow\uparrow}$. The terms $\propto \lambda_{I\uparrow\downarrow}$ and $\propto \lambda_{I\downarrow\uparrow}$ are necessary for the rotational symmetry preservation and are usually discarded in static variational calculations for paramagnetic and magnetic states with specified global spin quantization axis.  In the correlator \eqref{eq:GutzwIler_correlator} we have disregarded doublon-holon creation and annihilation terms ($\propto \lvert 0 \rangle_{I} {}_{I}\langle \uparrow\downarrow \rvert$ and $\propto \lvert \uparrow\downarrow \rangle_{I} {}_{I} \langle 0 \rvert$) so that there are in total six independent correlator-parameters per orbital.

The problem of evaluating the correlated expectation value~\eqref{eq:variational_energy} simplifies substantially (cf. Ref.~\citenum{BunemannEPL2012} and Appendix~\ref{appendix:linked_cluster_theorem}) if subsidiary condition

\begin{align}
  \label{eq:contraints_pg2_definition}
  \hat{P}_{G, I}^2 \equiv 1 + x_{I} \hat{d}^{\prime}_{I},
\end{align}

\noindent
is introduced, where

\begin{align}
  \label{eq:contraints_pg2}
  \hat{d}^{\prime}_{I} \equiv & \hat{d}_I - {n}_{I\uparrow} \hat{n}_{I\downarrow} - \hat{n}_{I\uparrow} {n}_{I\downarrow} + S^+_I \hat{S}_I^- + S_I^- \hat{S}_I^++ \nonumber \\ &  n_{I\uparrow} n_{I\downarrow} - S_I^+ S_I^-
\end{align}

\noindent
is the modified double-occupancy operator with $\hat{d}_I \equiv \hat{n}_{I\uparrow} \hat{n}_{I\downarrow}$, $n_{I\sigma} \equiv \langle \Psi_0 | \hat{n}_{I\sigma} | \Psi_0\rangle$,  $S_I^+ \equiv \langle \Psi_0 | \hat{c}^\dagger_{I\uparrow} \hat{c}_{I\downarrow} | \Psi_0\rangle$, $S_I^- \equiv S_I^{+*}$, and $x_I$ is another variational parameter. Requirement~\eqref{eq:contraints_pg2_definition}, along with the definitions~\eqref{eq:GutzwIler_correlator} and \eqref{eq:contraints_pg2}, imposes five conditions for six $\boldsymbol{\lambda}$ variables, leaving out only one free parameter per orbital,

\begin{align}
  \label{eq:x_par_definition}
  x_{I} \equiv \frac{\lambda_{I0}^2 - 1}{n_{I\uparrow} n_{I\downarrow} - S_I^+ S_I^-}.
\end{align}

\noindent
Instead of working directly with the condition \eqref{eq:contraints_pg2_definition}, hereafter we employ the five constraints of the form

\begin{align}
  C^1_{I} \equiv & \langle \hat{P}_{G, I}^2 \rangle_0 \equiv 1, \label{eq:c1}\\
  C^2_{I} \equiv & \langle \hat{P}_{G, I} \hat{c}^\dagger_{I\uparrow}\hat{c}_{I\uparrow} \hat{P}_{G, I} \rangle_0 - \langle \hat{c}^\dagger_{I\uparrow}\hat{c}_{I\uparrow}\rangle_0 \equiv 0, \label{eq:c2}\\
  C^3_{I} \equiv & \langle \hat{P}_{G, I} \hat{c}^\dagger_{I\downarrow}\hat{c}_{I\downarrow} \hat{P}_{G, I} \rangle_0 - \langle \hat{c}^\dagger_{I\downarrow}\hat{c}_{I\downarrow}  \rangle_0 \equiv 0, \label{eq:c3}\\
  C^4_{I} \equiv & \mathrm{Re} (\langle \hat{P}_{G, I} \hat{c}^\dagger_{I\uparrow}\hat{c}_{I\downarrow} \hat{P}_{G, I} \rangle_0 - \langle \hat{c}^\dagger_{I\uparrow}\hat{c}_{I\downarrow}  \rangle_0 ) \equiv 0, \label{eq:c4}\\
  C^5_{I} \equiv & \mathrm{Im} (\langle \hat{P}_{G, I} \hat{c}^\dagger_{I\uparrow}\hat{c}_{I\downarrow} \hat{P}_{G, I} \rangle_0 - \langle \hat{c}^\dagger_{I\uparrow}\hat{c}_{I\downarrow}  \rangle_0 ) \equiv 0, \label{eq:c5}
\end{align}

\noindent
which may be given a transparent physical interpretation (cf. Secs.~\ref{subsec:local_approximation} and \ref{subsec:gutzwiller_approximation} below) and imply Eq.~\eqref{eq:contraints_pg2_definition}, as discussed in Appendix~\ref{appendix:constraint_representation}.

If we restrict ourselves to one- and two-site contributions in the Hamiltonian, evaluation of the correlated energy functional $E_G$ reduces to computing of the two kinds of averages: $\langle \hat{A}_{I} \rangle_G$ and $\langle \hat{A}_{I} \hat{B}_{J} \rangle_G$, where $\hat{A}_{I}$ and $\hat{B}_{J}$ are operators acting on orbitals $I$ and $J$, respectively ($I \neq J$ is assumed). A specialized version of linked-cluster theorem (cf. Appendix~\ref{appendix:linked_cluster_theorem}) may be used to efficiently evaluate them. One obtains

\begin{widetext}
\begin{align}
  \langle \hat{A}_I \rangle_G &= \langle{\hat{A}_I^{\prime}}\rangle_0 + \sum\limits_{k=0}^\infty \sideset{}{'}\sum_{J_1 \ldots J_k} \frac{x_{J_1} \ldots x_{J_k}}{k!} \langle \left(\hat{A}^{\prime}_{I} - \langle{\hat{A}_I^{\prime}}\rangle_0 \hat{P}_{G, I}^2 \right) \cdot \hat{d}^\prime_{J_1} \ldots \hat{d}^\prime_{J_k}\rangle_0^c, \label{eq:diagrammatics_one_site}
\end{align}

\noindent
and

\begin{align}
  \langle \hat{A}_I \hat{B}_J \rangle_G = & \langle{\hat{A}_I^{\prime}}\rangle_0 \langle{\hat{B}_J^{\prime}}\rangle_0 + \sum\limits_{k=0}^\infty \sideset{}{'}\sum_{J_1 \ldots J_k} \frac{x_{J_1} \ldots x_{J_k}}{k!} \langle \left(\hat{A}^{\prime}_{I} - \langle{\hat{A}_I^{\prime}}\rangle_0 \hat{P}_{G, I}^2 \right) \cdot \left(\hat{B}^{\prime}_{J} - \langle{\hat{B}_J^{\prime}}\rangle_0 \hat{P}_{G, J}^2 \right) \cdot \hat{d}^\prime_{J_1} \ldots \hat{d}^\prime_{J_k}\rangle_0^c + \nonumber \\ &
\langle{\hat{A}_I^{\prime}}\rangle_0 \cdot \left(\sum\limits_{k=0}^\infty \sideset{}{'}\sum_{J_1 \ldots J_k} \frac{x_{J_1} \ldots x_{J_k}}{k!} \langle \left(\hat{B}^{\prime}_{I} - \langle{\hat{B}_J^{\prime}}\rangle_0 \hat{P}_{G, I}^2 \right) \cdot \hat{d}^\prime_{J_1} \ldots \hat{d}^\prime_{J_k}\rangle_0^c \right) + \nonumber \\ &  \left(\sum\limits_{k=0}^\infty \sideset{}{'}\sum_{J_1 \ldots J_k} \frac{x_{J_1} \ldots x_{J_k}}{k!} \langle \left(\hat{A}^{\prime}_{I} - \langle{\hat{A}_I^{\prime}}\rangle_0 \hat{P}_{G, I}^2 \right) \cdot \hat{d}^\prime_{J_1} \ldots \hat{d}^\prime_{J_k}\rangle_0^c \right) \cdot \langle{\hat{B}_J^{\prime}}\rangle_0, \label{eq:diagrammatics_two_site}
\end{align}
\end{widetext}

\noindent
where the superscript ``$c$'' means that only the diagrams, in which all operators entering $\hat{d}^\prime_{J_\alpha}$  are \emph{connected} to external vertices at positions $I/J$, are retained. We have defined ``primed'', locally-correlated operators $\hat{A}_I^\prime \equiv \hat{P}_{G, I} \hat{A}_I \hat{P}_{G, I}$  and $\hat{B}_I^\prime \equiv \hat{P}_{G, I} \hat{B}_I \hat{P}_{G, I}$. Primed summations indicate that $J_1, \ldots, J_n$ are restricted to be all different and also distinct from the external indices, $I$ and $J$.

Equations~\eqref{eq:diagrammatics_one_site}-\eqref{eq:diagrammatics_two_site} constitute the basis for the \textbf{D}iagrammatic \textbf{E}xpansion of the \textbf{G}utzwiller \textbf{W}ave \textbf{F}unction method (DE-GWF), developed elsewhere,\cite{BunemannEPL2012,KaczmarczykNewJPhys2014} here extended to the spin-rotationally invariant form. Since the variational state serves only as a saddle-point solution within our expansion scheme, we restrict to two basic approximations detailed below. Incorporation of high-order diagrammatic terms should be discussed separately.

\subsubsection{Local-diagram approximation}
\label{subsec:local_approximation}

Within the local-diagram (LD) approximation, only the operators, acting on the same sites as the terms of the original Hamiltonian, are retained. Disregarding all non-local contributions allows us to evaluate all expectation values contributing to the variational energy, $E_\mathrm{var}$,  in a closed form. Namely, one arrives at

\begin{align}
  \langle \hat{A}_I \rangle_G \approx & \langle \hat{A}_I^\prime  \rangle_0, \label{eq:local_approx_one_site} \\
  \langle \hat{A}_I \hat{B}_J \rangle_G \approx & \langle \hat{A}^\prime_I \rangle_0 \langle \hat{B}^\prime_I \rangle_0  + \nonumber\\& \langle (\hat{A}^\prime_I - \langle \hat{A}^\prime_0 \rangle \hat{P}_{G, I}^2) (\hat{B}^\prime_J - \langle \hat{B}^\prime_0 \rangle \hat{P}_{G, J}^2) \rangle_0. \label{eq:local_approx_two_site}
\end{align}

\noindent
If $\langle \hat{A}^\prime_I \rangle_0 = \langle \hat{B}^\prime_I \rangle_0 = 0$ for the two-site interaction terms, Eq.~\eqref{eq:local_approx_two_site} simplifies further to $\langle \hat{A}_I \hat{B}_J \rangle_G \approx \langle \hat{A}_I^\prime \hat{B}_J^\prime \rangle_0$. This holds for the paramagnetic solution for the Hubbard model, considered explicitly in the present contribution. Importantly, within the LD approximation, $\langle \hat{c}^\dagger_{i\sigma} \hat{c}_{i\sigma^\prime} \rangle_G = \langle \hat{c}^\dagger_{i\sigma} \hat{c}_{i\sigma^\prime} \rangle_0$ as a consequence of the constraints \eqref{eq:c2}-\eqref{eq:c5}. This means that the correlations \emph{do not renormalize} local single-particle operators, such as particle-number operator or Zeemann term. The latter useful property is generally invalidated by higher-order diagrammatic contributions.

\subsubsection{$d = \infty$ (Gutzwiller) approximation}
\label{subsec:gutzwiller_approximation}

An approximation, providing expressions simpler then those given by LD approximation [cf. Eqs.~\eqref{eq:local_approx_one_site}-\eqref{eq:local_approx_two_site}], is based on the formal assumption of large spatial dimensionality $d \rightarrow \infty$ so that $1/d$ plays the role of a small parameter.  For a single-orbital model, such as the Hubbard model, it can be argued\cite{GebhardPhysRevB1990} that $d = \infty$ condition is implemented by adopting the following simplifications

\begin{align}
  \langle \hat{A}_I \rangle_G \approx & \langle \hat{P}_{G, I} \hat{A}_I \hat{P}_{G, I} \rangle_0, \label{eq:gutzwiller_approx_one_site} \\
  \langle \hat{c}^\dagger_{I\sigma} \hat{c}_{J\sigma^\prime} \rangle_G \approx & \langle \hat{P}_{G, I} \hat{c}^\dagger_{I\sigma} \hat{P}_{G, I} \hat{P}_{G, J} \hat{c}_{J\sigma^\prime} \hat{P}_{G, J} \rangle_0^\text{one line}, \label{eq:gutzwiller_approx_cc} \\
  \langle \hat{\mathbf{S}}_{I} \hat{\mathbf{S}}_{J} \rangle_G \approx & \langle \hat{P}_{G, I} \hat{\mathbf{S}}_{I} \hat{P}_{G, I} \hat{P}_{G, J} \hat{\mathbf{S}}_{J} \hat{P}_{G, J} \rangle_0^\text{two lines}, \label{eq:gutzwiller_approx_SS} \\
    \langle \hat{n}_{I} \hat{n}_{J} \rangle_G \approx & \langle \hat{P}_{G, I} \hat{n}_{I} \hat{P}_{G, I} \hat{P}_{G, J} \hat{n}_{J} \hat{P}_{G, J} \rangle_0^\text{two lines}, \label{eq:gutzwiller_approx_nn}
\end{align}

\noindent
where we list explicitly only selected expectation values of interest. The subscripts ``one line'' and ``two lines'' mean that one should retain only the diagrams containing no more than one and two non-local lines, respectively (non-local are those connecting sites $I$ and $J$). The $d = \infty$ approximation may be thus viewed as a truncated version of the local-diagram calculation, where the diagrams with the largest number of loops are discarded. Gutzwiller approximation inherits thus the property $\langle \hat{c}^\dagger_{i\sigma} \hat{c}_{i\sigma^\prime} \rangle_G = \langle \hat{c}^\dagger_{i\sigma} \hat{c}_{i\sigma^\prime} \rangle_0$.

\section{Fierz ambiguity and resummed $1/\mathcal{N}_f$ expansion}
\label{section:large_n_expansion}

Closed-form approximations for the free energy and dynamical structure factors for the collective modes may be obtained by extending the number of fermionic flavors from one to $\mathcal{N}_f \gg 1$ and treating $1/\mathcal{N}_f$ formally as a small parameter. This is effectively implemented by Hubbard-Stratonovich (HS) decoupling of the fermionic interaction Hamiltonian in terms of auxiliary fields. The latter procedure is, however, the source of notorious Fierz-ambiguity problem (cf. Sec.~\ref{section:introduction}), which in the leading expansion order may also result in violating the symmetry-related properties, such as the Goldstone's theorem. The former issue may be, to some extent, mitigated by inclusion of subleading terms in $1/\mathcal{N}_f$ expansion or via renormalization-group procedure.\cite{JaeckelPhysRevD2003} The latter, on the other hand, may be cured by a careful selection of the decoupling scheme or applying specialized symmetrization procedures.\cite{SchulzPhysRevLett1990} Importantly, by properly selecting the exact form of the standard Hubbard-Stratonovich decoupling, Gaussian-order fluctuations can be brought to the form consistent with the weak-coupling random-phase-approximation (RPA) results in selected scattering channels, but at the cost of disrupting the agreement in the other. An explicit example of the spin-rotationally-symmetric Hubbard-Stratonovich decoupling that leads to correct weak-coupling behavior of spin susceptibilities, but yields unphysical degeneracy between the energies of spin- and charge modes, is discussed in Ref.~\citenum{JohnPhysRevB1991}. Also, alternative ways to avoid the Fierz ambiguity have been proposed very recently in the context of the dynamical mean-field theory extension, and are based either on cluster calculations\cite{AyralPhysRevLett2017} or multi-channel decouplings.\cite{StepanovPhysRevB2019} In this section, we develop a variant of the Hubbard-Stratonovich transformation which results in an unbiased $1/\mathcal{N}_f$ expansion and, at weak-coupling, reduces to the RPA results in \emph{all} particle-hole scattering channels at the same time. As is shown in Sec.~\ref{section:variational+fluctuations}, this decoupling may also be naturally extended to the regime of strong correlations.

\subsection{Unbiased Hubbard-Stratonovich-type transformation}

We consider a general normal-ordered Hamiltonian with four-fermion interactions only, i.e.,

\begin{align}
  \label{eq:general_hamiltonian}
  \hat{\mathcal{H}} = \hat{T} + \hat{V} = &\sum_{\alpha\beta} (t_{\alpha\beta} - \delta_{\alpha\beta} \mu) \hat{c}^\dagger_\alpha \hat{c}_\beta + \nonumber\\&  \frac{1}{2} \sum_{\alpha\beta\gamma\rho} V_{\alpha\beta\gamma\rho} \hat{c}^\dagger_\alpha \hat{c}^\dagger_\beta \hat{c}_\rho \hat{c}_\gamma,
\end{align}

\noindent
where  $\hat{T}$ and $\hat{V}$ are the kinetic and interaction energies, respectively. The Greek indices $\alpha, \beta, \gamma, \rho$ accommodate all the degrees of freedom for given model (lattice, spin, orbital, etc.), and $\mu$ is chemical potential. By hermiticity, the matrix elements fulfill the conditions $t_{\alpha\beta} = t_{\beta\alpha}^{*}$ and $V_{\alpha\beta\gamma\rho} = V_{\gamma\rho\alpha\beta}^{*}$.

We define \emph{uncorrelated} kinetic and potential energies in the form

\begin{align}
  \label{eq:wick_decomposition_functionals}
  E_{0, \mathrm{kin}} &\equiv \langle \hat{T} \rangle_0 =  \sum_{\alpha\beta} (t_{\alpha\beta} - \mu \delta_{\alpha\beta}) \langle{\hat{c}^\dagger_\alpha \hat{c}_\beta}\rangle_0
\end{align}

\noindent
and

\begin{align}
  \label{eq:e0int}
  E_{0, \mathrm{int}} & \equiv \langle \hat{V}\rangle_0 = \nonumber\\& \frac{1}{2} \sum_{\alpha\beta\gamma\rho} V_{\alpha\beta\gamma\rho} \left( \langle{\hat{c}^\dagger_\alpha \hat{c}_\gamma}\rangle_0 \langle{\hat{c}^\dagger_\beta \hat{c}_\rho}\rangle_0 - \langle{\hat{c}^\dagger_\alpha \hat{c}_\rho}\rangle_0 \langle{\hat{c}^\dagger_\beta \hat{c}_\gamma}\rangle_0 \right) =  \nonumber\\&\frac{1}{2} \sum_{\alpha\beta\gamma\rho} \langle{\hat{c}^\dagger_\alpha \hat{c}_\beta}\rangle_0^{*} \mathcal{V}_{\alpha\beta\gamma\rho} \langle{\hat{c}^\dagger_\gamma \hat{c}_\rho}\rangle_0,
\end{align}

\noindent
respectively. The symmetrized vertex $\mathcal{V}_{\alpha\beta\gamma\rho} \equiv (-V_{\beta\gamma\rho\alpha} + V_{\beta\gamma\alpha\rho} + V_{\gamma\beta\rho\alpha} - V_{\gamma\beta\alpha\rho})/2$ satisfies the  condition  $(\mathcal{V}_{\gamma\rho\alpha\beta})^{*} = \mathcal{V}_{\alpha\beta\gamma\rho}$. Alternatively, the interaction matrix may be written in a manifestly hermitian form $\mathcal{V}_{\alpha\beta\gamma\rho} = \partial_{P^*_{\alpha\beta}} \partial_{P_{\gamma\rho}} E_0$. It is also useful to define the total uncorrelated energy functional $E_0 \equiv E_{0, \mathrm{kin}} + E_{0, \mathrm{int}}$. The subscript ``0'' means that the averages are evaluated in an uncorrelated (Hartree-Fock-type) state $| \Psi_0 \rangle$ so that Wick's theorem holds. The fact that $T = 0$ averages are used to define $E_{0, \mathrm{kin}}$, $E_{0, \mathrm{int}}$, and $E_0$ may suggest that the resulting approach is limited to zero temperature, which is not the case. Below we demonstrate that VWF+$1/\mathcal{N}_f$ is applicable at $T > 0$ as well.

We now propose the following transformation

\begin{align}
  \label{eq:hubbard_stratonovich}
  &\exp\left( -\int d\tau \hat{\mathcal{H}}(\bar{\boldsymbol{\eta}}, \boldsymbol{\eta}, \mu) \right) \propto \nonumber \\ &\lim_{\epsilon\rightarrow 0^+}\int \mathcal{D}\mathbf{P} \mathcal{D} \boldsymbol{\xi} \exp\Bigg( - \int d\tau E_{0, \mathrm{kin}}(\mathbf{P}, \mu) - \nonumber \\ & \frac{1}{2} \int d\tau E_{0, \mathrm{int}}(\mathbf{P}) - i \boldsymbol{\xi}^\dagger(\hat{\mathbf{P}} - \mathbf{P})  -\frac{\epsilon}{2} \int d\tau \boldsymbol{\xi}^\dagger \boldsymbol{\xi}\Bigg),
\end{align}

\noindent
where $\bar{\eta}_\alpha$ and $\eta_\alpha$ are the Grassmann fields. On the right-hand side, we have introduced Grassmann bilinears $\hat{P}_{\alpha\beta} \equiv \bar{\eta}_\alpha \eta_{\beta}$, as well as the complex fields $P_{\alpha\beta}$ and $\xi_{\alpha\beta}$ being fluctuating lines and Lagrange multipliers enforcing that the line fluctuations are compatible with Hamiltonian dynamics, respectively. The last term, proportional to infinitesimal positive $\epsilon$, is introduced to regularize the integral. In the following, $P_{\alpha\beta}$ and $P_{\beta\alpha}$ are not considered to be independent variables, but we impose identities $P_{\alpha\beta} \equiv P_{\beta\alpha}^{*}$, in accordance with the symmetry properties of their Grassmann correspondants, $\hat{P}_{\alpha\beta}$. In particular, the ``diagonal'' lines $P_{\alpha\alpha}$ are manifestly real. We have introduced a vector notation $\mathbf{P} = \{P_{\alpha\beta}\} \hspace{1em} \text{for all values of $\alpha$ and $\beta$}$; the remaining fields are handled in an analogous manner. In particular, $\boldsymbol{\xi}^\dagger \mathbf{P} \equiv \sum_{\alpha\beta} \xi^{*}_{\alpha\beta} P_{\alpha\beta} = \sum_{\alpha\beta} P^{*}_{\alpha\beta} \xi_{\alpha\beta} =  \mathbf{P}^\dagger \boldsymbol{\xi}$. The kinetic and potential energy functionals, entering  Eq.~\eqref{eq:hubbard_stratonovich}, may be now written explicitly as $E_{0, \mathrm{kin}}(\mathbf{P}) \equiv \sum_{\alpha\beta} t_{\alpha\beta} P_{\alpha\beta}$ and $E_{0, \mathrm{int}}(\mathbf{P}) = \frac{1}{2} \sum_{\alpha\beta\gamma\rho} P_{\alpha\beta}^{*} \mathcal{V}_{\alpha\beta\gamma\rho} P_{\gamma\rho}$. Remarkably, the interaction energy functional $E_{0, \mathrm{int}}$ enters Eq.~\eqref{eq:hubbard_stratonovich} with factor $\frac{1}{2}$. This is necessary to compensate for the double counting due to two inequivalent Wick contractions [cf. Eq.~\eqref{eq:e0int}]. Note that $E_{0, \mathrm{int}}(\mathbf{P})$ plays the role similar to the quadratic term in the usual Hubbard-Stratonovich transformation. The equivalence of the two sides of Eq.~\eqref{eq:hubbard_stratonovich} is shown in Appendix~\ref{appendix:prove_of_hs_decoupling}.

A methodological remark is in order at this point. In general, it may happen that the interaction energy functional, $E_{0, \mathrm{int}}$, is not positive definite even for simple models. For the case of Hubbard Hamiltonian with on-site repulsive interaction, $U$, one gets $E_{0, \mathrm{int}} = U \sum_i \langle{\hat{n}_{i\uparrow} \hat{n}_{i\downarrow}}\rangle_0 = U \sum_i \left(n_{i\uparrow} n_{i\downarrow} - S_i^+ S_i^-\right)$, with $n_{i\sigma} \equiv \langle\hat{n}_{i\sigma}\rangle_0$ and $S_i^+ = \langle\hat{c}^\dagger_{i\uparrow} \hat{c}_{i\downarrow}\rangle_0$, $S^-_i \equiv S^{+*}$, which is clearly not bounded from below for unrestricted line values. The order of integration in Eq.~\eqref{eq:hubbard_stratonovich} thus matters and care should be taken to integrate out $\boldsymbol{\xi}$ variables before integrating over $\mathbf{P}$-fields whenever ambiguities arise. This effectively enforces constraint for $\mathbf{P}$-field integration.

Making use of decoupling \eqref{eq:hubbard_stratonovich}, the generating functional for the model \eqref{eq:general_hamiltonian} may be written as

\begin{align}
  \label{eq:generating_functional}
  Z[\mathbf{J}] = & \int \mathcal{D}\bar{\boldsymbol{\eta}} \mathcal{D}\boldsymbol{\eta} \exp\Big(-\int d\tau \left\{\hat{\mathcal{H}} + \bar{\boldsymbol{\eta}} \partial_\tau \boldsymbol{\eta} - \mathbf{J}^\dagger \hat{\mathbf{P}} \right\} \Big) \nonumber\\&  \propto  \lim_{\epsilon\rightarrow 0^+}\int \mathcal{D}\bar{\boldsymbol{\eta}} \mathcal{D}\boldsymbol{\eta} \mathcal{D}\mathbf{P} \mathcal{D} \boldsymbol{\xi} \exp\left(-\mathcal{S}\right),
\end{align}

\noindent
with the system action defined by

\begin{align}
  \label{eq:uncorrelated_action_full}
  \mathcal{S} = &  \int d\tau \bar{\boldsymbol{\eta}} \partial_\tau \boldsymbol{\eta} + \int d\tau E_{0, \mathrm{kin}}(\mathbf{P}, \mu) + \frac{1}{2} \int d\tau E_{0, \mathrm{int}}(\mathbf{P}) + \nonumber \\&   i \boldsymbol{\xi}^\dagger(\hat{\mathbf{P}} - \mathbf{P})  + \frac{\epsilon}{2} \int d\tau \boldsymbol{\xi}^\dagger \boldsymbol{\xi} - \int d\tau \mathbf{J}^\dagger \hat{\mathbf{P}},
\end{align}

\noindent
where $\mathbf{J}$ is the auxiliary current taken to define the generating functional, $Z[\mathbf{J}]$. This expression for action \eqref{eq:uncorrelated_action_full} is used in the subsequent $1/\mathcal{N}_f$ expansion and analysis.

\subsection{Resummed $1/\mathcal{N}_f$ expansion}

We now turn to the $1/\mathcal{N}_f$ expansion for the model Hamiltonian~\eqref{eq:general_hamiltonian}, starting from the action~\eqref{eq:uncorrelated_action_full}. First, we introduce fermionic flavor index, $s = 1, \ldots, \mathcal{N}
_f$ and formally create $\mathcal{N}_f$ copies of Grassmann fields,  $\boldsymbol{\eta} \rightarrow \boldsymbol{\eta}^s$. The rescaling of the interaction matrix elements $V_{\alpha\beta\gamma\rho}$ by the factor $C_{\mathcal{N}_f} \equiv \frac{2 \mathcal{N}_f - 1}{\mathcal{N}_f^2}$ is then carried out to assure that the solution is non-trivial in the large-$\mathcal{N}_f$ limit ($\mathcal{C}_f \sim 2/\mathcal{N}_f$). The prescription $\hat{V} \rightarrow \hat{V}_{\mathcal{N}_f} \equiv C_{\mathcal{N}_f }\cdot \hat{V}$ thus ensures that $\hat{V}_{\mathcal{N}_f}$ interpolates smoothly between $\hat{V}_{\mathcal{N}_f} \propto 2 \hat{V}/{\mathcal{N}_f}$ for $\mathcal{N}_f \gg 1$ and $\hat{V}_{\mathcal{N}_f} = \hat{V}$ for $\mathcal{N}_f = 1$, and distinguishes our approach from the usual choice $C_{\mathcal{N}_f} = 1/\mathcal{N}_f$ (see, e.g., Ref.~\citenum{JakovacChapter2016}). At the same time, we may rescale the dummy Hubbard-Stratonovich fields accordingly to $\mathbf{P} \rightarrow \mathcal{N}_f \mathbf{P}$ and the infinitesimal term $\epsilon \rightarrow \epsilon/\mathcal{N}_f$. Since kinetic- and potential energy functional are homogeneous functions of $\mathbf{P}$-fields of degree one and two, respectively, we thus obtain a simple rescaling $E_{0, \mathrm{kin}} \rightarrow \mathcal{N}_f E_{0, \mathrm{kin}}$ and $E_{0, \mathrm{int}} \rightarrow \mathcal{N}_f^2 C_{\mathcal{N}_f} E_{0, \mathrm{int}}$. In effect, we arrive at the action of the form

\begin{align}
  \label{eq:uncorrelated_large_n}
  \mathcal{S} = &  \int d\tau \bar{\boldsymbol{\eta}}^s \left(\partial_\tau + (i \boldsymbol{\xi}^\dagger - \mathbf{J}^\dagger) \hat{\boldsymbol{\mathcal{O}}} \right)\boldsymbol{\eta}^s + \nonumber\\& \int d\tau \Big( \mathcal{N}_f E_{0, \mathrm{kin}}(\mathbf{P}, \mu) + \mathcal{N}_f^2 \frac{C_{\mathcal{N}_f}}{2} E_{0, \mathrm{int}}(\mathbf{P}) - \nonumber\\& \hspace{5em} i \boldsymbol{\xi}^\dagger\mathbf{P}  + \frac{\epsilon}{2} \boldsymbol{\xi}^\dagger \boldsymbol{\xi} \Big) = \nonumber \\ & \int d\tau \bar{\boldsymbol{\eta}}^s \left(\partial_\tau + (i \boldsymbol{\xi}^\dagger - \mathbf{J}^\dagger) \hat{\boldsymbol{\mathcal{O}}} \right)\boldsymbol{\eta}^s - \frac{1}{2} \int d\tau E_{0, \mathrm{int}}(\mathbf{P}) + \nonumber\\& \mathcal{N}_f \int d\tau \left( E_0(\mathbf{P}, \mu) - i \boldsymbol{\xi}^\dagger\mathbf{P}  + \frac{\epsilon}{2} \boldsymbol{\xi}^\dagger \boldsymbol{\xi} \right),
\end{align}

\noindent
where we have introduced vector of matrices $\hat{\boldsymbol{\mathcal{O}}}$, with elements  $\hat{\mathcal{O}}_{\alpha\beta}$, defined by the condition $\hat{P}^s_{\alpha\beta} \equiv \bar{\boldsymbol{\eta}}^s \hat{\mathcal{O}}_{\alpha\beta} \boldsymbol{\eta}^s$.

The present formulation has two desired properties: \emph{(i)} it ensures that for physical situation of $\mathcal{N}_f = 1$ the Hubbard-Stratonovich decoupling reproduces exactly the original Hamiltonian (this becomes apparent after comparing Eq.~\eqref{eq:uncorrelated_large_n} with Eq.~\eqref{eq:uncorrelated_action_full} for $\mathcal{N}_f = 1$), and \emph{(ii)} in the regime of small interaction, the large-$\mathcal{N}_f$ solution is consistent with the RPA calculation. Moreover, by construction, the proposed decoupling is unbiased as it does not favor any of the fermionic scattering channels. Since the introduced rescaling factor $C_{\mathcal{N}_f}$ involves both $\mathcal{O}(\mathcal{N}_f^{-1})$ and $\mathcal{O}(\mathcal{N}_f^{-2})$ terms, it may be viewed physically as a resummation of the interaction analogous to those performed in finite-temperature quantum field theories.\cite{JakovacPhysRevD2005} The $\mathcal{O}(1/\mathcal{N}_f^2)$ term is not included at the level of the saddle-point solution, but it generates a two-point interaction vertex in the next-to-leading order, affecting the structure of expansion in all consecutive orders. 

By integrating over fermion fields, the effective action~\eqref{eq:uncorrelated_large_n} reduces to

\begin{align}
  \label{eq:effective_action_hf}
  \mathcal{S}_\mathrm{eff} = & - \mathcal{N}_f \mathrm{Tr} \ln \left(-\partial_\tau - (i \boldsymbol{\xi}^\dagger - \mathbf{J}^\dagger) \hat{\boldsymbol{\mathcal{O}}} \right) + \nonumber\\& \mathcal{N}_f \int d\tau \left( E_0(\mathbf{P}, \mu) - i \boldsymbol{\xi}^\dagger\mathbf{P}  + \frac{\epsilon}{2} \boldsymbol{\xi}^\dagger \boldsymbol{\xi} \right) - \nonumber\\& \frac{1}{2} \int d\tau E_\mathrm{int}(\mathbf{P}).
\end{align}

\noindent
The large-$\mathcal{N}_f$ solution is found as a saddle point of the first two terms on the right-hand-side of Eq.~\eqref{eq:effective_action_hf}, with external currents $\mathbf{J}$ set to zero and the last $O(1)$ term ignored. The saddle point equations must be supplemented with the additional requirement that the total number of electrons is equal to $N_e$, providing the condition for chemical potential, $\mu$. In effect, one obtains

\begin{align}
  &{P}_{\alpha\beta}^{(0)} = \hat{G}_0^{\beta\alpha}(0^{-}),   \label{eq:saddle_point_hf_1} \\
  &i \xi_{\alpha\beta}^{(0)} = \frac{\partial E_0(\mathbf{P}^{(0)}, \mu)}{\partial P_{\alpha\beta}^{(0)*}},   \label{eq:saddle_point_hf_2} \\
  &\sum_{\alpha} P_{\alpha\alpha} = N_e,   \label{eq:saddle_point_hf_3}
\end{align}

\noindent
which is equivalent to the Hartree-Fock solution. We have used the superscript ``$(0)$'' to mark large-$\mathcal{N}_f$ value and introduced bare  imaginary-time Green's function

\begin{align}
\hat{G}^{\alpha\beta}_0(\tau - \tau^\prime) = - \frac{\mathrm{Tr} \mathcal{T}_\tau \hat{c}_\alpha(\tau) \hat{c}_\beta^\dagger(\tau^\prime) \exp(-\hat{\mathcal{H}}_\mathrm{eff})}{\mathrm{Tr} \exp(-\hat{\mathcal{H}}_\mathrm{eff})}, \label{eq:hf_bare_green_function}
\end{align}

\noindent
defined through the effective quasiparticle Hamiltonian, which now takes the form

\begin{align}
  \hat{\mathcal{H}}_\mathrm{eff} = \sum_{\alpha\beta} i \xi_{\alpha\beta}^{(0)*} \hat{c}^\dagger_\alpha \hat{c}_\beta = \sum_{\alpha\beta} \frac{\partial E_0(\mathbf{P}^{(0)}, \mu)}{\partial P_{\alpha\beta}^{(0)}} \hat{c}^\dagger_\alpha \hat{c}_\beta.     \label{eq:hf_bare_effective_hamiltonian}
\end{align}

\noindent
The right-hand-side of Eq.~\eqref{eq:hf_bare_effective_hamiltonian} has been obtained with the use of the saddle-point condition~\eqref{eq:saddle_point_hf_2}.

Thermodynamic corrections to the saddle point solution may now be systematically studied by decomposing the fields into the static saddle-point and dynamic fluctuation parts as $\boldsymbol{\xi}(\tau) \rightarrow \boldsymbol{\xi}^{(0)} + \delta\boldsymbol{\xi}(\tau)$, $\mathbf{P}(\tau) \rightarrow \mathbf{P}^{(0)} + \delta \mathbf{P}(\tau)$, and expanding the action to quadratic terms in fluctuations and currents. One obtains then

\begin{widetext}
\begin{align}
  \mathcal{S}_\mathrm{eff} = & \mathcal{N}_f \mathcal{S}_\mathrm{eff}^{(0)} + \frac{\mathcal{N}_f}{2} \int d\tau d \tau^\prime
  \begin{array}{c}
    (\delta\mathbf{P}^\dagger, \delta\boldsymbol{\xi}^\dagger) \\
    {}
  \end{array}
  \left(
  \begin{array}{cc}
     \hat{\mathcal{V}}(\tau-\tau^\prime) & -i \delta(\tau-\tau^\prime) \\
    -i \delta(\tau-\tau^\prime)  & \epsilon \delta(\tau-\tau^\prime) + \hat{\chi}_0(\tau, \tau^\prime) \\
  \end{array}
  \right)
  \left(
  \begin{array}{c}
    \delta\mathbf{P} \\
    \delta\boldsymbol{\xi}
  \end{array}
  \right) - \frac{\mathcal{N}_f}{2} \int \mathbf{J}^\dagger \hat{\chi}_0 \mathbf{J} + \nonumber \\ &
i\mathcal{N}_f \int d\tau d\tau^\prime  \delta\boldsymbol{\xi}^\dagger(\tau) \hat{\chi}_0(\tau, \tau^\prime)\mathbf{J}(\tau^\prime) - \mathcal{N}_f \int d\tau \mathbf{J}^\dagger(\tau) \mathbf{P}^{(0)} - \frac{1}{2} \int d\tau E_\mathrm{int}(\mathbf{P}),
\end{align}

\end{widetext}

\noindent
where

\begin{align}
  \label{eq:zero_order_effective_action_hf}
  \mathcal{S}_\mathrm{eff}^{(0)} = &  -\sum_n \mathrm{Tr} \ln \hat{G}_0(i\omega_n)^{-1} + \nonumber \\& \beta \left( E(\mathbf{P}^{(0)}) - i \boldsymbol{\xi}^{(0)\dagger}\mathbf{P}^{(0)}  + \frac{\epsilon}{2} \boldsymbol{\xi}^{(0)\dagger} \boldsymbol{\xi}^{(0)} \right)
\end{align}

\noindent
is the saddle-point action and the dynamic susceptibility

\begin{align}
  \label{eq:bare_susc_hf}
  \hat{\chi}_0^{\alpha_1\beta_1\alpha_2\beta_2}(\tau, \tau^\prime) = & \langle \mathcal{T}_\tau \hat{c}^\dagger_{\alpha_1}(\tau) \hat{c}_{\beta_1}(\tau) \hat{c}^\dagger_{\beta_2}(\tau^\prime) \hat{c}_{\alpha_2}(\tau^\prime)  \rangle_0^c = \nonumber\\&  - G_0^{\beta_1\beta_2}(\tau-\tau^\prime) G_0^{\alpha_2\alpha_1}(\tau^\prime-\tau)
\end{align}

\noindent
is of the Lindhard form. Finally, by integrating out the Gaussian fluctuations and taking $\epsilon \rightarrow 0^+$ limit, one arrives at the final form of the effective action

\begin{align}
  \label{eq:hf_action_gaussian}
  \mathcal{S}_\mathrm{eff} = & \mathcal{N}_f \mathcal{S}_\mathrm{eff}^{(0)} + \frac{1}{2} \sum_n \mathrm{Tr} \ln \left( 1 + \hat{\mathcal{V}} \hat{\chi}_0(i\omega_n) \right) - \frac{\beta}{2} E_\mathrm{int}(\mathbf{P}^{(0)}) - \nonumber\\ & \frac{\mathcal{N}_f}{2 \beta} \sum_n \mathbf{J}^\dagger(i\omega_n) (1 + \hat{\chi}_0(i\omega_n) \hat{\mathcal{V}})^{-1} \hat{\chi}_0(i\omega_n) \mathbf{J}(i\omega_n) \nonumber\\& - \mathcal{N}_f \mathbf{J}^\dagger(i\omega_n = 0) \mathbf{P}^{(0)}
\end{align}

\noindent
to the order $\mathcal{O}(1)$. We have introduced Fourier-transformed fields and susceptibilities according to

\begin{align}
  \mathbf{J}(i\omega_n) =& \int d\tau \mathrm{e}^{i\omega_n \tau} \mathbf{J}(\tau), \\
  \hat{\chi}_0(i\omega_n) =& \int d(\tau-\tau^\prime) \mathrm{e}^{i\omega_n (\tau - \tau^\prime)} \hat{\chi}_0(\tau, \tau^\prime).
\end{align}

\noindent
The term $\propto \mathrm{Tr} \log$ in the first line of Eq.~\eqref{eq:hf_action_gaussian} is the thermodynamic correction due to collective excitations, whereas  $\frac{\beta}{2} E_\mathrm{int}(\mathbf{P}^{(0)})$  originates from our resummation procedure and is absent in standard $1/\mathcal{N}_f$ expansion. The term $\propto \mathbf{J}^\dagger(i\omega_n = 0) \mathbf{P}^{(0)}$  represents elastic (Bragg) contribution that disappears if the saddle-point value of the line is zero, i.e., $\mathbf{P}^{(0)} = 0$.

\begin{table}
  \centering
  \caption{Summary of the approximation schemes and acronyms used throughout the paper. The \textbf{S}tatistically-\textbf{C}onsistent \textbf{G}utzwiller \textbf{A}pproximation (SGA) for the energy functional $E_G$ is equivalent to retaining only those diagrams that prevail in the infinite-lattice-coordination limit ($d \rightarrow \infty$), see Sec.~\ref{subsec:gutzwiller_approximation}. The \textbf{L}ocal-\textbf{D}iagram approximation (LD) incorporates local diagrams with arbitrary number of loops (cf. Sec.~\ref{subsec:local_approximation}). Each approximation may be considered in ``$\lambda_d$'',  ``$x$'', or ``$f$'' flavor that differ by selection and handling of the sixth constraint (see the text).}
  \begin{tabular}[c]{ccc}
    \hline
    Acronym & $E_G$ truncation & 6-th constraint \\
    \hline\hline
    $\mathrm{SGA}_{\lambda_d}+1/\mathcal{N}_f$ & $d = \infty$ & $\lambda_d - \bar{\lambda}_d \equiv 0$ \\
    $\mathrm{SGA}_x+1/\mathcal{N}_f$ & $d = \infty$ & $\langle \hat{d} \rangle_0 x_{I} - (\lambda_{I0}^2 - 1) \equiv 0$ \\
    $\mathrm{SGA}_f+1/\mathcal{N}_f$ & $d = \infty$ & fluctuating $\lambda_d$ \\
    $\mathrm{LD}_{\lambda_d}+1/\mathcal{N}_f$ & all local diagrams & $\lambda_d - \bar{\lambda}_d \equiv 0$ \\
    $\mathrm{LD}_x+1/\mathcal{N}_f$ & all local diagrams & $\langle \hat{d} \rangle_0 x_{I} - (\lambda_{I0}^2 - 1) \equiv 0$ \\
    $\mathrm{LD}_f+1/\mathcal{N}_f$ & all local diagrams & fluctuating $\lambda_d$ \\
    \hline
\end{tabular}
\label{tab:acronyms}
\end{table}

\section{General variational-wave-function approach combined with $1/\mathcal{N}_f$ expansion}
\label{section:variational+fluctuations}

We are now in position to introduce the full VWF+$1/\mathcal{N}_f$ approach, combining diagrammatic VWF method with the resummed $1/\mathcal{N}_f$ expansion, outlined in Secs.~\ref{section:variational_approach} and \ref{section:large_n_expansion}, respectively. The idea is based on performing the \emph{second resummation}, in addition to that discussed in Sec.~\ref{section:large_n_expansion}, so that the resulting expansion fulfills the following conditions: \emph{(i)} it agrees with the VWF solution at the saddle-point level, \emph{(ii)} is not biased in favor of  any of the scattering channels, \emph{(iii)} reproduces the exact generating functional \eqref{eq:generating_functional} for the model Hamiltonian $\hat{\mathcal{H}}$ in the physical $\mathcal{N}_f = 1$ limit, \emph{(iv)} is capable of describing correlated collective excitations (e.g., charge, spin) already in the leading non-trivial order, and \emph{(v)} reduces to RPA in the weak-coupling limit.

To achieve the goals \emph{(i)}-\emph{(v)}, we argue that $1/\mathcal{N}_f$ expansion based on the generalized action appropriate for correlated system

\begin{widetext}
\begin{align}
  \label{eq:sga_1n_action}
  \mathcal{S}_\mathrm{var}[\mathbf{P}, \boldsymbol{\xi}, \boldsymbol{\rho}, \bar{\boldsymbol{\eta}}, \boldsymbol{\eta}, \mathbf{J}, \mu] = & \int d\tau \bar{\boldsymbol{\eta}}^s \left( \partial_\tau + (i \boldsymbol{\xi}^\dagger - \mathbf{J}^\dagger)\hat{\boldsymbol{\mathcal{O}}} \right) \boldsymbol{\eta}^s + \mathcal{N}_f \int d\tau \left( E_\mathrm{var}(\mathbf{P}, \boldsymbol{\lambda}, \mu) - i \boldsymbol{\xi}^\dagger\mathbf{P} \right) 
+ \nonumber \\ & \int d\tau E_{0, \mathrm{kin}}(\mathbf{P}, \mu) + \frac{1}{2} \int d\tau E_{0, \mathrm{int}}(\mathbf{P}) - \int d\tau E_\mathrm{var}(\mathbf{P}, \boldsymbol{\lambda}, \mu) - \nonumber \\ &
i \mathcal{N}_f \int d\tau \boldsymbol{\rho}^T\mathbf{C}(\mathbf{P}, \boldsymbol{\lambda}) + \frac{\mathcal{N}_f}{2} \kappa \int d\tau \mathbf{C}(\mathbf{P}, \boldsymbol{\lambda})^T \mathbf{C}(\mathbf{P}, \boldsymbol{\lambda}) - \ln \left|\mathrm{det} \frac{\partial \mathbf{C}}{\partial \boldsymbol{\lambda}}\right|
\end{align}
\end{widetext}

\noindent
indeed fulfills the requirements \emph{(i)}-\emph{(v)}. For brevity of notation, in the above action we have deliberately omitted the $\epsilon$-terms that should be reintroduced whenever necessary. The structure of the first two lines is reminiscent of the uncorrelated action of Eq.~\eqref{eq:uncorrelated_large_n}, albeit with two modifications. First, in the $\mathcal{O}(\mathcal{N}_f)$ block, the uncorrelated energy functional $E_0(\mathbf{P}, \mu)$ has been substituted with the correlated one, $E_\mathrm{var}(\mathbf{P}, \boldsymbol{\lambda}, \mu)$. To prove the concept, here we limit ourselves to calculations for the Gutzwiller functional $E_\mathrm{var} \equiv E_G$. Second, the resummed $\mathcal{O}(1)$ terms [second line of Eq.~\eqref{eq:sga_1n_action}] have been adjusted accordingly so that the action \eqref{eq:uncorrelated_action_full} is retrieved for $\mathcal{N}_f = 1$. Now, since the correlated energy depends not only on the line-fields $\mathbf{P}$, but also on the correlator parameter vector $\boldsymbol{\lambda}$, the constraints introduced in Section~\ref{section:variational_approach} need to be incorporated \emph{at the dynamical level}. This is implemented in the third line of Eq.~\eqref{eq:sga_1n_action} by means of an additional set of time-dependent Lagrange-multiplier fields, $\boldsymbol{\rho}$. The unconventional term $\propto \kappa \mathbf{C}^T\mathbf{C}$ in the last line, controlled by a positive parameter $\kappa$, is of no physical significance due to the constraints $\mathbf{C} = \mathbf{0}$, except that it regularizes the series expansion, as elaborated below. Finally, the $\ln \left| \mathrm{det} \frac{\partial \mathbf{C}}{\partial \boldsymbol{\lambda}}\right|$ term, with $\frac{\partial \mathbf{C}}{\partial \boldsymbol{\lambda}}$ being the Jacobian of the constraints, is necessary to compensate for unphysical interaction terms generated by the fluctuations of the correlator parameters. This contribution is highly non-linear and it may be equivalently represented as a functional integral over Faddeev-Popov ghost fields\cite{FaddeevPhysLettB1967,FaddeevBook2018} to facilitate calculations beyond the leading order.

To make the approach rigorous, one needs to demand additionally that the $\boldsymbol{\lambda}$-fields have no their own dynamics (i.e., they adjust passively to fluctuations of the line-fields, $\mathbf{P}$), as well as assume uniqueness of the constraints for given $\mathbf{P}$-field configuration. Note that, as long as only small fluctuations are considered, weaker condition of local uniqueness in the vicinity of the correlated saddle-point solution is sufficient. The most straightforward way to achieve this goal is to ensure that the number of inequivalent constraints matches the number of $\boldsymbol{\lambda}$-fields. In Sec.~\ref{section:variational_approach}, we have already introduced five constraints for six $\boldsymbol{\lambda}$ parameters per orbital so that one more is still needed. As an additional, sixth constraint, we will hereafter consider either \emph{(i)} $C_I^{6} = \lambda_{d, I} - \bar{\lambda}_{d, I}$ or \emph{(ii)} $C_I^{6} = ({n_{I\uparrow} n_{I\downarrow} - S_I^+ S_I^-}) x_{I} - (\lambda_{I0}^2 - 1)$, where $\bar{\lambda}_{d, I}$ and $x_I$ are understood to be \emph{static (time-independent) variational parameters}, to be determined by minimization of the free energy functional. The choices \emph{(i)} and \emph{(ii)} define two different resummation schemes that will be hereafter distinguished by the subscripts $\lambda_d$ and $x$, respectively (resulting in the $\mathrm{VWF}_{\lambda_d}$+$1/\mathcal{N}_f$ and $\mathrm{VWF}_{x}$+$1/\mathcal{N}_f$ variants of the method) and their selection should be made based on physical arguments and benchmarks against other techniques. The variant \emph{(i)} is based on the assumption that the dynamics in the doublon sector is much faster than that of the collective excitations. This is physically well-justified on the hole-doped side of the phase diagram, where the doubly occupied sites may be effectively avoided in the large-$U$ regime. The empty-sites (holons) will, however, be strongly coupled to singly-occupied sited by means of projected hopping terms. Note that the situation changes qualitatively in the electron-doped case, where doublons cannot be avoided, but holons become substantially suppressed. The $\mathrm{VWF}_{\lambda_d}$+$1/\mathcal{N}_f$ scheme, based on fixing $\lambda_d$, is thus not explicitly invariant with respect to particle-hole transformation and it should be substituted with an analogous $\mathrm{VWF}_{\lambda_0}$+$1/\mathcal{N}_f$ approximation ($C_i^6 \equiv \lambda_{0, I} - \bar{\lambda}_{0, I} \equiv 0$ condition) on the electron-doped side. In the present contribution we consider predominantly hole-doped situation and hence employ $\mathrm{VWF}_{\lambda_d}$+$1/\mathcal{N}_f$ in most of the calculations, unless stated otherwise. The scheme \emph{(ii)}, on the other hand, is motivated by the tadpole cancellation condition \eqref{eq:contraints_pg2}, now generalized to the dynamical level. It turns out that $\mathrm{VWF}_{x}$+$1/\mathcal{N}_f$ is automatically particle-hole symmetric.  Our benchmark calculations, detailed below, reveal that, up to a moderate coupling, the large-$\mathcal{N}_f$ solutions obtained within both schemes are quantitatively consistent, whereas \emph{(i)} provides a more reliable charge dynamics at strong coupling. Finally, we also propose the third solution that does not require defining an additional constraint, and has certain practical advantages over the former two. This is achieved by treating the field $\lambda_{d, I}(\tau)$ as an external parameter of the generating functional, fixed throughout the calculation, so that $Z = Z[\mathbf{J}, \lambda_d]$ becomes a functional of external currents and $\lambda_d$. The remaining five fluctuating $\boldsymbol{\lambda}$-fields are then determined by the five constraints, already introduced in Sec.~\ref{section:variational_approach}. Remarkably, for physical $\mathcal{N}_f = 1$, $Z[\mathbf{J}, \lambda_d]$ becomes actually independent of $\lambda_d$ which can be shown by integrating out consecutively the auxiliary fields (see Eq.~\eqref{eq:appendix_correlated_z_proof} in Appendix~\ref{appendix:action_correlated_1/n}). Such a dependence appears, however, in the large-$\mathcal{N}_f$ limit and it can be eliminated by integrating over $\lambda_d$ so that $Z[\mathbf{J}] = \int \mathcal{D}\lambda_d Z[\mathbf{J}, \lambda_d]$ (integration is performed over the domain of $\lambda_d$). The scheme, defined in this way will be hereafter referred to as $\mathrm{VWF}_{f}$+$1/\mathcal{N}_f$, where the subscript $f$ indicates fluctuating $\lambda_d$. This procedure makes $\mathrm{VWF}_{f}$+$1/\mathcal{N}_f$ unbiased with respect to particle-hole transformation. We have found that, in the hole-doped regime, that $\mathrm{VWF}_{f}$+$1/\mathcal{N}_f$ and $\mathrm{VWF}_{\lambda_d}$+$1/\mathcal{N}_{f}$ variants yield very close results for the model parameters considered here, providing argument in favor of consistency between schemes.

In summary, we have introduced three variants of $\mathrm{VWF}$+$1/\mathcal{N}_{f}$ method, differing by the detailed handling of the constraints, and distinguished by subscripts $\lambda_d$, $x$, and $f$. Each of them may be further considered in two versions, SGA and LD, depending on how the diagrammatic energy functional is truncated (see Secs.~\ref{subsec:local_approximation} and \ref{subsec:gutzwiller_approximation}). The resulting six schemes and acronyms used throughout the paper are summarized in Table~\ref{tab:acronyms}. Detailed discussion and comparison of the variants will be performed below. A formal derivation, as well as extended discussion of Eq.~\eqref{eq:sga_1n_action} appropriateness, are deferred to Appendix~\ref{appendix:action_correlated_1/n}.

\subsection{Correlated large-$\mathcal{N}_f$ limit}

We now proceed to analyze the saddle-point (large-$\mathcal{N}_f$) VWF+$1/\mathcal{N}_f$ solution by retaining only the leading  $\mathcal{O}(\mathcal{N}_f)$ terms and those involving Grassmann variables, in the full analogy to the discussion of Sec.~\ref{section:large_n_expansion}.  After integrating out fermions, one arrives at the effective correlated action

\begin{align}
  \label{eq:1}
  \mathcal{S}_{\mathrm{var}, \mathrm{eff}} [\mathbf{P}, &\boldsymbol{\xi}, \boldsymbol{\rho}, \mathbf{J}, \mu] =  \mathcal{N}_f \int d\tau \Bigg\{ E_\mathrm{\mathrm{var}}(\mathbf{P}, \boldsymbol{\lambda}, \mu) - i \boldsymbol{\xi}^\dagger \mathbf{P} - \nonumber\\& i \boldsymbol{\rho}^T \mathbf{C}(\mathbf{P}, \boldsymbol{\lambda}) + \frac{\kappa}{2}\mathbf{C}(\mathbf{P}, \boldsymbol{\lambda})^T \mathbf{C}(\mathbf{P}, \boldsymbol{\lambda}) \Bigg\} - \nonumber \\ & - \mathcal{N}_f \mathrm{Tr} \log \left[ -\partial_\tau - (i\boldsymbol{\xi}^\dagger - \mathbf{J}^\dagger) \hat{\boldsymbol{\mathcal{O}}}_i \right] + \mathcal{O}(1),
\end{align}

\noindent
so that the large-$\mathcal{N}_f$ saddle point equations read

\begin{align}
  &\mathbf{P}^{(0)}_{\alpha\beta}   = \hat{G}_0^{\beta\alpha}(0^-), \label{eq:saddle_point_correlated_1} \\
  & i \xi^{(0)*}_{\alpha\beta} = \frac{\partial E_\mathrm{var}}{\partial P_{\alpha\beta}} - i \boldsymbol{\rho}^T \frac{\partial \mathbf{C}}{\partial P^{(0)}_{\alpha\beta}}, \label{eq:saddle_point_correlated_2}\\
  & -\sum_\alpha \frac{\partial E_\mathrm{var}(\mathbf{P}^{(0)}, \boldsymbol{\lambda}^{(0)}, \mu^{(0)})}{\partial \mu^{(0)}} = N_e, \label{eq:saddle_point_correlated_3}\\  
  &\mathbf{C}(\mathbf{P}^{(0)}, \boldsymbol{\lambda}^{(0)}) \equiv \mathbf{0}, \label{eq:saddle_point_correlated_4}\\
  &\frac{\partial E_\mathrm{var}}{\partial \lambda^{(0)l}} - i \rho^\alpha \frac{\partial C^\alpha}{\partial \lambda^{(0)l}} = 0. \label{eq:saddle_point_correlated_5}
\end{align}

\noindent
Equations~\eqref{eq:saddle_point_correlated_1}-\eqref{eq:saddle_point_correlated_3} replace those for the uncorrelated expansion [Eqs.~\eqref{eq:saddle_point_hf_1}-\eqref{eq:saddle_point_hf_3}]. The essential difference is that the correlated energy $E_\mathrm{var}$ rather than $E_0$ enters Eq.~\eqref{eq:saddle_point_correlated_3}. The Green's function in Eq.~\eqref{eq:saddle_point_correlated_1} is defined in the same manner as in the uncorrelated case, but using the correlated energy functional, i.e.,

\begin{align}
  \hat{\mathcal{H}}_\mathrm{eff}^\mathrm{corr} = i \boldsymbol{\xi}^{(0)\dagger} \hat{\mathbf{P}} = \sum_{\alpha\beta} \left( \frac{\partial E_\mathrm{var}}{\partial P_{\alpha\beta}^{(0)}} - i \boldsymbol{\rho}^T \frac{\partial \mathbf{C}}{\partial P^{(0)}_{\alpha\beta}} \right) \hat{c}^\dagger_\alpha \hat{c}_\beta.     \label{eq:correlted_effective_hamiltonian}
\end{align}

\noindent
Note that Eq.~\eqref{eq:correlted_effective_hamiltonian} defines still a single-particle Hamiltonian, similarly as Eq.~\eqref{eq:hf_bare_effective_hamiltonian}, but it is evaluated on top of the \emph{correlated} variational state. As discussed previously, $\hat{\mathcal{H}}_\mathrm{eff}^\mathrm{corr}$ is an object controlling the energy dispersion of correlated single-quasiparticle excitations.\cite{BunemannPhysRevB2003,FidrysiakJPhysCondensMatter2018}

We point out that the left-hand-side of Eq.~\eqref{eq:saddle_point_correlated_3} for the chemical potential cannot be in general evaluated explicitly as opposed to the uncorrelated situation. Nonetheless, for truncated series expansions, described in \ref{subsec:local_approximation} and \ref{subsec:gutzwiller_approximation}, Eq.~\eqref{eq:saddle_point_correlated_3} becomes identical to Eq.~\eqref{eq:saddle_point_hf_3}. In the case of Gutzwiller energy functional ($E_\mathrm{var} = E_G$), Eqs.~\eqref{eq:saddle_point_correlated_1}-\eqref{eq:saddle_point_correlated_5} constitute the set of integral equations that are equivalent to those solved within the DE-GWF method.\cite{KaczmarczykPhysRevB2013}

\subsection{Effective scattering matrix and correlated collective modes}

Collective excitations around the correlated ground state may be studied by employing the procedure analogous to that described in Sec.~\ref{section:large_n_expansion} for the uncorrelated (Hartree-Fock) saddle-point. This leads to the Gaussian-fluctuation action

\begin{widetext}
\begin{align}
  \label{eq:action_expanded_de-gwf}
  \mathcal{S}_{\mathrm{var}, \mathrm{eff}}  &\approx  \mathcal{N}_f \mathcal{S}_{\mathrm{var}, \mathrm{eff}}^{(0)} + \frac{\mathcal{N}_f}{2} \int d\tau  (\delta\mathbf{P}^\dagger \delta\boldsymbol{\lambda}^\dagger \delta\boldsymbol{\rho}^\dagger)
  \left(
  \begin{array}{ccc}
    \hat{\mathcal{V}}_{P^{*}P} + \kappa \hat{\mathcal{V}}_{P^{*}\rho}\hat{\mathcal{V}}_{\rho P} & \hat{\mathcal{V}}_{P^{*} \lambda} + \kappa \hat{\mathcal{V}}_{P^{*}\rho}\hat{\mathcal{V}}_{\rho \lambda} & i \hat{\mathcal{V}}_{P^{*} \rho}\\
    \hat{\mathcal{V}}_{\lambda P} + \kappa \hat{\mathcal{V}}_{\lambda\rho}\hat{\mathcal{V}}_{\rho P}  & \hat{\mathcal{V}}_{\lambda\lambda} + \kappa \hat{\mathcal{V}}_{\lambda\rho}\hat{\mathcal{V}}_{\rho \lambda} & i \hat{\mathcal{V}}_{\lambda \rho}\\
    i\hat{\mathcal{V}}_{\rho P} &  i \hat{\mathcal{V}}_{\rho \lambda} & 0
  \end{array}
        \right)
        \left(
        \begin{array}{c}
          \delta\mathbf{P} \\
          \delta\boldsymbol{\lambda} \\
          \delta\boldsymbol{\rho}
        \end{array}
  \right) \nonumber \\
  &  -  \mathcal{N}_f \int d\tau \delta\boldsymbol{\xi}^\dagger \delta\mathbf{P}  - \frac{\mathcal{N}_f}{2} \int d\tau d\tau^\prime \Big(i \delta\boldsymbol{\xi}_i^\dagger(\tau) - \mathbf{J}^\dagger(\tau)\Big) \hat{\chi}_0 (\tau, \tau^\prime) \Big(i \delta\boldsymbol{\xi}(\tau^\prime) - \mathbf{J}(\tau^\prime)\Big) - \mathcal{N}_f \int d\tau \mathbf{J}^\dagger(\tau) \mathbf{P}^{(0)} + \mathcal{O}(1)
          ,
\end{align}

\end{widetext}

\noindent
where

\begin{align}
  \hat{\mathcal{V}}^{\alpha_1\beta_1,\alpha_2\beta_2}_{P^{*}P} \equiv & \partial_{P^{*}_{\alpha_1\beta_1}} \partial_{P_{\alpha_2\beta_2}} (E_G - i\boldsymbol{\rho}^T \mathbf{C}) |_0 \\
  \hat{\mathcal{V}}^{l, \alpha\beta}_{\lambda P} \equiv & \partial_{\lambda_l} \partial_{P_{\alpha\beta}}  (E_G - i\boldsymbol{\rho}^T \mathbf{C}) |_0\\
    \hat{\mathcal{V}}^{l_1, l_2}_{\lambda \lambda} \equiv & \partial_{\lambda_{l_1}} \partial_{\lambda_{l_2}}  (E_G - i\boldsymbol{\rho}^T \mathbf{C} ) |_0\\
  \hat{\mathcal{V}}^{l, \alpha\beta}_{\rho P} \equiv & -\partial_{P_{\alpha\beta}}  C^l |_0\\
  \hat{\mathcal{V}}^{l_1, l_2}_{\rho \lambda} \equiv & -\partial_{\lambda_{l_2}}  C^{l_1} |_0, \label{eq:scattering_matrix_element_l_rho_lambda}
\end{align}

\noindent
are scattering matrix elements, reflecting interactions between collective fields. As before, the subscript ``0'' indicates that the derivatives are evaluated at the saddle point.

The effective action of Eq.~\eqref{eq:action_expanded_de-gwf} can be given a physical interpretation by introducing the concept of \emph{effective scattering matrix}, $\hat{\mathcal{V}}_\mathrm{eff}$. The latter describes residual interactions between correlated collective excitations and is obtained by evaluating the integrals over the $\boldsymbol{\lambda}$ and $\boldsymbol{\rho}$ fields. As is shown in Appendix~\ref{appendix:action_correlated_1/n}, the resulting  $\hat{\mathcal{V}}_\mathrm{eff}$ is independent on $\kappa$. Inclusion of the terms $\propto \kappa$ may be thus regarded as a \emph{gauge transformation} that ensures convergence while evaluating the integral over $\boldsymbol{\lambda}$-fields. Explicitly, we obtain (cf. Appendix~\ref{appendix:action_correlated_1/n})

\begin{align}
  \label{eq:effective_scattering_matrix}
  \hat{\mathcal{V}}_\mathrm{eff} \equiv
    \hat{\mathcal{V}}_{P^{*} P} -
  \begin{array}{c}
    (\hat{\mathcal{V}}_{P^{*} \lambda}, i \hat{\mathcal{V}}_{P^{*} \rho})
  \end{array}
  \left(
  \begin{array}{cc}
    \hat{\mathcal{V}}_{\lambda\lambda} & i \hat{\mathcal{V}}_{\lambda\rho} \\
    i \hat{\mathcal{V}}_{\rho\lambda} & 0
  \end{array}
                                  \right)^{-1}
                                  \left(
\begin{array}{c}
  \hat{\mathcal{V}}_{\lambda P} \\
  i \hat{\mathcal{V}}_{\rho P}
\end{array}
  \right),
\end{align}

\noindent
so that now

\begin{align}
  &\mathcal{S}_{\mathrm{var}, \mathrm{eff}} = \mathcal{N}_f \mathcal{S}_{\mathrm{var}, \mathrm{eff}}^{(0)} + \frac{\mathcal{N}_f}{2} \int d\tau \delta \mathbf{P}^\dagger \mathcal{V}_\mathrm{eff} \delta \mathbf{P} - i \int d\tau \delta\boldsymbol{\xi}^\dagger \delta\mathbf{P} - \nonumber \\ &\frac{\mathcal{N}_f}{2}  \int d\tau d\tau^\prime \Big(i \delta\boldsymbol{\xi}_i^\dagger(\tau) - \mathbf{J}^\dagger(\tau)\Big) \hat{\chi} (\tau, \tau^\prime) \Big(i \delta\boldsymbol{\xi}(\tau^\prime) - \mathbf{J}(\tau^\prime)\Big) - \nonumber \\ &\mathcal{N}_f \int d\tau \mathbf{J}^\dagger \mathbf{P}^{(0)} + \mathcal{O}(1).
\end{align}

\noindent
This action may be used directly to construct the generating functional, as discussed in Sec.~\ref{section:large_n_expansion}. The final large-$\mathcal{N}_f$ form for the correlated case is thus

\begin{align}
  Z[\mathbf{J}]  \approx  \exp\Big( & -\mathcal{N}_f \mathcal{S}_\mathrm{eff}^{(0)} + \frac{\mathcal{N}_f}{2 \beta} \sum_n \mathbf{J}^\dagger(i\omega_n) \hat{\chi}(i\omega_n) \mathbf{J}(i\omega_n)  \nonumber\\ &- \mathcal{N}_f \int d\tau \mathbf{J}^\dagger \mathbf{P}^{(0)} \Big), \label{eq:vwf+1/n_functional}
\end{align}

\noindent
where the correlated dynamical susceptibility matrix reads

\begin{align}
  \label{eq:correlated_susceptibility}
  \hat{\chi}(i\omega_n) = (1 + \hat{\chi}_0(i\omega_n) \hat{\mathcal{V}}_\mathrm{eff})^{-1} \hat{\chi}_0(i\omega_n).
\end{align}

\noindent
In Eq.~\eqref{eq:correlated_susceptibility}, $\hat{\chi}_0(i\omega_n)$ represents correlated Lindhard susceptibility, evaluated starting from the correlated saddle point solution using effective single-particle Hamiltonian~\eqref{eq:correlted_effective_hamiltonian}.

This completes the formal aspects of the VWF+$1/\mathcal{N}_f$-method analysis. Selected computational aspects, involved in VWF+$1/\mathcal{N}_f$ calculations, are detailed in Appendix~\ref{appendix:computational_aspects}. Below, we apply and discuss in detail the leading-order solutions based on formula~\eqref{eq:vwf+1/n_functional}. A detailed numerical analysis and comparison with DQMC method are discussed next.

\section{Comparison with determinant quantum Monte-Carlo (DQMC) method}
\label{section:benchmark_Hubabrd_model}

\begin{figure}
  \centering
  \includegraphics[width=1\linewidth]{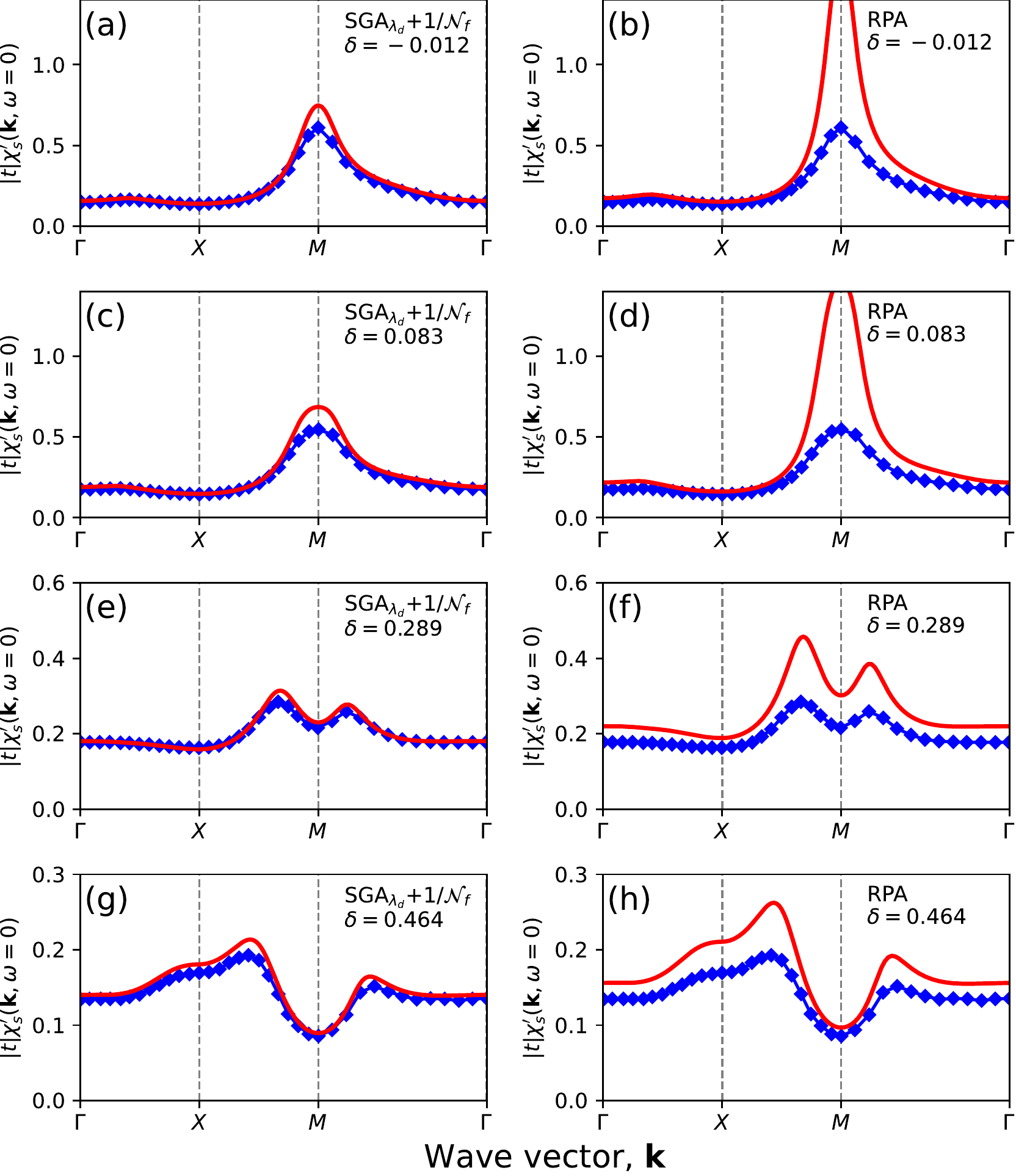}
  \caption{Static spin susceptibilities along the representative $\Gamma$-$X$-$M$-$\Gamma$ contour in the Brillouin zone for the two-dimensional Hubbard model with non-zero nearest- and next-nearest-neighbor hopping ($t < 0$ and $t^\prime = 0.2 |t|$, respectively) at weak coupling ($U/|t| = 2$) and for temperature $k_B T = 0.2 |t|$. Red solid lines in left panels represent $\mathrm{SGA}_{\lambda_d}$+$1/\mathcal{N}_f$ calculation, whereas in the right panels the corresponding random-phase approximation (RPA) results are displayed. The hole doping levels are set to $\delta=-0.017$ in panels (a)-(b), $\delta=0.083$ in (c)-(d), $\delta=0.289$ in (e)-(f), and $\delta=0.464$ in (g)-(h). Negative $\delta$ indicates electron doped system. Blue diamonds are the corresponding DQMC data of Ref.~\citenum{HillePhysRevResearch2020} for the same set of model parameters and temperature. The lattice size has been set to $100\times 100$ in the $\mathrm{SGA}_{\lambda_d}$+$1/\mathcal{N}_f$ calculation and no renormalization  factors have been applied to the susceptibilities.}
  \label{fig:weak_coupling_benchmark}
\end{figure}

The techniques providing quantitatively reliable description of the dynamics for large-sized strongly-correlated fermionic systems are scarce. Thus, to benchmark the results of our VWF+$1/\mathcal{N}_f$ approach, we resort here to the available determinant quantum Monte-Carlo (DQMC) data, regarded as a reasonable reference point. It should be emphasized though that DQMC and VWF+$1/\mathcal{N}_f$ have different usage scenarios and applicability regimes. Namely, VWF+$1/\mathcal{N}_f$ is an approximate tool for exploring real-time dynamics in the thermodynamic-limit, also in broken-symmetry state. On the other hand, DQMC provides a numerically exact description of the imaginary-time correlations for relatively small systems (typically $< 10^2$ sites) in the non-broken symmetry state. Since analytic continuation of the Monte-Carlo data is ill-conditioned and introduces poorly controlled errors (see, e.g., Ref.~\citenum{TripoltComPhysCommun2019}), such a benchmark amounts to comparison between two sets of approximate data points in correlated state.

\subsection{Static spin susceptibility}
\label{subsection:static_susceptibilities}

Before turning our attention to dynamical quantities, we  address first the static magnetic response that may be evaluated without any analytic continuation. In Fig.~\ref{fig:weak_coupling_benchmark}, the calculated $\mathrm{SGA}_{\lambda_d}$+$1/\mathcal{N}_f$  static spin susceptibilities (red solid lines in left panels) are compared with the DQMC data of Ref.~\citenum{HillePhysRevResearch2020} (blue diamonds) for the Hubbard model with nearest- and next-nearest neighbor hopping ($t<0$ and $t^\prime = 0.2|t|$) in the regime of weak on-site Coulomb repulsion, $U/|t| = 2$. For reference, we also display the corresponding random-phase-approximation (RPA) result in the right panels by solid lines. The temperature has been set to $k_B T = 0.2 |t|$ for $\mathrm{SGA}_{\lambda_d}$+$1/\mathcal{N}_f$, RPA, and DQMC techniques. The hole doping levels are set to $\delta=-0.017$ in panels (a)-(b), $\delta=0.083$ in (c)-(d), $\delta=0.289$ in (e)-(f), and $\delta=0.464$ in (g)-(h), covering the most substantial portion of the hole-side of the phase diagram. As is apparent from panels (a), (c), (e), and (f), the present approach semi-quantitatively matches the DQMC data, even without introducing renormalization factors. A mild boost of the spin susceptibility is, however, observed close to the $M$ point in the weakly-doped system and along the $X$-$M$ line for large doping. This effect is expected to be reduced by higher-order contributions in the $1/\mathcal{N}_f$ expansion. On the other hand, the RPA results, displayed in panels (b), (d), (f), and (h), substantially overestimate static spin response at all doping levels, and in particular close to the half-filling. This reflects unphysically strong tendency towards magnetic ordering within the mean field approach, even at weak-coupling.

\begin{figure}
  \centering
  \includegraphics[width=1\linewidth]{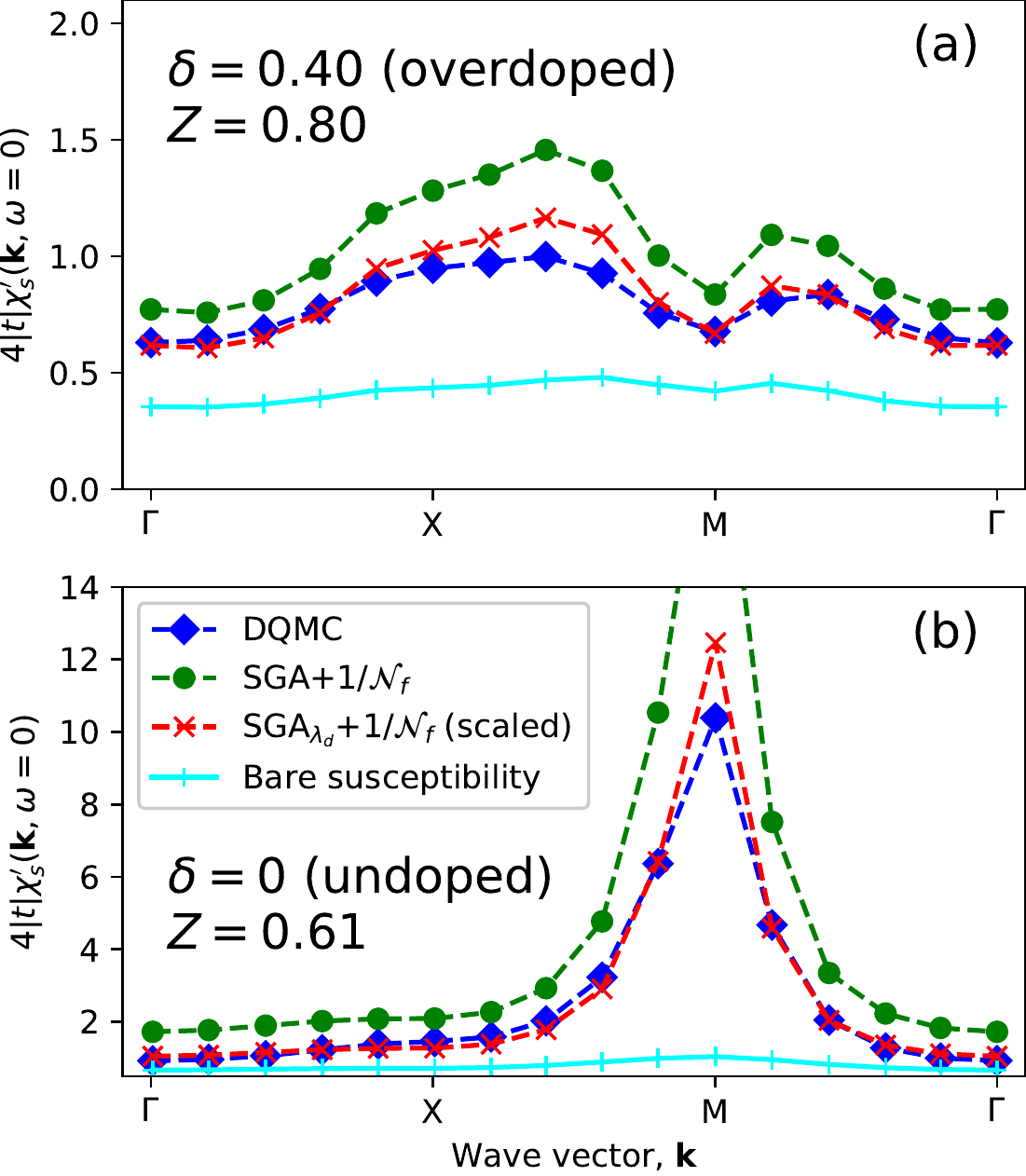}
  \caption{Static spin susceptibilities of the two-dimensional Hubbard model with nearest-neighbor hopping only, and the on-site repulsion  $U/|t| = 8$. Hole doping levels are set to $\delta = 0.40$ [panel (a)] and to $\delta = 0$ at half-filling [panel (b)]. Lattice size is taken as $10 \times 10$. Blue diamonds represent determinant quantum Monte-Carlo data of Ref.~\citenum{KungPhysRevB2017} at temperature $k_B T = |t|/3$, whereas green circles represent corresponding paramagnetic-phase $\mathrm{SGA}_{\lambda_d}$+$1/\mathcal{N}_f$ results for $k_B T = 0.333 |t|$ [panel (a)] and $k_B T = 0.35 |t|$ [panel (b)]. Red crosses are $\mathrm{SGA}_{\lambda_d}$+$1/\mathcal{N}_f$ scaled by respective renormalization factors $Z = 0.80$ for panel (a) and $Z = 0.61$ for panel (b) (cf. discussion in the text). The blue symbols represent correlated Lindhard susceptibilities, $\hat{\chi}_0(\mathbf{k}, \omega = 0)$. Lines are guides to the eye. The results correspond to the paradigmatic phase so that the transverse- and longitudinal-susceptibility components are equivalent. The maximum in (b) signals the tendency towards antiferromagnetic at $M$ point.}
  \label{fig:comparison_static-ld}
\end{figure}

We now turn to the strong-coupling situation and employ particle-hole-symmetric Hubbard model (only nearest-neighbor hopping, $t$, is retained and the on-site repulsion is set to $U/|t| = 8$). In Fig.~\ref{fig:comparison_static-ld}(a)-(b) we display calculated renormalized paramagnetic-state $\mathrm{SGA}_{\lambda_d}$+$1/\mathcal{N}_f$ (see Table~\ref{tab:acronyms}) and compare them with corresponding DQMC results of Ref.~\citenum{KungPhysRevB2017} (green circles and blue diamonds, respectively). Panels (a) and (b) correspond to the overdoped ($\delta = 0.40$) and undoped ($\delta = 0$) cases, respectively. The temperature has been set to $k_B T = |t|/3$ for DQMC, whereas $k_B T = 0.333 |t|$ and $k_B T = 0.35 |t|$ has been used for $\mathrm{SGA}_{\lambda_d}$+$1/\mathcal{N}_f$ in panels (a) and (b), respectively. The reason for selecting such high temperatures in the strong-coupling case is twofold: \emph{(i)} within DQMC it improves the statistics by reducing the sign problem and \emph{(ii)} it allows to stay clear of the spurious low-temperature broken-symmetry states within $\mathrm{SGA}_{\lambda_d}$+$1/\mathcal{N}_f$ (cf. Appendix~\ref{appendix:local_stability_of_phases} for a detailed discussion).

\begin{figure*}
    \onecolumngrid
  \centering
  \includegraphics[width=\linewidth]{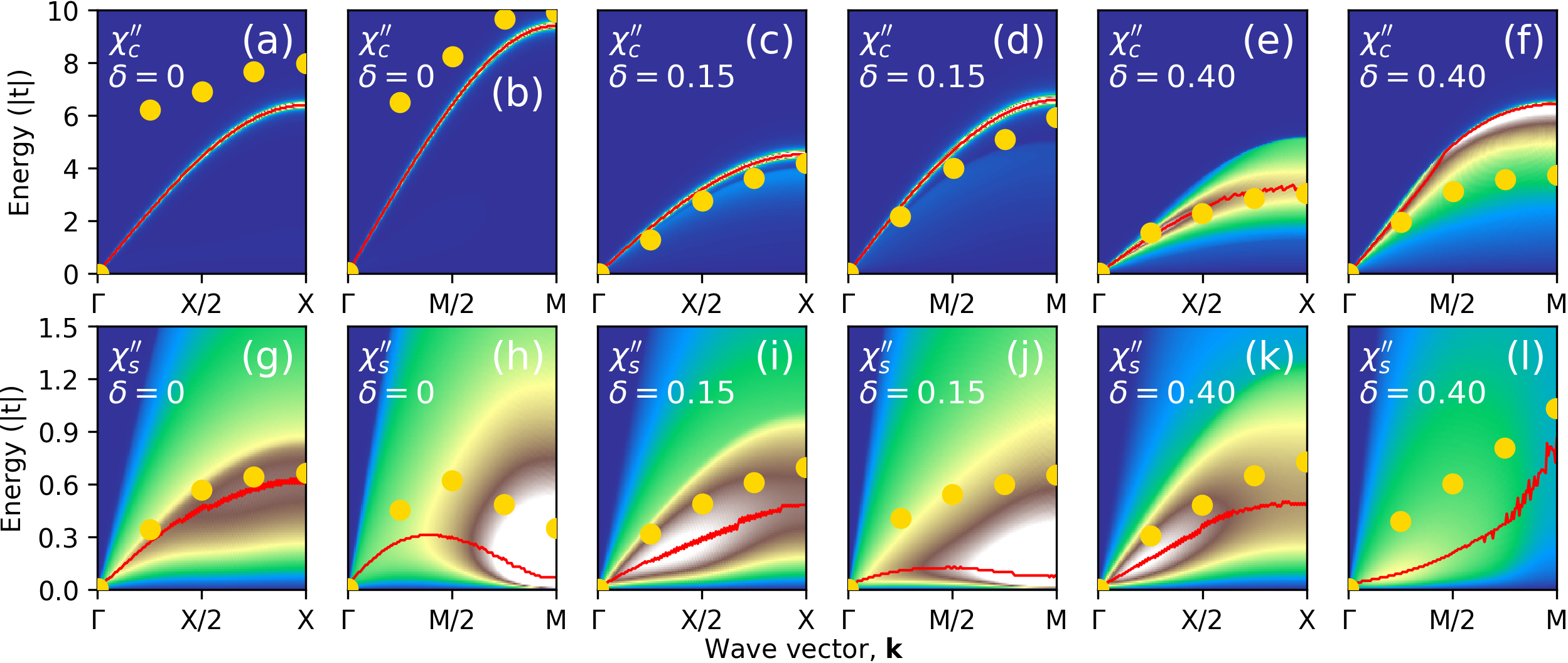}
  \caption{Comparison of imaginary parts of the dynamical charge- and spin susceptibilities (top and bottom panels, respectively) obtained within the $\mathrm{SGA}_{\lambda_d}$+$1/\mathcal{N}_f$ and DQMC methods for undoped ($\delta \rightarrow 0$), near-optimally-doped ($\delta = 0.15$), and overdoped ($\delta = 0.40$) for $300 \times 300$ square-lattice Hubbard model, each along two lines in $\mathbf{k}$-space. The model parameters are: $t < 0$, $t^\prime = 0.3 |t|$, $U = 8|t|$, temperature $k_B T = |t|/3$ (DQMC), $k_BT = 0.35 |t|$ ($\mathrm{SGA}_{\lambda_d}$+$1/\mathcal{N}_f$ at $\delta = 0$), and $k_BT = 0.333 |t|$ ($\mathrm{SGA}_{\lambda_d}$+$1/\mathcal{N}_f$ at $\delta > 0$). Color maps illustrate imaginary parts of $\mathrm{SGA}_{\lambda_d}$+$1/\mathcal{N}_f$ dynamical susceptibilities, whereas red lines follow maximal values of intensities for given wave vector. Yellow symbols are the corresponding maximum-intensity frequencies for analytically continued 64-site DQMC data of Ref.~\citenum{KungPhysRevB2015}. For definition of  $\mathrm{SGA}_{\lambda_d}$+$1/\mathcal{N}_f$, cf. Table~\ref{tab:acronyms}. Analytic continuation was performed as $i \omega_n \rightarrow w + i 0.02 |t|$.}
    \twocolumngrid
  \label{fig:benchmatk-ld}
\end{figure*}

As is apparent from Fig.~\ref{fig:comparison_static-ld}, the results coming from $\mathrm{SGA}_{\lambda_d}$+$1/\mathcal{N}_f$ follow closely the DQMC susceptibility profile along the representative $\Gamma$-$X$-$M$-$\Gamma$ $\mathbf{k}$-space contour (green circles and blue diamonds, respectively), yet they overestimate the latter processes quantitatively in a systematic manner. This behavior is expected to appear within the large-$\mathcal{N}_f$ calculation at strong coupling, since the effects of magnetic spectral-weight redistribution due to multi-magnon (multi-paramagnon) excitations are not included in the leading order. The latter tend to reduce static susceptibility and may be systematically calculated as $1/\mathcal{N}_f$ corrections within the framework of the model defined by Eq.~\eqref{eq:sga_1n_action}. However, this analysis goes beyond the scope of the present contribution and we  take another route by introducing an extra wave-vector-independent spin-susceptibility renormalization factor, ${\chi}_s \rightarrow Z {\chi}_s$. The latter procedure is fully analogous to that employed in linear spin-wave theory calculations, where the values of renormalization factors are known. We have found
$Z \approx 0.61$ for undoped system ($\delta = 0$) and $Z \approx 0.80$ for the overdoped case ($\delta = 0.40$). The $\mathrm{SGA}_{\lambda_d}$+$1/\mathcal{N}_f$ results, rescaled by respective renormalization factors, $Z$, are marked in Fig.~\ref{fig:comparison_static-ld} by red crosses, yielding a semi-quantitative agreement with the DQMC data. The renormalization factors for the undoped case can be directly compared with available results for the square-lattice Heisenberg model, where the spin-wave-theory calculation provides $Z_\chi = 0.4844$,\cite{CanaliPhysRevB1993} whereas within the series expansion method $Z_\chi = 0.52$.\cite{SinghPhysRevB1989} is obtained. Both results are consistent with our estimate for $\delta = 0$ within $\lesssim 30\%$ margin, even though we consider magnetically-disordered state at finite temperature and for finite Hubbard $U$ rather than the N\'{e}el state at $T = 0$ and in strong-coupling limit. As to the overdoped case, where reference results from spin-models are not available, an increase of $Z$ is observed. This type of behavior is expected and we attribute it to reduction of the nonlinear fluctuation corrections due to a loss of the spin-coherence deep in metallic state. Finally, blue symbols represent the correlated Lindhard susceptibility, $\hat{\chi}_0(\mathbf{k}, \omega)$, evaluated at $\omega = 0$ [cf. Eq.~\eqref{eq:correlated_susceptibility}], which is rather featureless for both dopings, in disagreement with the DQMC data. The residual scattering is thus the driving force of a substantial magnetic response enhancement.

\subsection{Dynamic collective modes}
\label{subsection:dynamic_collective_modes}

We now proceed to the collective-mode dynamics. In Fig.~\ref{fig:benchmatk-ld} we compare the $\mathrm{SGA}_{\lambda_d}$+$1/\mathcal{N}_f$ results (cf. Table~\ref{tab:acronyms}) with DQMC for the Hubbard model with nearest- and next-nearest-neighbor hoppings at the strong-coupling. The set of parameters, common for both methods, is $t < 0$, $t^\prime = 0.3 |t|$, $U = 8|t|$, and temperature $k_B T = |t|/3$ (DQMC),  $k_B T = 0.35|t|$ ($\mathrm{SGA}_{\lambda_d}$+$1/\mathcal{N}_f$ at $\delta = 0$), and $k_B T = 0.333|t|$ ($\mathrm{SGA}_{\lambda_d}$+$1/\mathcal{N}_f$ at $\delta > 0$).  Remarkably, for this parameter range, the weak-coupling RPA approach fails outright as it does not yield a locally-stable paramagnetic state. In previous RPA studies, this problem was bypassed by taking nonphysically small $U \sim 1.5|t|$.\cite{GuariseNatCommun2014} The top panels of Fig.~\ref{fig:benchmatk-ld} exhibit imaginary part of the dynamical charge susceptibility $\chi^{\prime\prime}_c$ along $\Gamma$-$X$ and $\Gamma$-$M$ lines, and are evaluated for three hole-doping levels: undoped $\delta = 0$ for (a)-(b), near-optimally-doped $\delta = 0.15$ for (c)-(d), and overdoped $\delta = 0.40$ for (e)-(f). The color map represents the paramagnetic-state $\mathrm{SGA}_{\lambda_d}$+$1/\mathcal{N}_f$ intensity obtained for a $300 \times 300$ lattice (blue and white colors correspond to low- and high-intensities, respectively), whereas red lines follow maximum values of calculated intensities. Yellow symbols are the corresponding DQMC maximum-intensity frequencies obtained for the 64-site lattice.\cite{KungPhysRevB2015} Lower panels represent longitudinal spin susceptibilities, obtained using the same set of parameters as the corresponding top panels (note different energy scales for charge- and spin excitations).

As follows from Fig.~\ref{fig:benchmatk-ld}(a)-(f), the characteristic charge excitation energies agree well between the two techniques, with the exception of $\Gamma$-$X$ direction for undoped systems [panel (a)] and $\Gamma$-$M$ direction for overdoped case [panel (f)]. To understand those discrepancies at both ends, one should take into consideration not only the peak positions, but also widths of the features in the dynamic charge susceptibilities. By inspecting DQMC intensity profiles [Fig. 3(d1)-(d2) of Ref. 69], it is apparent that the energy-width of charge excitations increases along the nodal $\Gamma$-$M$ direction, whereas the opposite tendency is seen for the anti-nodal $\Gamma$-$X$ axis, as the doping changes from $\delta=0$ to $0.4$. In turn, panels (a) and (f), correspond precisely to the situation, where charge modes are incoherent and not well defined. A detailed analysis of the charge-mode lineshapes across the Brillouin zone for both techniques is thus desired to make a robust comparison at those doping levels. Interestingly, $\mathrm{SGA}_{\lambda_d}$+$1/\mathcal{N}_f$ method yields sharp peaks in the charge susceptibility for undoped and intermediately-doped systems (yet a considerable damping appears for the overdoped case), in contrast to relatively broad features in DQMC profiles at \emph{all} doping levels. It is not obvious at this point whether the latter discrepancy should be interpreted as an artifact of a large-$\mathcal{N}_f$ limit, or an inaccuracy related to analytic continuation within DQMC that may smear-out or miss sharp spectral features (see, e.g., Ref.~\citenum{TripoltComPhysCommun2019}). The lattice size in both cases is also vastly different. One may speculate though, that multi-particle scattering channels for those excitations may open up at higher orders in the $1/\mathcal{N}_f$ expansion, leading to finite broadening. Those aspects should be analyzed separately. Finally, we note that the charge modes remain gapless at the $\Gamma$-point within both DQMC and VWF+$1/\mathcal{N}_f$, yet the spectral-weight decreases as one moves towards the Brillouin-zone center. In order to open the plasmon gap at the $\Gamma$-point and allow for a direct interpretation of the charge-excitation data for high-$T_c$ copper oxides, the present model should be supplemented with long-range Coulomb interactions.

Panels (g)-(l) of Fig.~\ref{fig:benchmatk-ld} show calculated imaginary parts of the dynamical longitudinal spin susceptibility $\chi^{\prime\prime}_s$ for the same doping levels as for the top panels (a)-(f). Whereas a semi-quantitative agreement between $\mathrm{SGA}_{\lambda_d}$+$1/\mathcal{N}_f$ and DQMC data is observed for the anti-nodal ($\Gamma$-$X$) direction at all hole concentrations [panels (g), (i), and (k)], notable differences occur for the nodal ($\Gamma$-$M$) line. For the intermediate- and highly-doped cases the discrepancies along $\Gamma$-$M$ line may be, once again, interpreted in terms of a broad DQMC line shapes. Indeed, by inspecting Fig.~2(d1)-(d2) of Ref.~\citenum{KungPhysRevB2015}, we find out that the features in the magnetic response along the nodal direction become extremely broad for $\delta > 0.1$, with widths by far exceeding the peak energies. On the other hand, the anti-nodal paramagnons retain their coherence down to the overdoped regime. Once again, matching of the peak positions alone is thus not a decisive benchmark along $\Gamma$-$M$ direction for the doped systems. On the other hand, the discussion for the $\Gamma$-$X$ direction, where both approaches yield well-defined paramagnons, is more reliable in this context. Parenthetically, such a strongly-anisotropic damping, revealed by both techniques, is consistent with recent experiments on high-$T_c$ cuprates (cf. Sec.~\ref{section:introduction}) and may be semiquantitatively described within the present VWF+$1/\mathcal{N}_f$ approach.\cite{FidrysiakPhysRevB2020} What concerns the undoped ($\delta = 0$) case, the $\mathrm{SGA}_{\lambda_d}$+$1/\mathcal{N}_f$ overestimates the paramagnon damping which leads to underestimated paramagnon energies. Remarkably though, a non-monotonic behavior of the intensity maximum in Fig.~\ref{fig:benchmatk-ld}(h) and (j), seen within DQMC, is also observed within $\mathrm{SGA}_{\lambda_d}$+$1/\mathcal{N}_f$. This is related to strong antiferromagnetic correlations that cause transfer of the magnetic spectral weight to lower energies near the $M$ point.

\begin{figure}
  \centering
  \includegraphics[width=\linewidth]{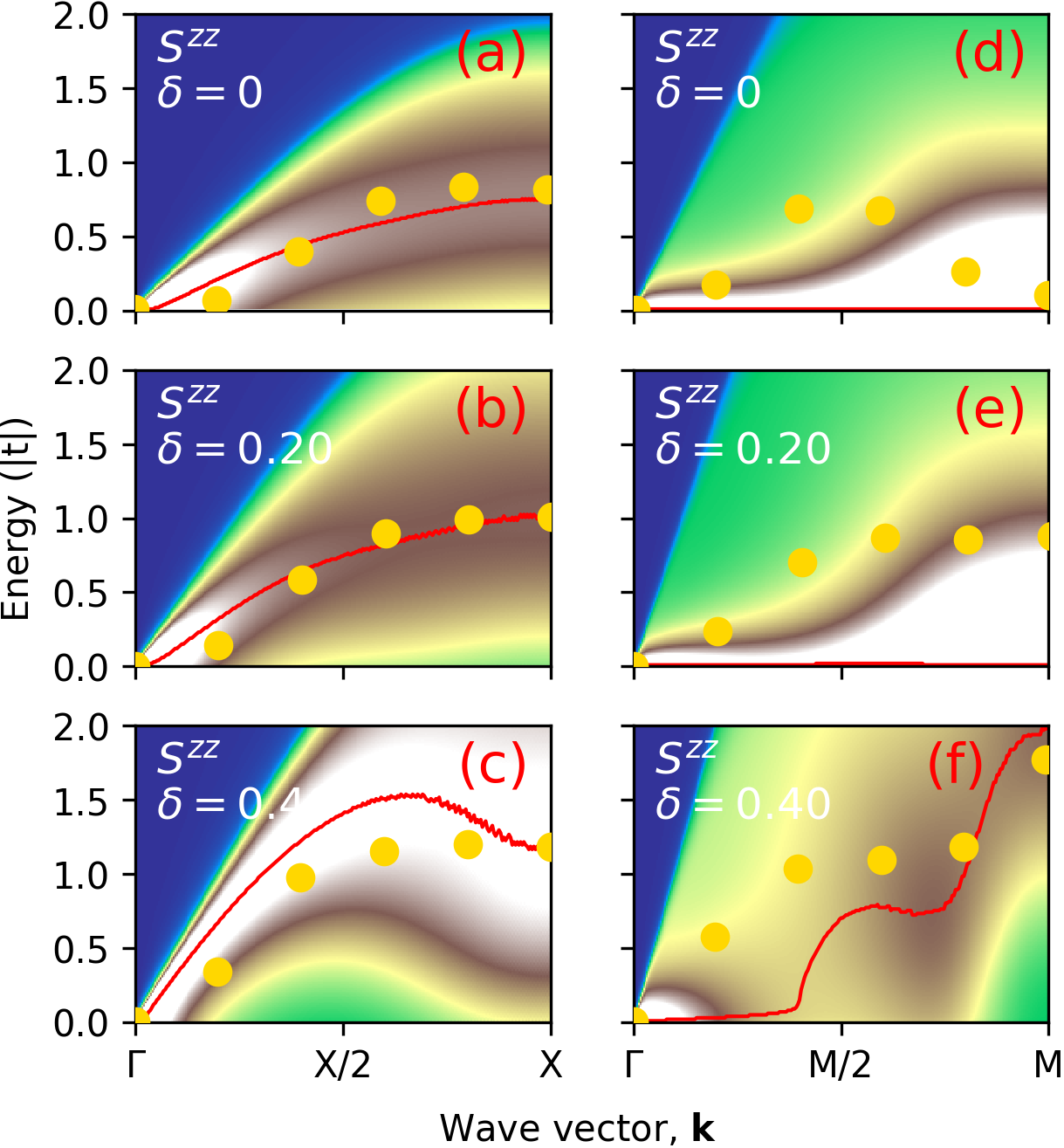}
  \caption{Comparison of imaginary parts of the dynamical longitudinal spin structure factor, $S^{zz}(\mathbf{k}, \omega)$, obtained within the $\mathrm{SGA}_{\lambda_d}$+$1/\mathcal{N}_f$ and DQMC methods for undoped ($\delta = 0$), near-optimally-doped ($\delta = 0.20$), and overdoped ($\delta = 0.40$) square-lattice Hubbard model. The model parameters are: $t < 0$, $t^\prime = 0$, $U = 8|t|$, and temperature $k_B T = |t|/3$ [within $\mathrm{SGA}_{\lambda_d}$+$1/\mathcal{N}_f$ we actually set $k_BT = 0.35 |t|$ for panels (a)-(b) and $k_BT = 0.333 |t|$ for panels (c)-(f); cf. discussion in the text]. Color maps represent imaginary parts of $\mathrm{SGA}_{\lambda_d}$+$1/\mathcal{N}_f$ dynamical susceptibilities, evaluated for $500 \times 500$ lattice, whereas red lines follow maximal values of $S^{zz}(\mathbf{k}, \omega)$. Yellow symbols are the maximum-intensity frequencies for analytically continued $10 \times 10$ lattice DQMC data of Ref.~\citenum{KungPhysRevB2017}. The analytic continuation of $\mathrm{SGA}_{\lambda_d}$+$1/\mathcal{N}_f$ data was performed as $i \omega_n \rightarrow \omega + i \epsilon$ with $\epsilon = 0.04 |t|$.}
  \label{fig:benchmatk-ld-ph_symm}
\end{figure}

In Fig.~\ref{fig:benchmatk-ld-ph_symm} we perform a similar analysis of the maximum-intensity profiles for the particle-hole symmetric Hubbard model ($t^\prime = 0$) and compare the results with the DQMC data of Ref.~\citenum{KungPhysRevB2017}. The model parameters are taken as $U = 8|t|$, $k_B T = |t|/3$ (DQMC), $k_B T = 0.333|t|$ ($\mathrm{SGA}_{\lambda_d}$+$1/\mathcal{N}_f$ for $\delta > 0$), and $k_B T = 0.35|t|$ ($\mathrm{SGA}_{\lambda_d}$+$1/\mathcal{N}_f$ for $\delta = 0$). The slightly elevated temperature for $\delta = 0$ is necessary to avoid an instability toward antiferromagnetic state. Since Ref.~\citenum{KungPhysRevB2017} provides dynamic structure factor in the form

\begin{align}
  \label{eq:dynamic_structure_factor}
  S(\mathbf{k}, \omega) = -\frac{2}{\exp(-\omega/k_BT) - 1} \chi_s^{\prime\prime}(\mathbf{k}, \omega),
\end{align}

\noindent
that differs from the imaginary part of the dynamical longitudinal spin susceptibility by the detailed balance factor, we compute this quantity here as well. Fig.~\ref{fig:benchmatk-ld-ph_symm} shows the comparison of $S(\mathbf{k}, \omega)$, obtained within $\mathrm{SGA}_{\lambda_d}$+$1/\mathcal{N}_f$  with the DQMC result for three doping levels, $\delta = 0$ [panels (a)-(b)], $\delta = 0.2$ [panels (c)-(d)], and $\delta = 0.4$ [panels (e)-(f)]. In full analogy to the particle-hole asymmetric case, the agreement between maximum-intensity profiles along $\Gamma$-$X$ direction is semiquantitative. Both methods yield systematic hardening of magnetic excitations with the increasing doping, more pronounced than for particle-hole non-symmetric case ($t^\prime \neq 0$). This is a counter-intuitive result originating from strong electronic correlations. In the $\Gamma$-$M$ direction, $\mathrm{SGA}_{\lambda_d}$+$1/\mathcal{N}_f$ structure factors for undoped and intermediately-doped case [panels (d) and (e)] exhibit features indicative of the close proximity to antiferromagnetic instability, which is not as pronounced in DQMC. This results in a transfer of the $\mathrm{SGA}_{\lambda_d}$+$1/\mathcal{N}_f$ magnetic spectral weight to low energies, which is further enhanced by the detailed balance factor, divergent for $\omega \rightarrow 0$, cf. Eq.~\eqref{eq:dynamic_structure_factor}. In effect, close to half-filling, the $\mathrm{SGA}_{\lambda_d}$+$1/\mathcal{N}_f$ intensity exhibits maxima for near-zero energies, in contrast to DQMC yielding substantially softened, yet non-zero peak-energies. For the overdoped system [panel (f)] both methods exhibit consistent step-like feature along the $\Gamma$-$M$ line.

\section{Emergence of robust collective excitations at strong coupling}
\label{section:robust_collective_excitations}

Experimental observation of robust propagating spin- and charge excitations in several families of correlated-electron materials calls for identification of the mechanism leading to the enhanced coherence of the collective-modes. Present semi-analytic techniques that have been successful in describing those excitations are based on Hubbard operators and thus apply mostly to the strong-coupling ($t$-$J$-model) limit.\cite{FoussatsPhysRevB2002} Our VWF+$1/\mathcal{N}_f$ approach covers both Hubbard- and $t$-$J$-type models and may be directly used to study the evolution of collective excitations from weak- to strong-coupling limits. In this section we perform such an analysis, employing Hubbard model with $t < 0$, $t^\prime = 0.3 |t|$, $k_BT = 0.35 |t|$, hole-doping $\delta = 0.2$, and a variable on-site Coulomb repulsion, $U$. We also compare the characteristics of various truncation schemes, listed in Table~\ref{tab:acronyms}, as a function of the interaction strength.

\begin{figure}
  \centering
  \includegraphics[width=\linewidth]{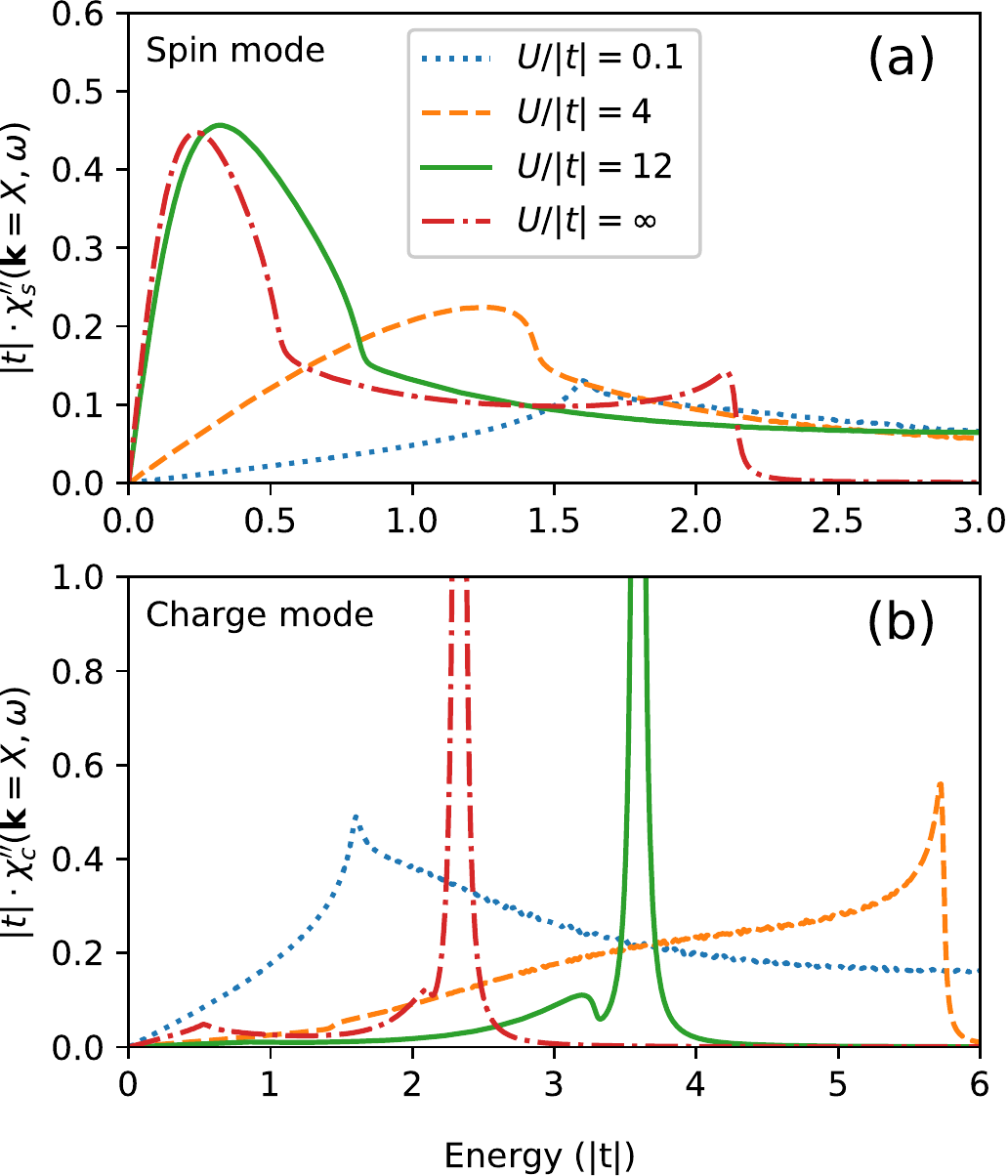}
  \caption{Imaginary parts of (a) spin and (b) charge dynamical susceptibilities as a function of the on-site Coulomb repulsion, $U$, for the Hubbard model with non-zero next- and next-nearest neighbor hopping, taken at $X$ point in the Brillouin zone. The employed model parameters are: $t < 0$, $t^\prime = 0.3 |t|$, $k_BT = 0.35 |t|$, and hole-doping $\delta = 0.2$. The calculations have been performed for a $300 \times 300$ square lattice using $\mathrm{SGA}_{\lambda_d}$+$1/\mathcal{N}_f$ approximation (cf. Table~\ref{tab:acronyms}). }
  \label{fig:susc_vs_u_ld}
\end{figure}

In Fig.~\ref{fig:susc_vs_u_ld} we plot calculated imaginary parts of dynamical spin- and charge susceptibilities at the antinodal $X$ point (panels (a) and (b), respectively) for $U/|t| = 0.1, 4, 12$, and $\infty$. At weak-coupling, magnetic response in panel (a) is broad and featureless, but it systematically acquires coherence as $U$ is increased. Around $U = |8|t$, a resonance-like feature emerges from particle-hole continuum and shifts towards lower energies for $U \rightarrow \infty$ (already at $U/|t| = 4$, an emerging mode with energy  $\sim |t|$ may be noticed). This can be related to well-defined paramagnons in metallic state of high-$T_c$ copper oxides. In Ref.~\citenum{FidrysiakPhysRevB2020} we have been able to reproduce semiquantitatively RIXS data for $\mathrm{La_{2-\mathit{x}}Sr_\mathit{x}CuO_4}$ and $\mathrm{(Bi,Pb)_2(Sr,La)_2CuO_{6+\delta}}$ with the VWF+$1/\mathcal{N}_f$ framework. In contrast to paramagnons, the peak-energy of charge excitations [panel (b)] exhibits a non-monotonic dependence on the interaction strength. For small values of $U$, charge dynamics is highly incoherent with the peak intensity shifting upward with increasing $U$. Already at intermediate interaction strengths, a sharp charge mode emerges form the continuum.  With further increase of the Hubbard $U$, the charge mode energy decreases in order to finally saturate at finite value for $U \rightarrow \infty$. This behavior is not observed in weak-coupling calculations and thus should be interpreted as a footprint of strong electronic correlations. Importantly, the latter mechanism of generating discrete charge peak is distinct from long-range Coulomb repulsion that is known to strongly affect the plasmon dispersion in extended multilayer models.\cite{KresinPhysRevB1988} Note the substantial downward shift of charge mode energies in the range $U/|t| = 12$ and $U/|t| = \infty$ (solid green- and red dash-dot lines in panel (b), respectively). The finite-$U$ effects are thus not negligible which calls for a reexamination of the $t$-$J$-model applicability to charge-modes in high-$T_c$ copper oxides in favor of more general models, e.g., $t$-$J$-$U$ or $t$-$J$-$U$-$V$.\cite{SpalekPhysRevB2017} This last suggestion is particularly relevant, since those extended models lead to a good overall and semiquantitative description of the equilibrium properties of the cuprates.\cite{ZegrodnikPhysRevB2020,ZegrodnikPhysRevB2019} Those effects should be treated separately.

\begin{figure}
  \centering
  \includegraphics[width=\linewidth]{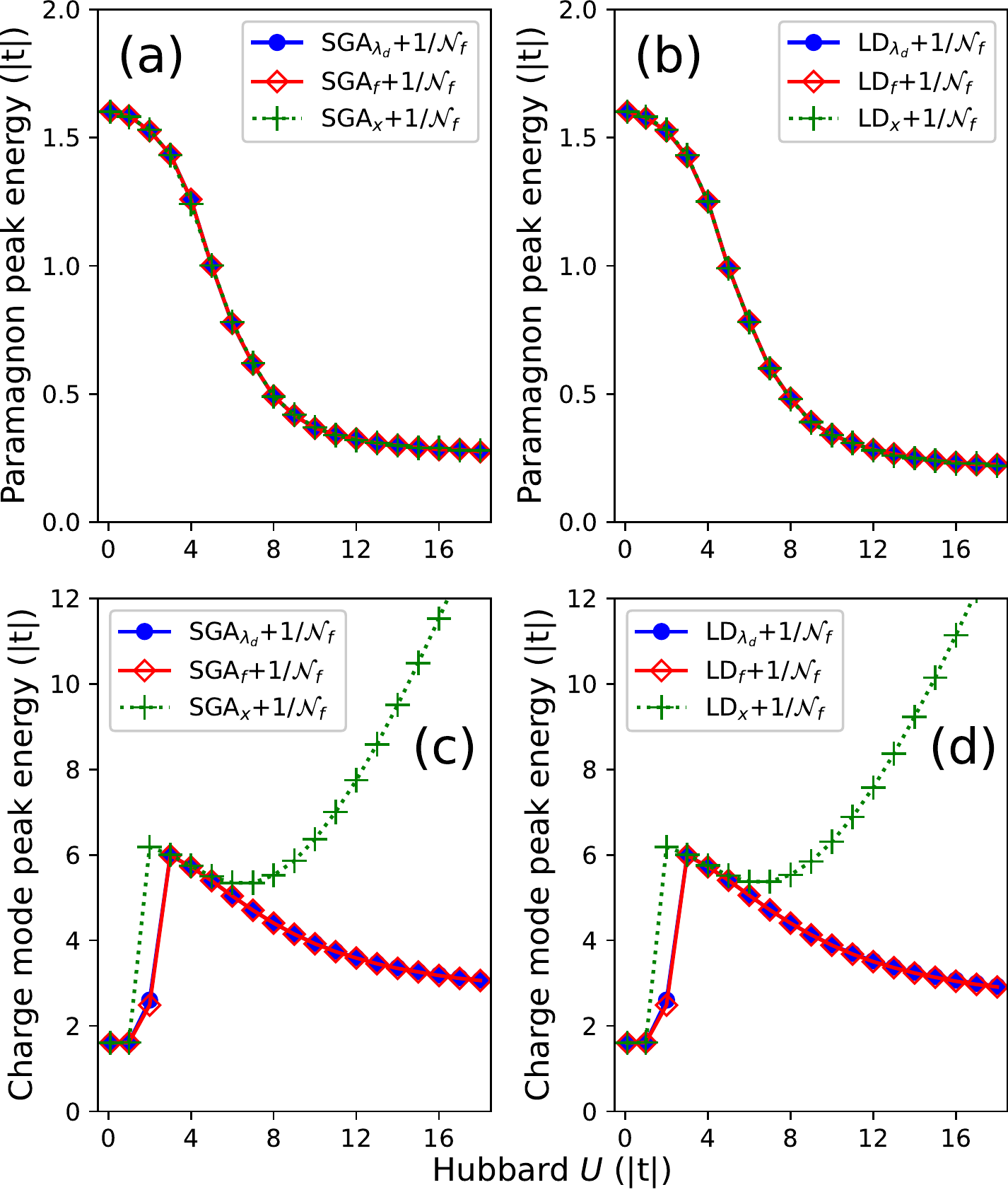}
  \caption{Comparison of the peak energies for spin- [top panels, (a)-(b)] and charge [bottom panels, (c)-(d)] dynamical susceptibilities for the Hubbard model as a function of the Hubbard $U$. The remaining parameters are: $t < 0$, $t^\prime = 0.3|t|$, $k_B T = 0.35 |t|$, hole doping $\delta = 0.2$, and lattice size is taken as $300 \times 300$. All susceptibilities are evaluated at the $X$ point, and four symbols described inside the panels (a)-(d) correspond to six schemes listed in Table~\ref{tab:acronyms}. Lines are guides to the eye.}
  \label{fig:susc_vs_u_comparison}
\end{figure}

In Fig.~\ref{fig:susc_vs_u_comparison}, a more detailed account of the crossover effects in collective-mode characteristics is presented. Top panels [(a)-(b)] and bottom panels [(c)-(d)] show the dependence of the peak energies on Hubbard-$U$ magnitude for spin- and charge excitations, respectively. The symbols, listed in the legend, correspond to six detailed VWF+$1/\mathcal{N}_f$ variants, summarized in Table~\ref{tab:acronyms}. As is apparent from panels (a) and (b), the paramagnon peak-energies nearly collapse onto each other for all variants of VWF+$1/\mathcal{N}_f$. This indicates that neither the doublon excitation sector nor the additional graphs included in full ``local diagram'' (LD) approximation affect noticeably spin dynamics. From panels (b)-(c), it follows that $\mathrm{SGA}_{\lambda_d}+1/\mathcal{N}_f$, $\mathrm{LD}_{\lambda_d}+1/\mathcal{N}_f$, $\mathrm{SGA}_{f}+1/\mathcal{N}_f$, and $\mathrm{LD}_{f}+1/\mathcal{N}_f$ schemes provide nearly the same charge-mode energies that decrease with $U$ at strong coupling and saturate for $U \rightarrow \infty$. This behavior is qualitatively consistent with the results for the charge dynamics in the $t$-$J$ model, obtained previously within the Hubbard-operator language in the large-$\mathcal{N}_f$ limit.\cite{FoussatsPhysRevB2002} The situation changes qualitatively for the case of $\mathrm{SGA}_{x}+1/\mathcal{N}_f$ and $\mathrm{LD}_{x}+1/\mathcal{N}_f$ approximations that both yield an abrupt jump of charge mode energy scale and monotonically increasing peak energies in the strong-coupling regime. The discontinuous jump is a consequence of a shoulder building up at the threshold of particle-hole continuum that rapidly grows in intensity for $U/|t| \gtrsim 2$. The monotonic increase of mode energy for $\mathrm{SGA}_{x}+1/\mathcal{N}_f$ and $\mathrm{LD}_{x}+1/\mathcal{N}_f$ at strong coupling is likely an unphysical behavior that we attribute to poor handling of the doublon excitations within those schemes at low expansion order. Physically, doublon dynamics is expected to occur on much faster timescale then the collective charge mode which may be effectively accounted for by freezing-out $\lambda_d$ Lagrange-multiplier or integrating its fluctuations out. This is achieved within $\mathrm{SGA}_{\lambda_d}$+$1/\mathcal{N}_f$ and $\mathrm{LD}_{\lambda_d}$+$1/\mathcal{N}_f$ ($\lambda_d \equiv \bar{\lambda}_d$ condition), and $\mathrm{SGA}_{f}$+$1/\mathcal{N}_f$ and $\mathrm{LD}_{f}$+$1/\mathcal{N}_f$ (fluctuating $\lambda_d$), but not within $\mathrm{SGA}_{x}$+$1/\mathcal{N}_f$ and $\mathrm{LD}_{x}$+$1/\mathcal{N}_f$ truncations. The choice of a  detailed expansion scheme does not matter as the model is solved exactly, but it generates noticable differences at the saddle-point and Gaussian-fluctuation levels. Finally, it is reassuring that, in all cases, the higher-order diagrams included in LD approximation affect only marginally the collective mode energy relative to the less demanding SGA approach so that SGA-based approximations may be sufficient in most applications. As is shown in Appendix~\ref{appendix:local_stability_of_phases}, those additional diagrams lead to a minor enhancement of the magnetic susceptibility value.

\section{Discussion and outlook}
\label{section:discussion}

Several qualitative comments may be formulated based on the analysis of Secs.~\ref{section:benchmark_Hubabrd_model}-\ref{section:robust_collective_excitations}. First, as is apparent from Figs.~\ref{fig:benchmatk-ld} and \ref{fig:benchmatk-ld-ph_symm}, the agreement between VWF+$1/\mathcal{N}_f$ and available DQMC data is quantitative in the experimentally relevant case of coherent collective-mode dynamics. Second, the position of the peak-maxima in dynamical susceptibilities, as obtained within the two methods, tend to differ quantitatively for the doping levels and Brillouin-zone contours, for which collective modes become excessively broad. In the latter situation, such a benchmark should be considered as overly simplified and thus insufficient. Based on the our present simulations and available experimental data for the cuprates, one can infer that low-order VWF+$1/\mathcal{N}_f$ calculation tends to overestimate paramagnon damping in the regime of incoherent magnetic dynamics.\cite{FidrysiakPhysRevB2020}

An important additional remark should be made on the reported collective-mode energy values. The peak maxima of the dynamical susceptibilities that are commonly associated with a characteristic collective mode energies in DQMC studies, do not generally reflect the physical dispersion of those excitations (either spin or charge). This problem has been addressed in detail recently in the context of extracting paramagnon energies from RIXS experiments.\cite{LamsalPhysRevB2016} Namely, one should carefully distinguish between several frequency scales appearing in the problem. Assuming a harmonic-oscillator model, dynamical susceptibilities may be expressed in the standard form

\begin{align}
  \label{eq:chi_prime_oscillator}
  \chi^{\prime\prime}(\mathbf{k}, \omega) \propto \frac{\gamma(\mathbf{k}) \omega}{(\omega^2 - \omega_0(\mathbf{k})^2)^2 + 4\gamma(\mathbf{k})^2 \omega^2} + \chi^{\prime\prime}_\mathrm{incoh}(\mathbf{k}, \omega).
\end{align}

\noindent
The susceptibility depends on the wave-vector-dependent damping rate, $\gamma(\mathbf{k})$, and bare frequency, $\omega_0(\mathbf{k})$. Yet, another energy appears naturally as the maximum of $\chi^{\prime\prime}(\mathbf{k}, \omega)$ for given $\mathbf{k}$, which may be roughly identified with $\omega_0(\mathbf{k})$ for $\gamma \ll \omega_0$. In the above, $\chi^{\prime\prime}_\mathrm{incoh}$ accommodates all incoherent and non-resonant contributions, such as particle-hole and multi-magnon continua. Whereas $\omega_0(\mathbf{k})$ appears explicitly in the formula \eqref{eq:chi_prime_oscillator}, the energy scale of physical relevance is the so-called propagating frequency, $\omega_p(\mathbf{k}) \equiv \sqrt{\omega_0^2(\mathbf{k}) - \gamma(\mathbf{k})^2}$ for $\gamma(\mathbf{k}) < \omega_0(\mathbf{k})$ and $\omega_p(\mathbf{k}) = 0$ otherwise. The value $\omega_p(\mathbf{k})$ reflects the real part of the collective quasiparticle pole and is of physical relevance. In effect, \emph{the mode may become overdamped even if} $\omega_0(\mathbf{k})$ \emph{is large}. Ideally, comparing various techniques should thus be based on the \emph{propagating energies} rather than peak energies. The value of $\omega_p(\mathbf{k})$ may be estimated based on VWF+$1/\mathcal{N}_f$ calculation.\cite{FidrysiakPhysRevB2020} We are not aware of any quantitative study of propagating frequencies within DQMC for the Hubbard model. The discrepancies between the propagating and bare frequencies are most pronounced in the physically interesting case of intermediately- to highly-doped systems.

We now discuss the relation of our VWF+$1/\mathcal{N}_f$ technique to variety of other approaches, with which is shares common features. The first is Kotliar-Ruckenstein (KR) slave-boson method that takes Gutzwiller approximation as the saddle point,\cite{KotilarPhysRevLett1986} and may be systematically supplemented with the fluctuation effects at higher expansion orders.\cite{LavagnaPhysRevB1990} The KR method has been successfully applied to model Hamiltonians (see, e.g., Ref.~\citenum{LillyPhysRevLett1990}), but this approach requires a careful treatment of emerging gauge symmetry and, akin to the plain $1/\mathcal{N}_f$ expansion, suffers from ambiguities that need to be handled on case by case basis.\cite{LiPhysRevB1989} The essential advantage of our VWF+$1/\mathcal{N}_f$ over slave-bosons is that it allows to go systematically beyond the Gutzwiller approximation ($d=\infty$ case) already at the saddle-point level. This is carried out by incorporating higher-order diagrammatic corrections  (cf. Sec.~\ref{section:variational_approach}).\cite{KaczmarczykNewJPhys2014,FidrysiakJPhysCondensMatter2018} Since the $d=\infty$ solution is known to exhibit artifacts that are not present at finite dimensions,\cite{vanDongenPhysRevB1989} this is a qualitative rather than technical improvement. Here we have restricted ourselves to the lowest-order non-trivial extension beyond $d=\infty$ limit that we dub ``local diagrammatic'' approximation (cf. Table~\ref{tab:acronyms}). Non-local diagrammatic contributions are more technically demanding\cite{KaczmarczykNewJPhys2014,FidrysiakJPhysCondensMatter2018} and should be analyzed separately. A related approach in the spin-rotationally invariant form has been analyzed recently,\cite{RieglerPhysRevB2020} and the noncollinear magnetic and fluctuations in the lowest order calculated.

Another family of approximations that may be connected to VWF+$1/\mathcal{N}_f$ relies on the Hubbard-operator representation of the Hamiltonian, for which a special diagrammatic technique has been developed\cite{IzyumovJPCM1991,IzyumovPhysRevB1992,OvchinnikovBook2004}. Resummation of diagrams leads to generalized random-phase approximation.\cite{IzyumovJPCM1992} Also, a different approach\cite{FoussatsPhysRevB2002} based on path-integral $1/N$ treatment of Hubbard operators  has been recently applied to the extended $t$-$J$ models to study charge modes in high-$T_c$ cuprates.\cite{GrecoPhysRevB2016,GrecoJPSJ2017,GrecoPhysRevB2020} The technique provides good agreement with RIXS charge-mode energies if interlayer hopping and long-range Coulomb interactions are included in the Hamiltonian. In this context, a qualitative advantage of VWF+$1/\mathcal{N}_f$ is its straightforward applicability both to the $t$-$J$ model and the finite-$U$ Hubbard-type models on the same footing, whereas Hubbard-operator-based techniques become excessively complex in the latter case. As we demonstrated in Sec.~\ref{section:robust_collective_excitations}, finite-$U$ effects have a non-trivial impact on the charge mode. Also, it has been long understood that finite-$U$ effects impact substantially the high-energy magnetic excitations in Hubbard-type models, particularly at the magnetic Brillouin-zone boundary.\cite{PeresPhysRevB2002} The magnetic dispersion along those contours may be accurately reproduced by taking $U/|t| \lesssim 8$, which is far from the $t$-$J$-model limit. Even though it is possible to describe high-energy spin-wave dispersion using $t$-$J$ Hamiltonians, the long-range exchange interactions and cyclic exchange terms need to be then incorporated into the Hamiltonian,\cite{PengNatPhys2017,ChaoPhysRevB1978} suggesting that the strong-coupling limit is not an adequate starting point to address high-energy paramagnons in the cuprates. In addition, in contrast to the Hubbard-operator based techniques, VWF+$1/\mathcal{N}_f$ allows for a systematic improvement of the saddle point solution by increasing the dimension of the variational space.

Another relevant technique is time-dependent Gutzwiller approximation\cite{SeiboldPhysRevLett2001,SeiboldPhysRevB2004,SeiboldPhysRevB2005} (dubbed Gutzwiller + random-phase-approximation, GA+RPA), which has been applied to study spin dynamics in lattice systems.\cite{MarkiewiczPhysRevB2010,BunemannNJP2013} An extension of this concept by means of adaptive canonical transformation has been proposed recently.\cite{WysokinskiPhysRevB2017,WysokinskiPhysRevB2017_2} The VWF+$1/\mathcal{N}_f$ and GA+RPA methods are similar in the sense that both represent a dynamical extension of the variational wave function, but the realization of this goal is achieved in different manner. Namely, VWF+$1/\mathcal{N}_f$ is a path-integral method applicable at finite-temperature, whereas GA+RPA is a $T=0$ formalism based on the equations of motion for the density-matrix. Also, in contrast to GA+RPA, VWF+$1/\mathcal{N}_f$ provides a small parameter ($1/\mathcal{N}_f$) that allows for a systematic incorporation of higher-order thermodynamic corrections, e.g., those due to multi-paramagnon effects. The latter renormalize various characteristics of the paramagnon spectrum and become increasingly important as the half-filling is approached, as we pointed out in Sec.~\ref{section:benchmark_Hubabrd_model}. It is not clear to us how to construct such a systematic extension within GA+RPA.

At the end, we elaborate on limitations and possible extension of our VWF+$1/\mathcal{N}_f$ method. First of all, $1/\mathcal{N}_f$ expansion in higher orders is computationally demanding, since it relies on evaluating of high-dimensional loop integrals. Here we have limited ourselves to the discussion of dynamical susceptibilities around correlated large-$\mathcal{N}_f$ limit. With an efficient implementation, the $\mathcal{O}(1)$ fluctuation corrections to thermodynamics should be still accessible without major investment in supercomputing. Including the latter should allow to eliminate spurious ordered phases at high temperatures by restoring the Mermin-Wagner theorem, and to explore the regime of low temperatures down to $T = 0$. Second, VWF+$1/\mathcal{N}_f$ suffers from a rapid proliferation of the number of lines, $P_{\alpha\beta}$, both as a consequence of incorporating diagrammatic corrections to the variational energy and including longer-range hopping and Coulomb integrals into the Hamiltonian. The dimension of the effective scattering matrix becomes thus substantial. Finally, in the present contribution we have completely disregarded superconducting fluctuations, represented by anomalous lines $S_{\alpha\beta} \propto \langle c^\dagger_\alpha c^\dagger_\beta \rangle_0$ \cite{KaczmarczykNewJPhys2014}. The latter may be incorporated by appropriate generalization of our version of the Hubbard-Stratonovich transformation, introduced in Sec.~\ref{section:large_n_expansion}. However, this would lead to doubling of the effective Hamiltonian dimension and result in further proliferation of lines. Therefore, VWF+$1/\mathcal{N}_f$ may be expected to be applicable to relative small (few-orbital) superconducting extended lattice systems. This extension, particularly to encompass superconducting fluctuations, should be the subject of a separate study.

One should note that the present formulation may be also applied to the discussion of single-particle properties with inclusion of the Gaussian fluctuations in the correlated state. As our approach is formulated  in the spirit of RPA, one can expect a strong renormalization of one-particle dynamics.
Finally, the approach can be extended to the discussion of magnetic and superconducting phases. In the former case, the approach should be complementary to that of Hertz,\cite{HertzPhysRevB1976} Moriya,\cite{MoriyaBook1985} and Millis\cite{MillisPhysRevB1993} (for review see, e.g., Ref.~\citenum{LohneysenRevModPhys2007}). In the latter case we should be able to apply the method for the description of unique ferromagnetic superconductors $\mathrm{UGe_2}$, $\mathrm{URhGe}$, $\mathrm{UCoGe}$, and $\mathrm{UIr}$ out of which the first, $\mathrm{UGe_2}$, can be described semiquantitatively by the variational approach,\cite{KadzielawaMajorPhysRevB2018} whereas in the remaining members of the family the effects of spin fluctuations are systematically enhanced.\cite{TateiwaPhysRevB2017} Obviously, the complexity of present approach will increase remarkably due to the additional order parameter appearance and associated with them quantum fluctuations in a multiorbital structure of the uranium compounds.

\textit{Note added.}---In a very recent paper,\cite{HillePhysRevResearch2020} a modified version of the functional renormalization group (fRG) scheme has been formulated and compared quantitatively with DQMC. As far as the static spin susceptibility is concerned, the method provides similar results to ours. However, no detailed analysis of the collective mode dynamics is carried out. We thank Referee for bringing Ref.~\citenum{HillePhysRevResearch2020} to our attention.

\section*{Acknowledgments}
The discussion with Krzysztof Wohlfeld is gratefully acknowledged. Suggestions from our colleagues, Micha{\l} Zegrodnik and Andrzej Biborski have been also valuable. This work  was  supported  by  Grant  OPUS  No.  UMO-2018/29/B/ST3/02646   from   Narodowe   Centrum   Nauki (NCN). The numerical analysis was carried out on dedicated supercomputer EDABI and was partly funded by the Priority Research Area SciMat under the program Excellence Initiative -- Research University at the Jagiellonian University in Krak\'ow.

\appendix

\section{Representation of the constraints}
\label{appendix:constraint_representation}

Here we show that the constraints \eqref{eq:c1}-\eqref{eq:c5} enforce the tadpole-cancellation condition \eqref{eq:contraints_pg2_definition}. First, we argue that in Eqs.~\eqref{eq:c2}-\eqref{eq:c5} one can commute $\hat{P}_G$ to the left, i.e., those equations may be represented as

\begin{align}
  \label{appendix:p_g_commute}
  \langle \hat{P}_{G, I}^2 \hat{c}^\dagger_{I\sigma} \hat{c}_{I\sigma^\prime} \rangle_0 = \langle \hat{c}^\dagger_{I\sigma} \hat{c}_{I\sigma^\prime}\rangle_0.
\end{align}

\noindent
For the extended off-diagonal correlator considered here, this is a non-trivial statement, since $\hat{P}_{G, I}$ does not commute with $\hat{c}^\dagger_{I\sigma} \hat{c}_{I\sigma^\prime}$ in general. Yet, one can perform a unitary transformation $\tilde{\hat{c}}^\dagger_\sigma \rightarrow \hat{\mathcal{U}}_{\sigma\sigma^\prime} \hat{c}^\dagger_{\sigma^\prime}$ so that $\langle{\tilde{\hat{c}}^\dagger_\sigma \tilde{\hat{c}}_{\sigma^\prime}}\rangle_0 = \delta_{\sigma\sigma^\prime} \langle \tilde{\hat{n}}_\sigma\rangle_0$. Eqs.~\eqref{eq:c2}-\eqref{eq:c5} may be then jointly written as $\langle {{\hat{P}}}_{G, I} \tilde{\hat{c}}^\dagger_{I\sigma} \tilde{\hat{c}}_{I\bar{\sigma}} {{\hat{P}}}_{G, I} \rangle_0 = \delta_{\sigma\sigma^\prime} \langle \tilde{\hat{n}}_\sigma\rangle_0$. For the off-diagonal terms, one gets

\begin{align}
  \langle {\hat{{P}}}_{G, I} \hat{\tilde{c}}^\dagger_{I\sigma} \hat{\tilde{c}}_{I\bar{\sigma}} {\hat{{P}}}_{G, I} \rangle_0 = \tilde{\lambda}^{\bar{\sigma}\sigma}_I \cdot \left(\sum_\alpha \tilde{\lambda}^{\alpha\alpha}_I \langle |\tilde{\alpha}\rangle_I{}_I\langle\tilde{\alpha}| \rangle_0\right) = 0,
\end{align}

\noindent
where the rotated correlator parameters $\lambda_I^{\sigma\sigma^\prime}$ are obtained from the condition

\begin{align}
    \hat{P}_{G, I} \equiv  \tilde{\lambda}^0_{I} \lvert \tilde{0}\rangle_{I}{}_{I}\langle \tilde{0}\rvert + \sum_{\sigma\sigma^\prime}\tilde{\lambda}^{\sigma\sigma^\prime}_{I} \lvert\tilde{\sigma}\rangle_{I}{}_{I}\langle \tilde{\sigma}^\prime\rvert  + \tilde{\lambda}^{d}_{I} \lvert\tilde{\uparrow\downarrow}\rangle_{I}{}_{I}\langle \tilde{\uparrow\downarrow}\rvert.
\end{align}

\noindent
As long as the electron density is non-zero and (as we have assumed) $\tilde{\lambda}^{\alpha\alpha}_{I} > 0$, the expression in the bracket is positive and we have $\tilde{\lambda}^{\bar{\sigma}\sigma}_I = 0$. The density matrix and $\lambda^{\sigma\sigma^\prime}_I$ are thus simultaneously diagonalizable for the field configurations satisfying the constraints. In that diagonal basis ($\lambda_I^{\sigma\bar{\sigma}} = 0$ and $\langle{\hat{c}^\dagger_{I\sigma} \hat{c}_{I\bar{\sigma}}}\rangle$ = 0) it is easy to verify that $\hat{P}_{G, I}$ may be commuted to the left. Finally, Eq.~\eqref{appendix:p_g_commute} along with Eq.~\eqref{eq:c1} effectively enforces \eqref{eq:contraints_pg2_definition}. This can be seen on diagrammatic level, where  it implies that diagrams with a single tadpole attached to $\hat{P}_{G, I}^2$ cancel out and $\langle\hat{P}_{G, I}^2\rangle_0 = 1$. Furthermore, we have also checked directly in simulations that the constraints~\eqref{eq:c1}-\eqref{eq:c5} and \eqref{eq:contraints_pg2_definition} lead to the same saddle-point values of $\boldsymbol{\lambda}$-parameters.

\onecolumngrid

\section{Linked cluster theorem for rotationally-invariant correlators}
\label{appendix:linked_cluster_theorem}

Here we justify the decomposition of Eqs.~\eqref{eq:diagrammatics_one_site} and \eqref{eq:diagrammatics_two_site}, employing arguments analogous to those developed in Ref.~\citenum{BunemannEPL2012}, carefully generalized to the rationally invariant off-diagonal correlators and general (either isotropic or broken-symmetry) states. For brevity, we introduce joint position and orbital indices, $I \equiv (i, l)$, and locally-correlated operators $\hat{A}^{(\alpha)^\prime}_I \equiv \hat{P}_{G,I} \hat{A}^{(\alpha)}_{I} \hat{P}_{G,I}$, indexed by $\alpha = 1, \ldots, n$. The starting point is the partial result that under the assumption $\langle{\hat{A}^{(\alpha)^\prime}_I}\rangle_0 = 0$ for each $\alpha$, one has

\begin{align}
  \label{appendix:linked_cluster_theorem_eq}
\frac{\langle \hat{P}_G \hat{A}^{(1)}_{I_1} \ldots \hat{A}^{(n)}_{I_n} \hat{P}_G\rangle_0}{\langle \hat{P}_G^2\rangle_0} = \sum\limits_{k=0}^\infty \sideset{}{'}\sum_{J_1 \ldots J_k} \frac{x_{J_1} \ldots x_{J_k}}{k!} \langle \hat{A}^{(1)\prime}_{I_1} \ldots \hat{A}^{(n)\prime}_{I_n} \hat{d}^\prime_{J_1} \ldots \hat{d}^\prime_{J_k}\rangle_0^c,
\end{align}

\noindent
where the superscript ``$c$'' indicates that only the diagrams with all internal vertices $\hat{d}_{J_\alpha}$ connected to the external vertices, $\hat{A}^{(\alpha)}_{I_\alpha}$, are retained in the Wick's decomposition . Primed summation means that the indices $J_\alpha$ are constrained to be all different and also take values distinct from $I_1, \ldots, I_n$. To show Eq.~\eqref{appendix:linked_cluster_theorem_eq}, we transform the numerator of the left-hand-side

\begin{align}
  \label{label:appendix_transformation}
  & \langle \hat{P}_G \hat{A}^{(1)}_{I_1} \ldots \hat{A}^{(n)}_{I_n} \hat{P}_G\rangle_0 =  \langle \hat{A}^{(1)\prime}_{I_1} \ldots \hat{A}^{(n)\prime}_{I_n} \prod_{J \neq I_1, \ldots, I_n} P_{G, J}^2\rangle_0 = \sum\limits_{k=0}^\infty \sideset{}{'}\sum_{J_1 \ldots J_k} \frac{x_{J_1} \ldots x_{J_k}}{k!} \langle \hat{A}^{(1)\prime}_{I_1} \ldots \hat{A}^{(1)\prime}_{I_n} \hat{d}^\prime_{J_1} \ldots \hat{d}^\prime_{J_k}\rangle_0,
\end{align}

\noindent
where we have made use of the correlator property $\hat{P}_{G, I}^2 = 1 + x_I \hat{d}^\prime_I$. Importantly, the summation restrictions in the second line of Eq.~\eqref{label:appendix_transformation} may be relaxed if diagrammatic expansion is slightly altered, i.e.,

\begin{align}
    \label{label:appendix_transformation_2}
  \langle \hat{P}_G \hat{A}^{(1)}_{I_1} \ldots \hat{A}^{(n)}_{I_n} \hat{P}_G\rangle_0 = \sum\limits_{k=0}^\infty \sum_{J_1 \ldots J_k} \frac{x_{J_1} \ldots x_{J_k}}{k!} \langle \hat{A}^{(1)\prime}_{I_1} \ldots \hat{A}^{(1)\prime}_{I_n} \hat{d}_{J_1} \ldots \hat{d}_{J_k}\rangle_0^\mathrm{modified},
\end{align}

\noindent
where the subscript ``modified'' means that, in the Wick's expansion, we reject  diagrams involving local loops with a leg attached to internal vertices $\hat{d}_{J_{\alpha}}$ (loops attached solely to external vertices $\hat{A}^{(\alpha)}_{I_\alpha}$ are allowed). Note lack of ``primed'' double-occupancy operators and primed summation in Eq.~\eqref{label:appendix_transformation_2}.

At this point it may be clarified why the assumption $\langle{\hat{A}^{(1)\prime}_{I_n}}\rangle_0 = 0$ is needed. Namely, this condition ensures that all ``modified'' diagrams involving the new operators $\hat{d}_{I_1}, \ldots, \hat{d}_{I_n}$, inserted at the sites already occupied by external vertices, cancel out automatically.  To show that, we point out that $\langle{\hat{A}^{(1)\prime}_{I_n}}\rangle_0 = 0$ implies that non-zero contributions originate solely from diagrams in which $\hat{A}^{(1)\prime}_{I_n}$ is connected to some other operator by at least one fermionic line. The expectation values of the type $\langle \hat{A}^{(1)\prime}_{I_1} \ldots \hat{A}^{(1)\prime}_{I_n} \hat{d}_{J_1} \ldots \hat{d}_{J_k}\rangle_0^\mathrm{modified}$ then vanish identically if any of the indices $J_1, \ldots, J_k$. is equal to an external vertex position (in the range $I_1, \ldots, I_n$). Indeed, without loss of generality, let us assume that $J_1 = I_1$ so that $\langle \hat{A}^{(1)\prime}_{I_1} \hat{c}^\dagger_{I_1 \uparrow} \hat{c}_{I_1 \uparrow} \hat{c}^\dagger_{I_1 \downarrow} \hat{c}_{I_1 \downarrow}  \cdot \text{other operators}\rangle_0^\mathrm{modified}$. In the wick decomposition of this expression one always encounters pairs of truncations of the type

\begin{align}
  \contraction{\langle}{\hat{A}^{(1)\prime}_{I_1}}{}{\langle  \hat{A}^{(1)\prime}_{I_1} \hat{c}^\dagger_{I_1 \uparrow}  \hat{c}_{I_1 \uparrow} \hat{c}^\dagger_{I_1 \downarrow} \hat{c}_{I_1 \downarrow}  \cdot \text{other operators}\rangle_0}
  \contraction[1em]{\langle  \hat{A}^{(1)\prime}_{I_1} \hat{c}^\dagger_{I_1 \uparrow}}{\hat{c}^\dagger}{}{\langle  \hat{A}^{(1)\prime}_{I_1} \hat{c}^\dagger_{I_1 \uparrow}  \hat{c}_{I_1 \uparrow} \hat{c}^\dagger_{I_1 \downarrow} \hat{c}_{I_1 \downarrow}  \cdot \text{other operators}\rangle_0}
 \langle  \hat{A}^{(1)\prime}_{I_1} \hat{c}^\dagger_{I_1 \uparrow}  \hat{c}_{I_1 \uparrow} \hat{c}^\dagger_{I_1 \downarrow} \hat{c}_{I_1 \downarrow}  \cdot \text{other operators}\rangle_0^\mathrm{modified}
\end{align}

and

\begin{align}
  \contraction{\langle  \hat{A}^{(1)\prime}_{I_1} \hat{c}^\dagger_{I_1 \uparrow} }{\hat{c}^\dagger}{}{\langle  \hat{A}^{(1)\prime}_{I_1} \hat{c}^\dagger_{I_1 \uparrow}  \hat{c}_{I_1 \uparrow} \hat{c}^\dagger_{I_1 \downarrow} \hat{c}_{I_1 \downarrow}  \cdot \text{other operators}}
  \contraction[1em]{}{\hat{A}^{(1)\prime}}{\langle  \hat{A}^{(1)\prime}_{I_1} \hat{c}^\dagger_{I_1 \uparrow}  \hat{c}_{I_1 \uparrow} \hat{c}^\dagger_{I_1 \downarrow} \hat{c}_{I_1 \downarrow}  \cdot}{operators}
 \langle  \hat{A}^{(1)\prime}_{I_1} \hat{c}^\dagger_{I_1 \uparrow}  \hat{c}_{I_1 \uparrow} \hat{c}^\dagger_{I_1 \downarrow} \hat{c}_{I_1 \downarrow}  \cdot \text{other operators}\rangle_0^\mathrm{modified}
\end{align}

\noindent
that cancel each other. For the sake of argument, we have assumed that the line connects the operator $\hat{c}_{I_1}$ in $\hat{A}^{(1)\prime}_{I_1}$; other cases may be handled in an analogous manner. Note that operator $\hat{c}_{I_1}$ in $\hat{A}^{(1)\prime}_{I_1}$ cannot be directly connected to $\hat{c}_{I_1}^\dagger$ in $\hat{d}_{I_1}$ due to modified prescription of diagrammatic expansion.

Now, since each diagram in \eqref{label:appendix_transformation_2} may be unambiguously separated into part connected to external vertices and the remaining one (disconnected component), one obtains

\begin{align}
    \label{label:appendix_transformation_3}
  &\langle \hat{P}_G \hat{A}^{(1)}_{I_1} \ldots \hat{A}^{(n)}_{I_n} \hat{P}_G\rangle_0 = \sum\limits_{k=0}^\infty \sum_{J_1 \ldots J_k} \frac{x_{J_1} \ldots x_{J_k}}{k!} \sum \limits_{l = 0}^k {k \choose k-l} \langle \hat{A}^{(1)\prime}_{I_1} \ldots \hat{A}^{(1)\prime}_{I_n} \hat{d}_{J_1} \ldots \hat{d}_{J_l}\rangle_0^{c, \mathrm{modified}} \cdot \langle \hat{d}_{J_{l+1}} \ldots \hat{d}_{J_k} \rangle^{\mathrm{modified}} = \nonumber \\ & \left(\sum\limits_{k=0}^\infty \sum_{J_1 \ldots J_k} \frac{x_{J_1} \ldots x_{J_k}}{k!} \langle \hat{A}^{(1)\prime}_{I_1} \ldots \hat{A}^{(1)\prime}_{I_n} \hat{d}_{J_1} \ldots \hat{d}_{J_k}\rangle_0^{c, \mathrm{modified}}\right) \cdot \left(\sum\limits_{k=0}^\infty \sum_{J_1 \ldots J_k} \frac{x_{J_1} \ldots x_{J_k}}{k!} \langle \hat{d}_{J_1} \ldots \hat{d}_{J_k}\rangle_0^\mathrm{modified}\right).
\end{align}

\noindent
In the second line of Eq.~\eqref{label:appendix_transformation_3}, the last term in bracket is equal to $\langle P_G^2\rangle_0$ and thus is canceled by denominator of the left-hand side of Eq.~\eqref{appendix:linked_cluster_theorem_eq}. As the final step, one should reintroduce restriction summations in Eq.~\eqref{label:appendix_transformation_3} in order to restore usual diagrammatic expansion. This yields Eq.~\eqref{appendix:linked_cluster_theorem_eq}.

\subsection*{Expansion for one-site operators}

We now consider general operator $\hat{A}_I$ and corresponding $\hat{A}_I^{\prime} \equiv \hat{P}_{G, I} \hat{A}_I \hat{P}_{G, I}$ with not necessarily vanishing expectation value, i.e., $\langle{\hat{A}_I^{\prime}}\rangle_0 \neq 0$. One then has

\begin{align}
  \langle \hat{A}_I \rangle_G \equiv & \frac{\langle \hat{P}_G \hat{A}_I \hat{P}_G \rangle_0}{\langle \hat{P}_G^2\rangle} = \langle{\hat{A}_I^{\prime}}\rangle_0 + \frac{\langle \hat{P}_G (\hat{A}_I - \langle{\hat{A}_I^{(\alpha)\prime}}\rangle_0) \hat{P}_G \rangle_0}{\langle \hat{P}_G^2\rangle} = \nonumber \\ & \langle{\hat{A}_I^{(\alpha)\prime}}\rangle_0 + \sum\limits_{k=0}^\infty \sideset{}{'}\sum_{J_1 \ldots J_k} \frac{x_{J_1} \ldots x_{J_k}}{k!} \langle \left(\hat{A}^{\prime}_{I} - \langle{\hat{A}_I^{\prime}}\rangle_0 \hat{P}_{G, I}^2 \right) \cdot \hat{d}^\prime_{J_1} \ldots \hat{d}^\prime_{J_k}\rangle_0^c
\end{align}

\noindent
which is equivalent to Eq.~\eqref{eq:diagrammatics_one_site}. Note that the linked cluster expansion \eqref{appendix:linked_cluster_theorem_eq} can be applied here since $\langle\hat{A}^{\prime}_{I} - \langle{\hat{A}_I^{\prime}} \rangle_0 \hat{P}_{G, I}^2 \rangle_0 = 0$. Analogous procedure may be used to derive Eq.~\eqref{eq:diagrammatics_two_site}; it is straightforward but cumbersome.
 
\section{Path integral decomposition of the interaction term}
\label{appendix:prove_of_hs_decoupling}

Here we will prove the Hubbard-Stratonovich decomposition, employed in Sec.~\ref{section:large_n_expansion}, i.e.,

\begin{align}
  \label{eq:appendix:hs}
  \exp\left({-\mathcal{F}(\hat{\mathbf{P}})}\right) \propto \lim_{\epsilon \rightarrow 0^+} \int \mathcal{D} \mathbf{P} \int \mathcal{D}\boldsymbol{\xi} \exp\left({-\int d\tau \mathcal{F}(\mathbf{P}) - i \int d\tau \boldsymbol{\xi}^\dagger(\hat{\mathbf{P}} - \mathbf{P}) - \frac{\epsilon}{2} \int d\tau \boldsymbol{\xi}^\dagger \boldsymbol{\xi}}\right),
\end{align}

\noindent
where $\hat{\mathbf{P}}$ are Grassmann bilinears, and $\mathbf{P}$ and $\boldsymbol{\xi}$ are corresponding fields (real for the ``diagonal``, $P_{\alpha\alpha}$/${\xi}_{\alpha\alpha}$, and complex for ``off-diagonal``, $P_{\alpha\beta}$/${\xi}_{\alpha\beta}$, entries; here $\alpha \neq \beta$). In physical terms, the imaginary-time-dependent vector $\mathbf{P}$ represent various channels of particle hole excitations, whereas the aim of $\boldsymbol{\xi}$ is to ensure that those excitations are consistent with Hamiltonian dynamics. The infinitesimal term $\epsilon$ has been introduced to make the integral over $\boldsymbol{\xi}$-fields convergent. Also, it should be possible to bound the functional $\mathcal{F}$ from below by a quadratic form in $\mathbf{P}$-fields to make the integral over $\mathbf{P}$ well-defined for sufficiently small $\epsilon$. This is always the case for models with only quartic fermion-fermion interactions, such as that given by the Hamiltonian~\eqref{eq:general_hamiltonian}, but might not be satisfied if the uncorrelated energy functional $E_0(\mathbf{P}, \mu)$ is substituted with $E_\mathrm{var}(\mathbf{P}, \boldsymbol{\lambda}, \mu)$, as is done within the VWF+$1/\mathcal{N}_f$ framework (cf. Sec.~\ref{section:variational+fluctuations}). In the latter case $\mathcal{F}$, one can introduce a modified functional $\mathcal{F}_\eta = \mathcal{F} + \eta \mathcal{F}^2$ and take $\eta \rightarrow 0^+$ limit at the end of calculation.

We first switch to real representation by means of the unitary transformation, $\tilde{\mathbf{P}} = \hat{\Gamma} \mathbf{P}$, defined as follows:

\begin{align}
  &\tilde{P}_{\alpha\beta}^\prime = P_{\alpha\beta} \hspace{10em} \text{for $\alpha = \beta$}, \nonumber \\
  &\left\{
  \begin{array}{l}
    \tilde{P}_{\alpha\beta}^\prime = \frac{1}{\sqrt{2}} (P_{\alpha\beta} + P_{\alpha\beta}^{*}) \\
    \tilde{P}_{\alpha\beta}^{\prime\prime} = \frac{1}{\sqrt{2} i} (P_{\alpha\beta} - P_{\alpha\beta}^{*}) \\
  \end{array}
\right. \hspace{3em} \text{for $\alpha \neq \beta$},
  \label{eq:real_to_complex_transformation}
\end{align}

\noindent
where $\tilde{P}_{\alpha\beta}^{\prime/\prime\prime}$ are manifestly real fields. Note that the imaginary components, $P_{\alpha\alpha}$ are identically zero. In the following, we understand that $\tilde{\mathbf{P}}$ is a vector composed of all fields ${P}^\prime_{\alpha\alpha}$, ${P}^\prime_{\alpha\beta}$, and ${P}^{\prime\prime}_{\alpha\beta}$ for $\alpha \neq \beta$. Similarly, we define $\tilde{\boldsymbol{\xi}} \equiv \hat{\Gamma} \boldsymbol{\xi}$ and $\hat{\tilde{\mathbf{P}}} \equiv \hat{\Gamma} \hat{\mathbf{P}}$.

Using Eq.~\eqref{eq:real_to_complex_transformation}, one gets

\begin{align}
  \int d\tau \mathcal{F}(\mathbf{P}) + i \int d\tau \boldsymbol{\xi}^\dagger(\hat{\mathbf{P}} - \mathbf{P}) + \frac{\epsilon}{2} \int d\tau \boldsymbol{\xi}^\dagger \boldsymbol{\xi} = \int d\tau \mathcal{F}(\hat{\Gamma}^\dagger\tilde{\mathbf{P}}) + i \int d\tau \tilde{\boldsymbol{\xi}}^T(\hat{\tilde{\mathbf{P}}} - \tilde{\mathbf{P}}) + \frac{\epsilon}{2} \int d\tau \tilde{\boldsymbol{\xi}}^T \tilde{\boldsymbol{\xi}}
\end{align}

\noindent
which can be used to transform the right-hand side (RHS) of Eq.~\eqref{eq:appendix:hs} as

\begin{align}
  \label{eq:appendix:hs_calculations}
&  \mathrm{RHS} =   \lim_{\epsilon \rightarrow 0^+} \int \mathcal{D} \tilde{\mathbf{P}} \int \mathcal{D}\tilde{\boldsymbol{\xi}} \exp\left({-\int d\tau \mathcal{F}(\hat{\Gamma}^\dagger\tilde{\mathbf{P}}) - i \int d\tau \tilde{\boldsymbol{\xi}}^T(\hat{\tilde{\mathbf{P}}} - \tilde{\mathbf{P}}) - \frac{\epsilon}{2} \int d\tau \tilde{\boldsymbol{\xi}}^T \tilde{\boldsymbol{\xi}}}\right) \propto \nonumber \\ & \lim_{\epsilon \rightarrow 0^+} \int \mathcal{D} \tilde{\mathbf{P}} \exp\left({-\int d\tau \mathcal{F}(\hat{\Gamma}^\dagger\tilde{\mathbf{P}}) - \frac{1}{2\epsilon} \int d\tau (\hat{\tilde{\mathbf{P}}} - \tilde{\mathbf{P}})^T(\hat{\tilde{\mathbf{P}}} - \tilde{\mathbf{P}})}\right) = \nonumber \\ &
\nonumber \\ & \lim_{\epsilon \rightarrow 0^+}\exp\left(- \frac{1}{2\epsilon} \int d\tau \hat{\tilde{\mathbf{P}}}^T \hat{\tilde{\mathbf{P}}}\right) \sum_{i_1, \ldots, i_M=0}^\infty \int \mathcal{D} \tilde{\mathbf{P}} \frac{\left(\int d\tau_1\hat{\tilde{P}}_{1}\tilde{P}_1\right)^{i_1}}{i_1! \epsilon^{i_1}}  \ldots \frac{\left(\int d\tau_M\hat{\tilde{P}}_{M}\tilde{P}_M\right)^{i_M}}{i_M! \epsilon^{i_M}} \exp\left({-\int d\tau \mathcal{F}(\hat{\Gamma}^\dagger\tilde{\mathbf{P}}) - \frac{1}{2\epsilon} \int d\tau \tilde{\mathbf{P}}^T \tilde{\mathbf{P}}}\right),
\end{align}

\noindent
where the $M$ is the total number of scalar field components. The transition from the second to the third line of Eq.~\eqref{eq:appendix:hs_calculations} has been performed by Taylor-expanding the exponential term

\begin{align}
  \label{eq:2}
\exp\left(\frac{1}{\epsilon} \int d\tau \hat{\tilde{\mathbf{P}}}^T \cdot \tilde{\mathbf{P}} \right) = \exp\left(\frac{1}{\epsilon} \sum_{i=1}^M \int d\tau \hat{\tilde{P_i}} \cdot \tilde{P_i} \right).
\end{align}

\noindent
The integral in the last term of Eq.~\eqref{eq:appendix:hs_calculations} can be evaluated by introducing the generating functional

\begin{align}
  Z[\tilde{\mathbf{J}}] \equiv & \int \mathcal{D} \tilde{\mathbf{P}} \exp\left(-\int d\tau \mathcal{F}(\hat{\Gamma}^\dagger\tilde{\mathbf{P}}) - \frac{1}{2\epsilon} \int d\tau \tilde{\mathbf{P}}^T \tilde{\mathbf{P}} + \frac{1}{\epsilon} \int d\tau \tilde{\mathbf{J}}^T \tilde{\mathbf{P}}\right) \propto \nonumber \\ & \int \mathcal{D} \tilde{\mathbf{P}} \exp\left(-\int d\tau \mathcal{F}(\epsilon \hat{\Gamma}^\dagger\tilde{\mathbf{P}} + \hat{\Gamma}^\dagger\tilde{\mathbf{J}}) - \frac{\epsilon}{2} \int d\tau \tilde{\mathbf{P}}^T \tilde{\mathbf{P}} + \frac{1}{2 \epsilon} \int d\tau \tilde{\mathbf{J}}^T \tilde{\mathbf{J}}\right),                                            
\end{align}

\noindent
where the second line has been obtained by a change of variables, $\tilde{\mathbf{P}} \rightarrow \epsilon \tilde{\mathbf{P}} + \tilde{\mathbf{J}}$. One thus gets

\begin{align}
  \mathrm{RHS} \propto & \lim_{\epsilon \rightarrow 0^+}\exp\left(- \frac{1}{2\epsilon} \int d\tau \hat{\tilde{\mathbf{P}}}^T \hat{\tilde{\mathbf{P}}}\right) \sum_{i_1, \ldots, i_N=0}^\infty \frac{\left( \int d\tau_1 \hat{\tilde{P}}_{1} \frac{\delta}{\delta \tilde{J}_1 }\right)^{i_1}}{i_1!} \ldots \frac{\left( \int d\tau_M \hat{\tilde{P}}_{M} \frac{\delta}{\delta \tilde{J}_M }\right)^{i_M}}{i_M!} Z[\tilde{\mathbf{J}}] \propto \nonumber \\ & \int \mathcal{D} \tilde{\mathbf{P}} \exp\left(-\int d\tau \mathcal{F}(\epsilon \hat{\Gamma}^\dagger\tilde{\mathbf{P}} + \hat{\Gamma}^\dagger\hat{\tilde{\mathbf{P}}}) - \frac{\epsilon}{2} \int d\tau \tilde{\mathbf{P}}^T \tilde{\mathbf{P}}\right) \overset{\epsilon \rightarrow 0^+}{\longrightarrow} \exp\left(-\mathcal{F}(\hat{\Gamma}^\dagger\hat{\tilde{\mathbf{P}}})\right) \rightarrow \exp\left(-\mathcal{F}(\hat{\mathbf{P}})\right),
\end{align}

\noindent
which concludes the reasoning.

One can now use Eq.~\eqref{eq:appendix:hs} to derive Eq.~\eqref{eq:hubbard_stratonovich} by taking $\mathcal{F}(\mathbf{P}) = E_{0, \mathrm{kin}}(\mathbf{P}) + \frac{1}{2} E_{0, \mathrm{int}}(\mathbf{P})$. Indeed, we get

\begin{align}
  \label{eq:appendix:proof_hs_tranformation}
  &\lim_{\epsilon\rightarrow 0^+}\int \mathcal{D}\mathbf{P} \mathcal{D} \boldsymbol{\xi} \exp\left( - \int d\tau E_{0, \mathrm{kin}}(\mathbf{P}) - \frac{1}{2} \int d\tau E_{0, \mathrm{int}}(\mathbf{P}) - i \int d\tau \boldsymbol{\xi}^\dagger(\hat{\mathbf{P}} - \mathbf{P})  -\frac{\epsilon}{2} \int d\tau \boldsymbol{\xi}^\dagger \boldsymbol{\xi}\right) \propto \nonumber \\ & \exp\left( - \int d\tau E_{0, \mathrm{kin}}(\hat{\mathbf{P}}) - \frac{1}{2} \int d\tau E_{0, \mathrm{int}}(\hat{\mathbf{P}})\right) = \exp\left(-\int d\tau \hat{\mathcal{H}}[\bar{\eta}, \eta]\right).
\end{align}

\noindent
The last inequality in Eq.~\eqref{eq:appendix:proof_hs_tranformation} follows directly from the structure of the Hamiltonian. For the kinetic part, one obtains $E_\mathrm{kin}(\hat{\mathbf{P}}) = \sum_{\alpha\beta} t_{\alpha\beta} \hat{P}_{\alpha\beta} = \sum_{\alpha\beta} t_{\alpha\beta}  \bar{\eta}_\alpha \eta_\beta$. In a similar manner, the potential-energy part may be transformed as

\begin{align}
  E_{0, \mathrm{int}}(\hat{\mathbf{P}}) = \frac{1}{2} \sum_{\alpha\beta\gamma\rho} \hat{P}_{\alpha\beta}^\dagger \mathcal{V}_{\alpha\beta\gamma\rho} \hat{P}_{\gamma\rho} = \frac{1}{2} \sum_{\alpha\beta\gamma\rho} \bar{\eta}_\beta \eta_\alpha \mathcal{V}_{\alpha\beta\gamma\rho} \bar{\eta}_\gamma \eta_\rho = \sum_{\alpha\beta\rho\gamma} V_{\alpha\beta\rho\gamma}  \bar{\eta}_\alpha \bar{\eta}_\beta \eta_\gamma \eta_\rho = 2 \cdot \hat{V}[\bar{\boldsymbol{\eta}}, \boldsymbol{\eta}],
\end{align}

\noindent
where we have made use of the definition of the summarized vertex, $\mathcal{V}_{\alpha\beta\gamma\rho} \equiv (-V_{\beta\gamma\rho\alpha} + V_{\beta\gamma\alpha\rho} + V_{\gamma\beta\rho\alpha} - V_{\gamma\beta\alpha\rho})/2$.
Finally, one gets $\int d\tau E_\mathrm{0, kin}(\hat{\mathbf{P}}) + \frac{1}{2} \int d\tau E_{0, \mathrm{int}}(\hat{\mathbf{P}}) = \int d\tau \hat{\mathcal{H}}[\bar{\eta}, \eta]$ The apparent double counting of interaction after substituting Grassmann bilinears into $E_{0, \mathrm{int}}$ originates from unbiased (all-channel) version of Hubbard-Stratonovich decomposition, as discussed in detail in Sec.~\ref{section:large_n_expansion}.

\section{Action for correlated $1/\mathcal{N}_f$ expansion and effective scattering matrix}
\label{appendix:action_correlated_1/n}

Here we show that after integrating out the fields $\boldsymbol{\lambda}$ and $\boldsymbol{\rho}$, the action~\eqref{eq:sga_1n_action} becomes equivalent to the resummed $1/\mathcal{N}_f$ action~\eqref{eq:effective_action_hf}. For $\mathcal{N}_f = 1$, the action~\eqref{eq:sga_1n_action} reproduces thus faithfully the thermodynamics of the original model, defined by the Hamiltonian $\mathcal{\hat{H}}$ so that our approach provides a systematic route to improvement of the VWF technique by incorporating consecutive $1/\mathcal{N}_f$ corrections around the correlated saddle point solution.

The proof is performed by consecutively integrating out the $\boldsymbol{\rho}$, $\boldsymbol{\lambda}$, $\boldsymbol{\xi}$, $\mathbf{P}$ By taking the action~\eqref{eq:sga_1n_action} with  $\mathcal{N}_f = 1$, one arrives at

\begin{align}
  \label{eq:appendix_correlated_z_proof}
  Z[\mathbf{J}] = & \int \mathcal{D} \bar{\boldsymbol{\eta}} \mathcal{D}\boldsymbol{\eta} \mathcal{D} \mathbf{P} \mathcal{D} \boldsymbol{\xi} \mathcal{D} \boldsymbol{\lambda}\mathcal{D} \boldsymbol{\rho} \exp\left(-\mathcal{S}_\mathrm{var}[\mathbf{P}, \boldsymbol{\xi}, \boldsymbol{\lambda}, \boldsymbol{\rho}, \bar{\boldsymbol{\eta}}, \boldsymbol{\eta}, \mathbf{J}, \mu]\right) \propto \nonumber \\ &
                                                                                                                                                                                                                                                                                                                                                                                                       \int \mathcal{D} \bar{\boldsymbol{\eta}} \mathcal{D}\boldsymbol{\eta} \mathcal{D} \mathbf{P} \mathcal{D} \boldsymbol{\xi}\mathcal{D} \boldsymbol{\lambda} \exp\left(-\mathcal{S}[\mathbf{P}, \boldsymbol{\xi}, \bar{\boldsymbol{\eta}}, \boldsymbol{\eta}, \mathbf{J}, \mu]\right) \delta(\mathbf{C}(\mathbf{P}, \boldsymbol{\lambda})) \left|\mathrm{det} \frac{\partial \mathbf{C}}{\partial \boldsymbol{\lambda}}\right| \exp(-\frac{1}{2} \kappa \int d\tau \mathbf{C}^T \mathbf{C}) \propto \nonumber \\ &                                                                                                                                             \int \mathcal{D} \bar{\boldsymbol{\eta}} \mathcal{D}\boldsymbol{\eta} \mathcal{D} \mathbf{P} \mathcal{D} \boldsymbol{\xi}\exp\left(-\mathcal{S}[\mathbf{P}, \boldsymbol{\xi}, \bar{\boldsymbol{\eta}}, \boldsymbol{\eta}, \mathbf{J}, \mu]\right) \propto \nonumber \\ & 
\int \mathcal{D} \bar{\boldsymbol{\eta}} \mathcal{D}\boldsymbol{\eta} \exp\left(-\int d\tau \bar{\boldsymbol{\eta}} \partial_\tau \boldsymbol{\eta} - \int d\tau \hat{\mathcal{H}}[\bar{\boldsymbol{\eta}}, \boldsymbol{\eta}, \mu]\right),
\end{align}

\noindent
where $\mathcal{S}_\mathrm{var}$ is the VWF+$1/\mathcal{N}_f$ action [Eq.~\eqref{eq:sga_1n_action}], $\mathcal{S}$ denotes the ``uncorrelated'' action [Eq.~\eqref{eq:uncorrelated_large_n}], and $\hat{\mathcal{H}}[\bar{\boldsymbol{\eta}}, \boldsymbol{\eta}, \mu]$ is the Hamiltonian expressed in terms of Grassmann fields. The Dirac delta imposing $\mathbf{C} = \mathbf{0}$ in the second line originates from integration over $\boldsymbol{\rho}$ fields. As a consequence, the term $\propto \kappa$ is equal to zero, making the observables generated based on $Z[\mathbf{J}]$ independent on $\kappa$, as stated in the main text. Finally, the $\mathcal{O}(1)$ constraint Jacobian term allows to explicitly evaluate integral over $\boldsymbol{\lambda}$ without generating any artifact $\mathbf{P}$-field interactions. The $\boldsymbol{\lambda}$-fields evolution is then fully determined by line-fields, $\mathbf{P}$, and may be obtained by solving the equations $\mathbf{C} = 0$. The transition between the last two lines of Eq.~\eqref{eq:appendix_correlated_z_proof} has been performed based on the results of Sec.~\ref{section:large_n_expansion} and Appendix~\ref{appendix:prove_of_hs_decoupling}. The approach thus reproduces the thermodynamics of the model, defined by $\hat{\mathcal{H}}$.

We now derive the effective scattering matrix, given by Eq.~\eqref{eq:effective_scattering_matrix}, taking the Gaussian-fluctuation action~\eqref{eq:action_expanded_de-gwf} as the departure point. First, for the reasons that will be made cellar below, it is essential to switch to real representation with the use of~\eqref{eq:real_to_complex_transformation}. By transforming the quadratic part, one obtains

\begin{align}
  \label{eq:quadratic_term_to_real_transfomration}
  \mathcal{S}_\mathrm{quadratic} = \frac{\mathcal{N}_f}{2} \int d\tau
  \begin{array}{c}
    (\delta\mathbf{P}^\dagger \delta\boldsymbol{\lambda}^\dagger \delta\boldsymbol{\rho}^\dagger) \\
    {} \\
    {}
    \end{array}
  \left(
  \begin{array}{ccc}
    \hat{\mathcal{V}}_{P^{*}P} + \kappa \hat{\mathcal{V}}_{P^{*}\rho}\hat{\mathcal{V}}_{\rho P} & \hat{\mathcal{V}}_{P^{*} \lambda} + \kappa \hat{\mathcal{V}}_{P^{*}\rho}\hat{\mathcal{V}}_{\rho \lambda} & i \hat{\mathcal{V}}_{P^{*} \rho}\\
    \hat{\mathcal{V}}_{\lambda P} + \kappa \hat{\mathcal{V}}_{\lambda\rho}\hat{\mathcal{V}}_{\rho P}  & \hat{\mathcal{V}}_{\lambda\lambda} + \kappa \hat{\mathcal{V}}_{\lambda\rho}\hat{\mathcal{V}}_{\rho \lambda} & i \hat{\mathcal{V}}_{\lambda \rho}\\
    i\hat{\mathcal{V}}_{\rho P} &  i \hat{\mathcal{V}}_{\rho \lambda} & \epsilon_\rho
  \end{array}
        \right)
        \left(
        \begin{array}{c}
          \delta\mathbf{P} \\
          \delta\boldsymbol{\lambda} \\
          \delta\boldsymbol{\rho}
        \end{array}
  \right) = \nonumber \\
  \frac{\mathcal{N}_f}{2} \int d\tau
  \begin{array}{c}
    (\delta\tilde{\mathbf{P}}^T \delta \tilde{\boldsymbol{\lambda}}^T \delta\tilde{\boldsymbol{\rho}}^T) \\
    {} \\
    {}
    \end{array}
  \left(
  \begin{array}{ccc}
    \hat{\tilde{\mathcal{V}}}_{PP} + \kappa \hat{\tilde{\mathcal{V}}}_{P\rho}\hat{\tilde{\mathcal{V}}}_{\rho P} & \hat{\tilde{\mathcal{V}}}_{P \lambda} + \kappa \hat{\tilde{\mathcal{V}}}_{P\rho}\hat{\tilde{\mathcal{V}}}_{\rho \lambda} & i \hat{\tilde{\mathcal{V}}}_{P \rho}\\
    \hat{\tilde{\mathcal{V}}}_{\lambda P} + \kappa \hat{\tilde{\mathcal{V}}}_{\lambda\rho}\hat{\tilde{\mathcal{V}}}_{\rho P}  & \hat{\tilde{\mathcal{V}}}_{\lambda\lambda} + \kappa \hat{\tilde{\mathcal{V}}}_{\lambda\rho}\hat{\tilde{\mathcal{V}}}_{\rho \lambda} & i \hat{\tilde{\mathcal{V}}}_{\lambda \rho}\\
    i\hat{\tilde{\mathcal{V}}}_{\rho P} &  i \hat{\tilde{\mathcal{V}}}_{\rho \lambda} & \epsilon_\rho
  \end{array}
        \right)
        \left(
        \begin{array}{c}
          \delta\tilde{\mathbf{P}} \\
          \delta\tilde{\boldsymbol{\lambda}} \\
          \delta\tilde{\boldsymbol{\rho}}
        \end{array}
  \right) \equiv (\delta\tilde{\mathbf{P}}^T \delta \tilde{\boldsymbol{\lambda}}^T \delta\tilde{\boldsymbol{\rho}}^T) \hat{\mathcal{V}}^{(2)}     \left(
        \begin{array}{c}
          \delta\tilde{\mathbf{P}} \\
          \delta\tilde{\boldsymbol{\lambda}} \\
          \delta\tilde{\boldsymbol{\rho}}
        \end{array}
  \right),
\end{align}

\noindent
where $\delta\tilde{\mathbf{P}}$,  $\delta \tilde{\boldsymbol{\lambda}}$, and $\delta\tilde{\boldsymbol{\rho}}$ are real and the transformed scattering matrix, $\hat{\tilde{\mathcal{V}}}^{(2)} = \hat{\tilde{\mathcal{V}}}^{(2)\prime} + i \hat{\tilde{\mathcal{V}}}^{(2)\prime\prime}$, is complex symmetric. We have also explicitly restored the infinitesimal term $\frac{\mathcal{N}_f \epsilon_\rho}{2}  \int d\tau \delta\tilde{\boldsymbol{\rho}}^T\delta\tilde{\boldsymbol{\rho}}$ with $\epsilon_\rho > 0$ that was discarded in Sec.~\ref{section:variational+fluctuations} for brevity of notation.

Let us begin the derivation by representing $\mathcal{S}_\mathrm{quadratic}$ as

\begin{align}
  \mathcal{S}_\mathrm{quadratic} = \frac{\mathcal{N}_f}{2} \int d\tau \delta \tilde{\mathbf{P}}^T \left(\hat{\tilde{\mathcal{V}}}_{PP} + \kappa \hat{\tilde{\mathcal{V}}}_{P\rho}\hat{\tilde{\mathcal{V}}}_{\rho P} \right) \tilde{\mathbf{P}} +
  \frac{\mathcal{N}_f}{2} \int d\tau
  \begin{array}{c}
    (\delta \tilde{\boldsymbol{\lambda}}^T \delta\tilde{\boldsymbol{\rho}}^T) \\
    {}
    \end{array}
  \left(
  \begin{array}{ccc}
     \hat{\tilde{\mathcal{V}}}_{\lambda\lambda} + \kappa \hat{\tilde{\mathcal{V}}}_{\lambda\rho}\hat{\tilde{\mathcal{V}}}_{\rho \lambda} & i \hat{\tilde{\mathcal{V}}}_{\lambda \rho}\\
    i \hat{\tilde{\mathcal{V}}}_{\rho \lambda} & \epsilon_\rho
  \end{array}
        \right)
        \left(
        \begin{array}{c}
          \delta\tilde{\boldsymbol{\lambda}} \\
          \delta\tilde{\boldsymbol{\rho}}
        \end{array}
  \right) + \nonumber \\
  \frac{\mathcal{N}_f}{2} \int d\tau
  \begin{array}{ccc}
     \Big(\delta\tilde{\mathbf{P}}^T \hat{\tilde{\mathcal{V}}}_{P \lambda} + \kappa \delta\tilde{\mathbf{P}}^T  \hat{\tilde{\mathcal{V}}}_{P\rho}\hat{\tilde{\mathcal{V}}}_{\rho \lambda} & i \delta\tilde{\mathbf{P}}^T \hat{\tilde{\mathcal{V}}}_{P \rho}\Big)\\ {}
  \end{array}
        \left(
        \begin{array}{c}
          \delta\tilde{\boldsymbol{\lambda}} \\
          \delta\tilde{\boldsymbol{\rho}}
        \end{array}
  \right) + 
  \frac{\mathcal{N}_f}{2} \int d\tau
  \begin{array}{c}
    (\delta \tilde{\boldsymbol{\lambda}}^T \delta\tilde{\boldsymbol{\rho}}^T) \\
    {} \\
    {}
    \end{array}
  \left(
  \begin{array}{ccc}
    \hat{\tilde{\mathcal{V}}}_{\lambda P} \delta\tilde{\mathbf{P}} + \kappa \hat{\tilde{\mathcal{V}}}_{\lambda\rho}\hat{\tilde{\mathcal{V}}}_{\rho P} \delta\tilde{\mathbf{P}} \\
    i\hat{\tilde{\mathcal{V}}}_{\rho P} \delta\tilde{\mathbf{P}}
  \end{array}
  \right),
  \label{eq:quadratic_real_represetation_rewritten}
\end{align}

\noindent
and integrating out $\delta \tilde{\boldsymbol{\lambda}}$ and $\delta\tilde{\boldsymbol{\rho}}$ fields. It is necessary to argue that one can apply the standard formula for evaluating Gaussian path integrals to the present case, based on completing the square and calculating functional determinant. This is difficult to prove in the complex-field representation of Eq.~\eqref{eq:quadratic_term_to_real_transfomration}, since the involved matrix is complex and non-symmetric. On the other hand, by employing the real representation of Eq.~\eqref{eq:quadratic_real_represetation_rewritten}, this can be shown by making an additional assumption that the real-part of the scattering matrix between $\boldsymbol{\lambda}$ and $\boldsymbol{\rho}$ fields, 

\begin{align}
  \label{eq:rho_lambda_scattering_matrix}
  \hat{\tilde{\mathcal{V}}}^{(2)} \equiv \left(
  \begin{array}{cc}
\hat{\tilde{\mathcal{V}}}_{\lambda\lambda} + \kappa \hat{\tilde{\mathcal{V}}}_{\lambda\rho}\hat{\tilde{\mathcal{V}}}_{\rho \lambda} & i \hat{\tilde{\mathcal{V}}}_{\lambda \rho}\\
i \hat{\tilde{\mathcal{V}}}_{\rho \lambda} & \epsilon_\rho
  \end{array}
        \right),
\end{align}

\noindent
is positive definite. Recall that $\hat{\mathcal{V}}_{\lambda\rho}$ is the Jacobian of the constraints ($\mathbf{C} = 0$) [cf. Eq.~\eqref{eq:scattering_matrix_element_l_rho_lambda}], being a real square matrix since the number of constraints equal to the number of time-dependent $\boldsymbol{\lambda}$-fields, as demanded in Sec.~\ref{section:variational+fluctuations}. Moreover, we require that the constraints $\mathbf{C} \equiv \mathbf{0}$ locally determine $\boldsymbol{\lambda}$-fields (cf. Sec.~\ref{section:variational+fluctuations}), which is implemented by the condition $\det \hat{\tilde{\mathcal{V}}}_{\lambda \rho} \neq 0$. Under those circumstances, to make the real part of $\hat{\tilde{\mathcal{V}}}^{(2)}$ positive definite is sufficient to assure that its left-upper block $\hat{\mathcal{V}}^\prime_{\lambda\lambda} + \kappa \hat{\mathcal{V}}_{\lambda\rho}\hat{\mathcal{V}}_{\rho \lambda}$ is positive definite (here  $\hat{\mathcal{V}}^\prime_{\lambda\lambda}$ denotes real-part of the matrix $\hat{\mathcal{V}}_{\lambda\lambda} \equiv \hat{\mathcal{V}}^\prime_{\lambda\lambda} + i \hat{\mathcal{V}}_{\lambda\lambda}^{\prime\prime}$). The latter condition is always met if sufficiently large positive $\kappa$ is taken, thus $\kappa$-parameter has been introduced to ensure convergence. By employing the above assumptions, $\hat{\tilde{\mathcal{V}}}^{(2)}$ may be brought to diagonal form by congruence as $\hat{\tilde{\mathcal{V}}}^{(2)} = \hat{K}( \mathbb{1} + i \hat{D} )\hat{K}^T$, where $\mathbb{1}$ is identity matrix,  $\hat{K}$ is real, and $\hat{D}$ is diagonal. This can be seen by application of Cholesky factorization to the real-part, $\hat{\tilde{\mathcal{V}}}^{(2)\prime} = \hat{L} \hat{L}^T$, and orthogonally diagonalizing the symmetric matrix $\hat{L}^{-1} \hat{\tilde{\mathcal{V}}}^{(2)\prime\prime} (\hat{L}^{-1})^T = \hat{Q} \hat{D} \hat{Q}^T$. Then one gets $\hat{K} = \hat{L} \hat{Q}$. Using this decomposition, it is straightforward to show that the Gaussian integral factorizes into a product of Gaussian integrals for which it is the usual formulas for integration hold.

Integrating out $\delta \tilde{\boldsymbol{\lambda}}$ and $\delta\tilde{\boldsymbol{\rho}}$ fields by completing to the square, which yields effective contribution depending only on the fluctuation of $\tilde{\mathbf{P}}$-fields, i.e.,

\begin{align}
  \label{eq:effective_quadratic_action}
  \mathcal{S}_\mathrm{quadratic} \rightarrow \mathcal{S}_{\mathrm{quadratic}, \mathrm{eff}} = \frac{\mathcal{N}_f}{2} \int d\tau \delta \tilde{\mathbf{P}}^T \hat{\tilde{\mathcal{V}}}_\mathrm{eff}^\kappa \tilde{\mathbf{P}} + \mathcal{O}(1),
\end{align}

\noindent
where $\mathcal{O}(1)$ terms originate from the determinant generated by evaluation of the Gaussian integral. In Eq.~\eqref{eq:effective_quadratic_action} we have introduced effective $\kappa$-dependent scattering matrix

\begin{align}
\hat{\tilde{\mathcal{V}}}_\mathrm{eff}^\kappa \equiv \hat{\tilde{\mathcal{V}}}_{PP} + \kappa \hat{\tilde{\mathcal{V}}}_{P\rho}\hat{\tilde{\mathcal{V}}}_{\rho P} -   \begin{array}{ccc}
     \Big(\hat{\tilde{\mathcal{V}}}_{P \lambda} + \kappa \hat{\tilde{\mathcal{V}}}_{P\rho}\hat{\tilde{\mathcal{V}}}_{\rho \lambda}, & i \hat{\tilde{\mathcal{V}}}_{P \rho}\Big)\\ {}
                                                                                                                                                                     \end{array}
    \left(
  \begin{array}{ccc}
     \hat{\tilde{\mathcal{V}}}_{\lambda\lambda} + \kappa \hat{\tilde{\mathcal{V}}}_{\lambda\rho}\hat{\tilde{\mathcal{V}}}_{\rho \lambda} & i \hat{\tilde{\mathcal{V}}}_{\lambda \rho}\\
    i \hat{\tilde{\mathcal{V}}}_{\rho \lambda} & \epsilon_\rho
  \end{array}
                                                 \right)^{-1}
  \left(
  \begin{array}{ccc}
    \hat{\tilde{\mathcal{V}}}_{\lambda P} + \kappa \hat{\tilde{\mathcal{V}}}_{\lambda\rho}\hat{\tilde{\mathcal{V}}}_{\rho P} \\
    i\hat{\tilde{\mathcal{V}}}_{\rho P}
  \end{array}
  \right).
  \label{eq:kappa-dependent-veff}
\end{align}

\noindent
Equation~\eqref{eq:kappa-dependent-veff} resembles Eq.~\eqref{eq:effective_scattering_matrix} of the main text, but it seemingly depends on the unphysical gauge parameter, $\kappa$. We now show that all contributions $\propto \kappa$ actually \emph{cancel out exactly} in Eq.~\eqref{eq:kappa-dependent-veff}, so that this expression is \emph{gauge-invariant}. Note that, whereas independence of the exact generating functional $Z[\mathbf{J}]$ on $\kappa$ follows directly from the series of transformations~\eqref{eq:appendix_correlated_z_proof}, the statement that $\kappa$-dependence is lost at the leading order of $1/\mathcal{N}_f$ expansion is non-trivial and requires justification. Provided that the determinant of the Jacobian $\det \hat{\tilde{\mathcal{V}}}_{\lambda \rho} \neq 0$, the inverse of $\hat{\mathcal{V}}^{(2)}$ on the right-hand-side of Eq.~\eqref{eq:kappa-dependent-veff} exists and may be evaluated using the block-matrix inverse formula\cite{LuCompMath2002} as

\begin{figure*}
    \onecolumngrid
  \centering
  \includegraphics[width=0.8\linewidth]{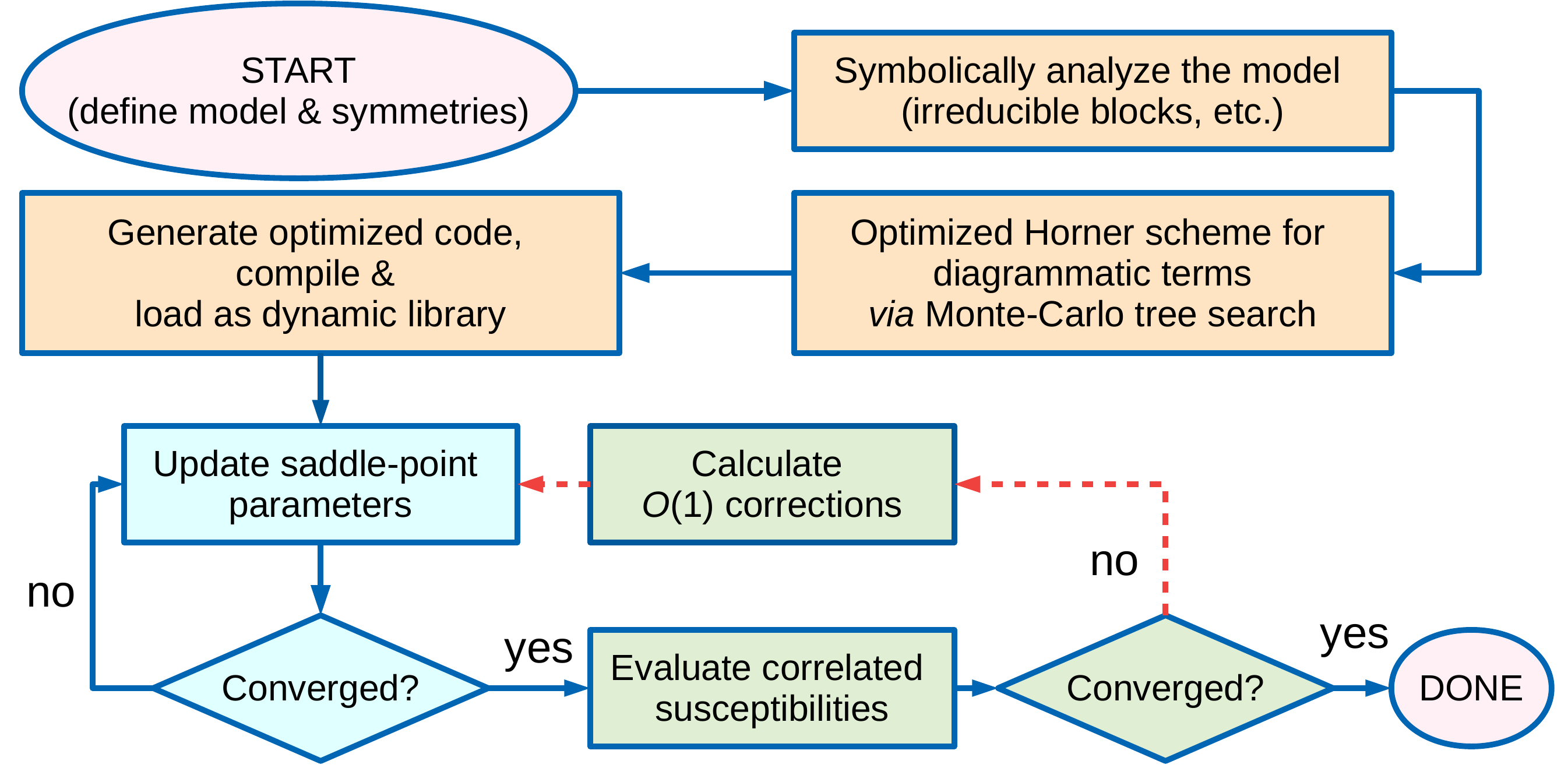} 
  \caption{A simplified flowchart of VWF+$1/\mathcal{N}_f$ approach to order $\mathcal{O}(1)$. Pink ellipses are entry-point and end-point of the algorithm and orange blocks represent preparation stage. The light-blue decision tree on the bottom-left represents the self-consistent diagrammatic VWF solution (cf. Sec.~\ref{section:variational_approach}), light-green block represent evaluation of dynamical corrections in the correlated state. The results, presented in this paper, have been obtained following blue lines. Red dashed lines reflect the $\mathcal{O}(1)$ contributions to thermodynamics and are yet to be implemented, which will allow to study the impact of  collective-modes on static quantities (cf. discussion of Sec.~\ref{section:discussion}).}
  \label{fig:flowchart}
    \twocolumngrid
\end{figure*}

\begin{align}
  \label{eq:block_matrix_inverse}
    \left(
  \begin{array}{ccc}
     \hat{\tilde{\mathcal{V}}}_{\lambda\lambda} + \kappa \hat{\tilde{\mathcal{V}}}_{\lambda\rho}\hat{\tilde{\mathcal{V}}}_{\rho \lambda} & i \hat{\tilde{\mathcal{V}}}_{\lambda \rho}\\
    i \hat{\tilde{\mathcal{V}}}_{\rho \lambda} & 0
  \end{array}
\right)^{-1} =
    \left(
  \begin{array}{ccc}
    0 & - i \hat{\tilde{\mathcal{V}}}_{\rho \lambda}^{-1} \\
   - i \hat{\tilde{\mathcal{V}}}_{\lambda \rho}^{-1} & \hat{\tilde{\mathcal{V}}}_{\lambda \rho}^{-1} \hat{\tilde{\mathcal{V}}}_{\lambda\lambda} \hat{\tilde{\mathcal{V}}}_{\rho \lambda}^{-1} + \kappa
  \end{array}
\right),
\end{align}

\noindent
where we have already taken $\epsilon_\rho \rightarrow 0^{+}$ limit. By working out the structure of Eq.~\eqref{eq:kappa-dependent-veff} using Eq.~\eqref{eq:block_matrix_inverse} it becomes apparent that all the terms proportional to $\kappa$ cancel out, and Eq.~\eqref{eq:kappa-dependent-veff} is equivalent Eq.~\eqref{eq:effective_scattering_matrix} if the fields and matrices are rotated-back to the complex representations with the use of transformation $\hat{\Gamma}$ (see Appendix~\ref{appendix:prove_of_hs_decoupling}). This concludes the derivation of the scattering matrix.

\twocolumngrid

\section{Computational aspects}
\label{appendix:computational_aspects}

In this Appendix we discuss selected computational aspects, involved in VWF+$1/\mathcal{N}_f$ calculations. In Fig.~\ref{fig:flowchart}, a simplified flowchart of VWF+$1/\mathcal{N}_f$ implementation to the order $\mathcal{O}(1)$, is shown. Pink ellipses are entry-point and end-point of the algorithm. The orange blocks represent preparation stage. First, symbolic analysis of the Hamiltonian and the corresponding action~\eqref{eq:sga_1n_action} is performed. In the second step, the diagrammatic expressions for both VWF+$1/\mathcal{N}_f$ action and auxiliary objects [e.g. the effective Hamiltonian, given by Eq.~\eqref{eq:correlted_effective_hamiltonian}] are simplified by means of the stochastic Monte-Carlo tree search algorithm (MCTS). In our workflow, we have used an external FORM symbolic manipulation program\cite{FORM_citation} to find optimized Horner schemes for diagrammatic sums. Finally, those expressions are  compiled and loaded as dynamic libraries, which provides a significant performance boost.

The decision tree on the bottom-left of Fig.~\ref{fig:flowchart} (light-blue color) represents the diagrammatic variational calculation, where the correlated large-$\mathcal{N}_f$ equations \eqref{eq:saddle_point_correlated_1}-\eqref{eq:saddle_point_correlated_5} are solved in a self-consistent manner. This is executed by means of an iterative procedure involving diagonalization of the effective Hamiltonian~\eqref{eq:correlted_effective_hamiltonian}, evaluating line fields, and solving Eqs.~\eqref{eq:saddle_point_correlated_4}-\eqref{eq:saddle_point_correlated_5} with respect to variational parameters, $\boldsymbol{\lambda}$, and Lagrange multipliers, $\boldsymbol{\rho}$. For the on-site Gutzwiller correlator, this segment is equivalent to DE-GWF\cite{KaczmarczykNewJPhys2014} (if diagrammatic sums are solved in real-space) or $\mathbf{k}$-DE-GWF \cite{FidrysiakJPhysCondensMatter2018} method (if calculations are carried out in $\mathbf{k}$-space by Monte-Carlo integration). The details of this part of the algorithm are presented elsewhere.\cite{KaczmarczykNewJPhys2014,FidrysiakJPhysCondensMatter2018}

The consecutive segment, composed of light-green blocks, is concerned with incorporating collective excitations around the correlated ground state. This is implemented by evaluating the  Gaussian-fluctuation action of Eq.~\eqref{eq:action_expanded_de-gwf}, as described in detail in Sec.~\ref{section:variational+fluctuations}. In the present contribution, we have restricted ourselves to discussion of dynamic spin- and charge- susceptibilities, starting from the variational state. Those steps are marked in Fig.~\ref{fig:flowchart} by blue lines. Evaluation of the $\mathcal{O}(1)$ terms, present in Eq.~\eqref{eq:action_expanded_de-gwf}, as well as $\mathrm{Tr} \log$ terms that arise from Gaussian integration, should be the undertaken in a separate study. Incorporating those $\mathcal{O}(1)$ corrections will generate the \emph{second self-consistent loop} (dashed red arrows in Fig.~\ref{fig:flowchart}). Implementation of this part should allow us to study multi-paramagnon effects on static thermodynamic properties.

\section{Stability of the paramagnetic state against fluctuations}
\label{appendix:local_stability_of_phases}

\begin{figure*}
    \onecolumngrid
  \centering
  \includegraphics[width=\linewidth]{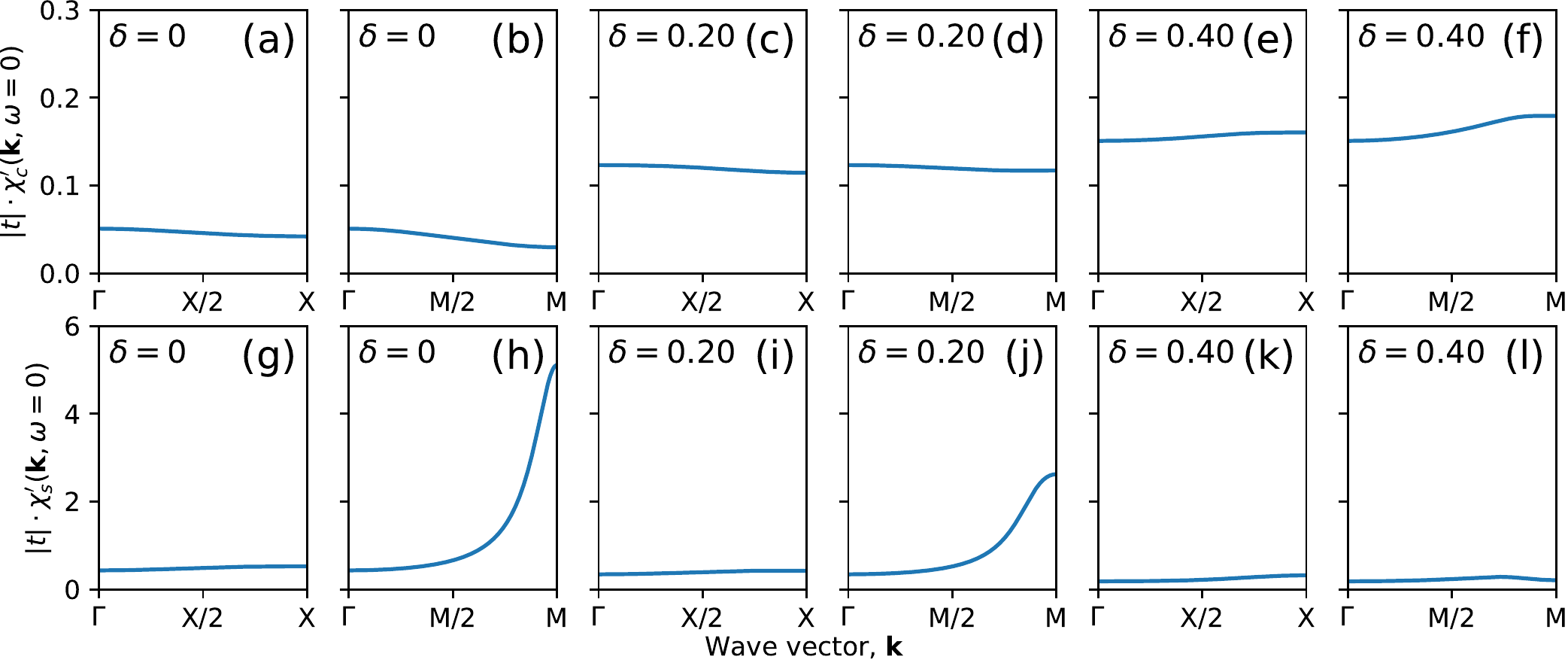}
  \caption{Stability analysis of the paramagnetic state for the case of zero next-nearest hopping. The panels show  calculated $\mathrm{SGA}_{\lambda_d}$+$1/\mathcal{N}_f$ static spin [panels (a)-(f)] and charge [panels (g)-(l)] susceptibility ($\chi_s^\prime(\mathbf{k}, \omega = 0)$ and $\chi_c^\prime(\mathbf{k}, \omega = 0)$, respectively) for the Hubbard model with the same parameters as those used to generate Fig.~\ref{fig:benchmatk-ld-ph_symm}, i.e., $t < 0$, $t^\prime = 0$, $U = 8|t|$, lattice size $500 \times 500$, and doping levels $\delta = 0, 0.2$, and $0.4$. The temperature has been set to $k_B T = 0.35 |t|$ for $\delta = 0$, and $k_B T = 0.333 |t|$ for $\delta = 0.2$ and $\delta = 0.4$. Hole doping levels are provided inside the panels.}
  \label{fig:benchmatk-ld-ph_symm-real}
  \twocolumngrid
\end{figure*}

\begin{figure*}
    \onecolumngrid
  \centering
  \includegraphics[width=\linewidth]{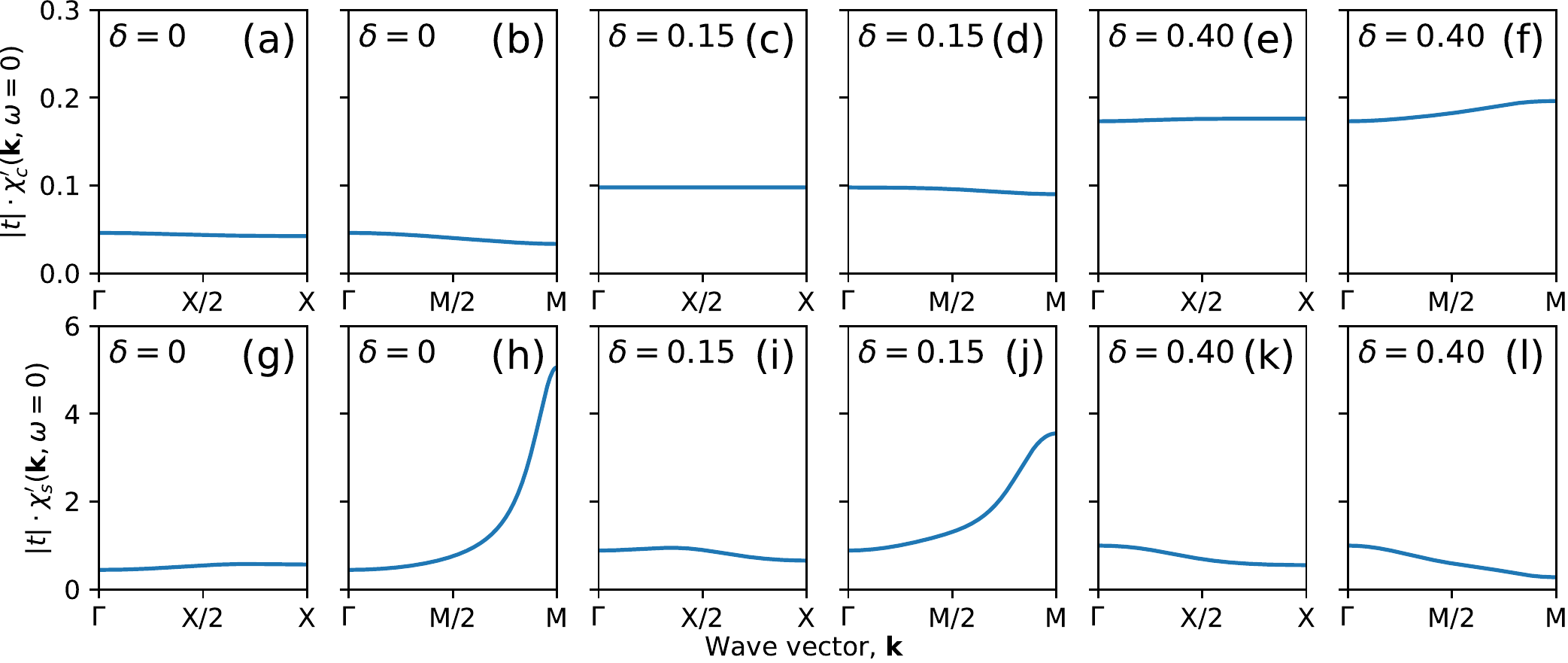}
  \caption{Stability analysis of the paramagnetic state for the case of non-zero next-nearest hopping. The panels show calculated  $\mathrm{SGA}_{\lambda_d}$+$1/\mathcal{N}_f$ static spin [panels (a)-(f)] and charge [panels (g)-(l)] susceptibility ($\chi_s^\prime(\mathbf{k}, \omega = 0)$ and $\chi_c^\prime(\mathbf{k}, \omega = 0)$, respectively) for the Hubbard model with the same parameters as those used to generate Fig.~\ref{fig:benchmatk-ld}, i.e., $t < 0$, $t^\prime = 0.3|t|$, $U = 8|t|$, lattice size $300 \times 300$, and hole doping  $\delta = 0, 0.15$, and $0.4$. The temperature has been set to $k_B T = 0.35 |t|$ for $\delta = 0$ and to $k_B T = 0.333 |t|$ otherwise. Hole doping levels are marked inside the panes.}
  \label{fig:benchmark-ld-real}
  \twocolumngrid
\end{figure*}

In this Appendix, we discuss the stability of $\mathrm{SGA}_{\lambda_d}$+$1/\mathcal{N}_f$ paramagnetic state against both charge- and spin excitations for the model parameters employed in Sec.~\ref{section:benchmark_Hubabrd_model} and \ref{section:robust_collective_excitations}. In Fig.~\ref{fig:benchmatk-ld-ph_symm-real}, we show static spin [panels (a-f)] and charge [panels (g-l)] susceptibility ($\chi_s^\prime(\mathbf{k}, \omega = 0)$ and $\chi_c^\prime(\mathbf{k}, \omega = 0)$, respectively) for the model parameters, doping levels, and temperatures the same as those used to generate Fig.~\ref{fig:benchmatk-ld}. As is apparent from Fig.~\ref{fig:benchmark-ld-real}, the static response is finite for all panels. This indicates \emph{local stability} of the paramagnetic phase against small perturbations with spatial modulation along high-symmetry directions. The charge susceptibility is strongly suppressed at half-filling and grows in magnitude with doping. It suggests that charge degrees of freedom are, to a degree, quenched at $\delta = 0$. The spin susceptibility exhibits an opposite tendency with a substantial enhancement near the antiferromagnetic $M$ point at $\delta = 0$. An analogous analysis for the particle-hole symmetric Hubbard model for the parameters equivalent to those employed for Fig.~\ref{fig:benchmatk-ld-ph_symm} in the main text, is summarized in Fig.~\ref{fig:benchmatk-ld-ph_symm-real}. Local stability against spin and charge excitations is apparent. We point out that a comprehensive analysis of magnetic-only instabilities in the Hubbard model within random-phase approximation and slave-boson technique was carried out previously,\cite{IgoshevPhysRevB2010,IgoshevJPCM2015} revealing variety of incommensurate ordering both at hole- and electron-doping sides. In the present study we have selected high temperatures to stay clear of those orderings. Lowering temperature to $k_B T = 0.333$ at $\delta = 0$ results in a divergence of spin susceptibility around $M$ point in both considered situations.

\begin{figure}
  \centering
  \includegraphics[width=\linewidth]{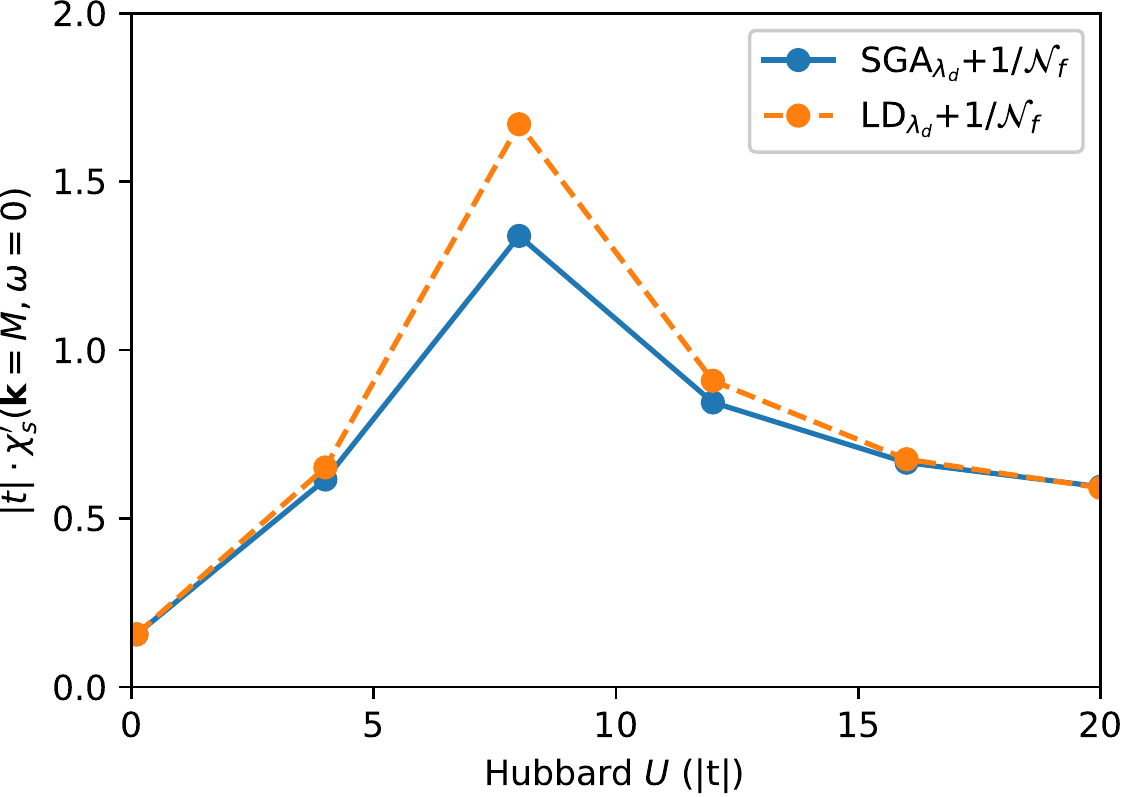} 
  \caption{Calculated static spin susceptibility at the $M$ point for the Hubbard model ($t < 0$, $t^\prime = 0.3|t|$, $k_B T = 0.35 |t|$, hole doping $\delta = 0.2$, and lattice $300 \times 300$) as a function of the on-site Coulomb repulsion $U$. Blue and yellow point correspond to $\mathrm{SGA}_{\lambda_f}$+$1/\mathcal{N}_f$ and $\mathrm{LD}_{\lambda_f}$+$1/\mathcal{N}_f$, respectively. Lines are guides to the eye. The maximum is located in the regime of metal-insulator transition $U \approx 8 |t|$.}
  \label{fig:susc_stability_var_u}
\end{figure}

In Fig.~\ref{fig:susc_stability_var_u} we show the calculated dependence of the static spin susceptibility on the on-site Coulomb repulsion, $U$, for the Hubbard model. We plot only the data for the point $M$ as the representative one in the context of magnetic instability. Model parameters are the same parameters as those employed to generate Fig.~\ref{fig:susc_vs_u_comparison}. Blue- and yellow points represent $\mathrm{SGA}_{\lambda_f}$+$1/\mathcal{N}_f$ and $\mathrm{LD}_{\lambda_f}$+$1/\mathcal{N}_f$ calculation, respectively. We have found, that the respective $\mathrm{SGA}_{x}$+$1/\mathcal{N}_f$ and $\mathrm{LD}_{x}$+$1/\mathcal{N}_f$ schemes (not shown in Fig.~\ref{fig:susc_stability_var_u}) yield the same values of the spin response for considered parameter set, yet minor differences between them appear in charge susceptibility channel. It is apparent that the susceptibility initially increases with $U$, whereas opposite behavior is seen above the crossover scale $U \sim 8 |t|$ of the order of bare bandwidth. The paramagnetic is thus locally stable against antiferromagnetic [$\mathbf{q} = (\pi, \pi)$] spin fluctuations from weak- to strong coupling. Interestingly, the $\mathrm{LD}_{\lambda_f}$+$1/\mathcal{N}_f$ yields systematically larger magnitude of static susceptibility than $\mathrm{SGA}_{\lambda_f}$+$1/\mathcal{N}_f$ truncation. This suggests that higher-order local correlations, captured by the diagrammatic contributions in local diagrammatic scheme, are non-negligible in the context of static quantities. 


%

\end{document}